\documentclass[longauth]{aa}

\usepackage{xcolor}
\usepackage{graphicx}
\usepackage{txfonts}
\usepackage{hyperref}
\usepackage{multirow}
\usepackage{amsmath,amssymb,amsfonts}
\usepackage{enumitem}

\newcommand{\orcit}[1]{\protect\href{https://orcid.org/#1}{\protect\includegraphics[width=8pt]{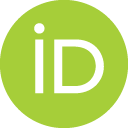}}}

\makeatletter
\renewcommand*\maketitle{%
  \thispagestyle{firstpage}
\begingroup
    \if@wideboxfn
    \setlength\bibindent{1.4\parindent}
    \else
    \setlength\bibindent{\parindent}
    \fi
    \renewcommand*\thefootnote{\@fnsymbol\c@footnote}%
    \renewcommand\@makefntext[1]{%
    \ifaa@longfn\hsize\textwidth\fi
    \noindent
    \hb@xt@\bibindent{\hss\@makefnmark\enspace}##1}
  \ifaa@twocolumn
  \begingroup
    \begin{aa@strip}
          \aa@maketitle
    \end{aa@strip}
    \@thanks            
  \endgroup
  \else
    \begingroup
      \let\thanks\footnote
      \aa@maketitle
    \endgroup
  \fi
\endgroup
  \setcounter{footnote}{0}
}
\makeatother

\makeatletter
\renewcommand*\aa@pageof{, page \thepage{} of \pageref*{LastPage}}
\makeatother

\newcommand{\secref}[1]{Section~\ref{#1}}
\newcommand{\tabref}[1]{Table~\ref{#1}}
\newcommand{\figref}[1]{Figure~\ref{#1}}

\def\deg{\ensuremath{^\circ}}
\def\sqdeg{\ensuremath{\,\rm deg}$^2$}

\def\arcsec{\ensuremath{''}}

\newcommand{\modulename}[1]{#1\xspace}
\newcommand{\apsis}{\modulename{Apsis}}
\newcommand{\smsgen}{\modulename{SMSgen}}
\newcommand{\dsc}{\modulename{DSC}}
\newcommand{\gspphot}{\modulename{GSP-Phot}}

\newcommand{\ugc}{\modulename{UGC}}
\newcommand{\oa}{\modulename{OA}}

\newcommand{\qsoc}{\modulename{QSOC}}
\newcommand{\tge}{\modulename{TGE}}

\renewcommand*\vec[1]{\ensuremath{\boldsymbol{#1}}}
\newcommand{\matfont}[1]{\ensuremath{\boldsymbol{\mathsf{#1}}}}
\newcommand{\mat}[1]{\matfont{#1}}

\newcommand{\prob}[2]{\ensuremath{{P^#1_#2}}}
\newcommand{\prior}[2]{\ensuremath{{\pi^#1_#2}}}

\newcommand{\gaia}{Gaia\xspace}
\newcommand{\gdr}[1]{\gaia~DR#1}
\newcommand{\teff}{\ensuremath{T_{\rm eff}}\xspace}
\newcommand{\bporrp}{BP/RP\xspace}
\newcommand{\gmag}{\ensuremath{G}\xspace}
\newcommand{\gbp}{\ensuremath{G_{\rm BP}}\xspace}
\newcommand{\grp}{\ensuremath{G_{\rm RP}}\xspace}
\newcommand{\gminr}{\ensuremath{G-G_{\rm RP}}\xspace}
\newcommand{\bming}{\ensuremath{G_{\rm BP}-G}\xspace}

\def\a0{\ensuremath{A_{\rm 0}}\xspace}

\newcommand{\bpminrp}{\ensuremath{G_\mathrm{BP}-G_\mathrm{RP}}\xspace}

\newcommand{\hip}{Hipparcos\xspace}
\newcommand{\tyc}{Tycho\xspace}
\newcommand{\tyctwo}{Tycho-2\xspace}

\makeatletter
\DeclareRobustCommand*{\fieldName}[1]{%
  \begingroup\@fieldName\scantokens{\texttt{\small {#1}}\noexpand}\endgroup}
\begingroup\lccode`\~=`\_\relax
   \lowercase{\endgroup\def\@fieldName{\catcode`\_=\active \let~\_}}
\makeatother
\newcommand{\linktodoc}{https://gea.esac.esa.int/archive/documentation/GDR3}

\newcommand{\linktoMainTable}[1]{\href{\linktodoc/Gaia_archive/chap_datamodel/sec_dm_main_source_catalogue/ssec_dm_#1.html}{\fieldName{#1}\xspace}}
\newcommand{\linktoAPTable}[1]{\href{\linktodoc/Gaia_archive/chap_datamodel/sec_dm_astrophysical_parameter_tables/ssec_dm_#1.html}{\fieldName{#1}\xspace}}
\newcommand{\linktoEGTable}[1]{\href{\linktodoc/Gaia_archive/chap_datamodel/sec_dm_extra--galactic_tables/ssec_dm_#1.html}{\fieldName{#1}\xspace}}

\newcommand{\linktoMainParam}[2]{\href{\linktodoc/Gaia_archive/chap_datamodel/sec_dm_main_source_catalogue/ssec_dm_#1.html\##1-#2}{\fieldName{#2}\xspace}}
\newcommand{\linktoAPParam}[2]{\href{\linktodoc/Gaia_archive/chap_datamodel/sec_dm_astrophysical_parameter_tables/ssec_dm_#1.html\##1-#2}{\fieldName{#2}\xspace}}
\newcommand{\linktoEGParam}[2]{\href{\linktodoc/Gaia_archive/chap_datamodel/sec_dm_extra--galactic_tables/ssec_dm_#1.html\##1-#2}{\fieldName{#2}\xspace}}

\newcommand{\linktosec}[3]{\href{\linktodoc/Data_analysis/chap_#1/sec_#1_#2/ssec_#1_#2_#3.html}{online documentation}}

\begin{document}

   \title{\gdr{3}: \apsis III -  Non-stellar content and source classification}
   \authorrunning{L. Delchambre et al.}

\author{
L.~                    Delchambre\orcit{0000-0003-2559-408X}
\inst{\ref{inst:0001}}
\and     C.A.L.~                  Bailer-Jones
\inst{\ref{inst:0002}}
\and         I.~                Bellas-Velidis
\inst{\ref{inst:0003}}
\and         R.~                       Drimmel\orcit{0000-0002-1777-5502}
\inst{\ref{inst:0004}}
\and         D.~                      Garabato\orcit{0000-0002-7133-6623}
\inst{\ref{inst:0005}}
\and         R.~                      Carballo\orcit{0000-0001-7412-2498}
\inst{\ref{inst:0006}}
\and         D.~                Hatzidimitriou\orcit{0000-0002-5415-0464}
\inst{\ref{inst:0007},\ref{inst:0003}}
\and       D.J.~                      Marshall\orcit{0000-0003-3956-3524}
\inst{\ref{inst:0009}}
\and         R.~                        Andrae\orcit{0000-0001-8006-6365}
\inst{\ref{inst:0002}}
\and         C.~                       Dafonte\orcit{0000-0003-4693-7555}
\inst{\ref{inst:0005}}
\and         E.~                       Livanou\orcit{0000-0003-0628-2347}
\inst{\ref{inst:0007}}
\and         M.~                     Fouesneau\orcit{0000-0001-9256-5516}
\inst{\ref{inst:0002}}
\and       E.L.~                        Licata\orcit{0000-0002-5203-0135}
\inst{\ref{inst:0004}}
\and     H.E.P.~                  Lindstr{\o}m
\inst{\ref{inst:0004},\ref{inst:0016},\ref{inst:0017}}
\and         M.~                      Manteiga\orcit{0000-0002-7711-5581}
\inst{\ref{inst:0018}}
\and         C.~                         Robin
\inst{\ref{inst:0019}}
\and         A.~                       Silvelo\orcit{0000-0002-5126-6365}
\inst{\ref{inst:0005}}
\and         A.~                Abreu Aramburu
\inst{\ref{inst:0021}}
\and       M.A.~                   \'{A}lvarez\orcit{0000-0002-6786-2620}
\inst{\ref{inst:0005}}
\and         J.~                       Bakker
\inst{\ref{inst:0041}}
\and         A.~                       Bijaoui
\inst{\ref{inst:0023}}
\and         N.~                     Brouillet\orcit{0000-0002-3274-7024}
\inst{\ref{inst:0024}}
\and         E.~                    Brugaletta\orcit{0000-0003-2598-6737}
\inst{\ref{inst:0025}}
\and         A.~                       Burlacu
\inst{\ref{inst:0026}}
\and         L.~                   Casamiquela\orcit{0000-0001-5238-8674}
\inst{\ref{inst:0024},\ref{inst:0028}}
\and         L.~                        Chaoul
\inst{\ref{inst:0029}}
\and         A.~                     Chiavassa\orcit{0000-0003-3891-7554}
\inst{\ref{inst:0023}}
\and         G.~                      Contursi\orcit{0000-0001-5370-1511}
\inst{\ref{inst:0023}}
\and       W.J.~                        Cooper\orcit{0000-0003-3501-8967}
\inst{\ref{inst:0032},\ref{inst:0004}}
\and       O.L.~                       Creevey\orcit{0000-0003-1853-6631}
\inst{\ref{inst:0023}}
\and         A.~                    Dapergolas
\inst{\ref{inst:0003}}
\and         P.~                    de Laverny\orcit{0000-0002-2817-4104}
\inst{\ref{inst:0023}}
\and         C.~                      Demouchy
\inst{\ref{inst:0037}}
\and       T.E.~                 Dharmawardena\orcit{0000-0002-9583-5216}
\inst{\ref{inst:0002}}
\and         B.~                    Edvardsson
\inst{\ref{inst:0039}}
\and         Y.~                    Fr\'{e}mat\orcit{0000-0002-4645-6017}
\inst{\ref{inst:0040}}
\and         P.~              Garc\'{i}a-Lario\orcit{0000-0003-4039-8212}
\inst{\ref{inst:0041}}
\and         M.~             Garc\'{i}a-Torres\orcit{0000-0002-6867-7080}
\inst{\ref{inst:0042}}
\and         A.~                         Gavel\orcit{0000-0002-2963-722X}
\inst{\ref{inst:0043}}
\and         A.~                         Gomez\orcit{0000-0002-3796-3690}
\inst{\ref{inst:0005}}
\and         I.~   Gonz\'{a}lez-Santamar\'{i}a\orcit{0000-0002-8537-9384}
\inst{\ref{inst:0005}}
\and         U.~                        Heiter\orcit{0000-0001-6825-1066}
\inst{\ref{inst:0043}}
\and         A.~          Jean-Antoine Piccolo\orcit{0000-0001-8622-212X}
\inst{\ref{inst:0029}}
\and         M.~                      Kontizas\orcit{0000-0001-7177-0158}
\inst{\ref{inst:0007}}
\and         G.~                    Kordopatis\orcit{0000-0002-9035-3920}
\inst{\ref{inst:0023}}
\and       A.J.~                          Korn\orcit{0000-0002-3881-6756}
\inst{\ref{inst:0043}}
\and       A.C.~                     Lanzafame\orcit{0000-0002-2697-3607}
\inst{\ref{inst:0025},\ref{inst:0052}}
\and         Y.~                      Lebreton\orcit{0000-0002-4834-2144}
\inst{\ref{inst:0053},\ref{inst:0054}}
\and         A.~                         Lobel\orcit{0000-0001-5030-019X}
\inst{\ref{inst:0040}}
\and         A.~                         Lorca
\inst{\ref{inst:0056}}
\and         A.~               Magdaleno Romeo
\inst{\ref{inst:0026}}
\and         F.~                       Marocco\orcit{0000-0001-7519-1700}
\inst{\ref{inst:0058}}
\and         N.~                          Mary
\inst{\ref{inst:0019}}
\and         C.~                       Nicolas
\inst{\ref{inst:0029}}
\and         C.~                     Ordenovic
\inst{\ref{inst:0023}}
\and         F.~                       Pailler\orcit{0000-0002-4834-481X}
\inst{\ref{inst:0029}}
\and       P.A.~                       Palicio\orcit{0000-0002-7432-8709}
\inst{\ref{inst:0023}}
\and         L.~               Pallas-Quintela\orcit{0000-0001-9296-3100}
\inst{\ref{inst:0005}}
\and         C.~                         Panem
\inst{\ref{inst:0029}}
\and         B.~                        Pichon\orcit{0000 0000 0062 1449}
\inst{\ref{inst:0023}}
\and         E.~                        Poggio\orcit{0000-0003-3793-8505}
\inst{\ref{inst:0023},\ref{inst:0004}}
\and         A.~                  Recio-Blanco\orcit{0000-0002-6550-7377}
\inst{\ref{inst:0023}}
\and         F.~                        Riclet
\inst{\ref{inst:0029}}
\and         J.~                       Rybizki\orcit{0000-0002-0993-6089}
\inst{\ref{inst:0002}}
\and         R.~                 Santove\~{n}a\orcit{0000-0002-9257-2131}
\inst{\ref{inst:0005}}
\and       L.M.~                         Sarro\orcit{0000-0002-5622-5191}
\inst{\ref{inst:0073}}
\and       M.S.~                    Schultheis\orcit{0000-0002-6590-1657}
\inst{\ref{inst:0023}}
\and         M.~                         Segol
\inst{\ref{inst:0037}}
\and         I.~                        Slezak
\inst{\ref{inst:0023}}
\and       R.L.~                         Smart\orcit{0000-0002-4424-4766}
\inst{\ref{inst:0004}}
\and         R.~                         Sordo\orcit{0000-0003-4979-0659}
\inst{\ref{inst:0078}}
\and         C.~                      Soubiran\orcit{0000-0003-3304-8134}
\inst{\ref{inst:0024}}
\and         M.~                  S\"{ u}veges\orcit{0000-0003-3017-5322}
\inst{\ref{inst:0080}}
\and         F.~                  Th\'{e}venin
\inst{\ref{inst:0023}}
\and         G.~                Torralba Elipe\orcit{0000-0001-8738-194X}
\inst{\ref{inst:0005}}
\and         A.~                          Ulla\orcit{0000-0001-6424-5005}
\inst{\ref{inst:0083}}
\and         E.~                          Utrilla
\inst{\ref{inst:0056}}
\and         A.~                     Vallenari\orcit{0000-0003-0014-519X}
\inst{\ref{inst:0078}}
\and         E.~                    van Dillen
\inst{\ref{inst:0037}}
\and         H.~                          Zhao\orcit{0000-0003-2645-6869}
\inst{\ref{inst:0023}}
\and         J.~                         Zorec
\inst{\ref{inst:0087}}
}
\institute{
     Institut d'Astrophysique et de G\'{e}ophysique, Universit\'{e} de Li\`{e}ge, 19c, All\'{e}e du 6 Ao\^{u}t, B-4000 Li\`{e}ge, Belgium\relax\label{inst:0001}
\and Max Planck Institute for Astronomy, K\"{ o}nigstuhl 17, 69117 Heidelberg, Germany\relax\label{inst:0002}\vfill
\and National Observatory of Athens, I. Metaxa and Vas. Pavlou, Palaia Penteli, 15236 Athens, Greece\relax\label{inst:0003}\vfill
\and INAF - Osservatorio Astrofisico di Torino, via Osservatorio 20, 10025 Pino Torinese (TO), Italy\relax\label{inst:0004}\vfill
\and CIGUS CITIC - Department of Computer Science and Information Technologies, University of A Coru\~{n}a, Campus de Elvi\~{n}a s/n, A Coru\~{n}a, 15071, Spain\relax\label{inst:0005}\vfill
\and Dpto. de Matem\'{a}tica Aplicada y Ciencias de la Computaci\'{o}n, Univ. de Cantabria, ETS Ingenieros de Caminos, Canales y Puertos, Avda. de los Castros s/n, 39005 Santander, Spain\relax\label{inst:0006}\vfill
\and Department of Astrophysics, Astronomy and Mechanics, National and Kapodistrian University of Athens, Panepistimiopolis, Zografos, 15783 Athens, Greece\relax\label{inst:0007}\vfill
\and IRAP, Universit\'{e} de Toulouse, CNRS, UPS, CNES, 9 Av. colonel Roche, BP 44346, 31028 Toulouse Cedex 4, France\relax\label{inst:0009}\vfill
\and Niels Bohr Institute, University of Copenhagen, Juliane Maries Vej 30, 2100 Copenhagen {\O}, Denmark\relax\label{inst:0016}\vfill
\and DXC Technology, Retortvej 8, 2500 Valby, Denmark\relax\label{inst:0017}\vfill
\and CIGUS CITIC, Department of Nautical Sciences and Marine Engineering, University of A Coru\~{n}a, Paseo de Ronda 51, 15071, A Coru\~{n}a, Spain\relax\label{inst:0018}\vfill
\and Thales Services for CNES Centre Spatial de Toulouse, 18 avenue Edouard Belin, 31401 Toulouse Cedex 9, France\relax\label{inst:0019}\vfill
\and ATG Europe for European Space Agency (ESA), Camino bajo del Castillo, s/n, Urbanizacion Villafranca del Castillo, Villanueva de la Ca\~{n}ada, 28692 Madrid, Spain\relax\label{inst:0021}\vfill
\and Universit\'{e} C\^{o}te d'Azur, Observatoire de la C\^{o}te d'Azur, CNRS, Laboratoire Lagrange, Bd de l'Observatoire, CS 34229, 06304 Nice Cedex 4, France\relax\label{inst:0023}\vfill
\and Laboratoire d'astrophysique de Bordeaux, Univ. Bordeaux, CNRS, B18N, all{\'e}e Geoffroy Saint-Hilaire, 33615 Pessac, France\relax\label{inst:0024}\vfill
\and INAF - Osservatorio Astrofisico di Catania, via S. Sofia 78, 95123 Catania, Italy\relax\label{inst:0025}\vfill
\and Telespazio for CNES Centre Spatial de Toulouse, 18 avenue Edouard Belin, 31401 Toulouse Cedex 9, France\relax\label{inst:0026}\vfill
\and GEPI, Observatoire de Paris, Universit\'{e} PSL, CNRS, 5 Place Jules Janssen, 92190 Meudon, France\relax\label{inst:0028}\vfill
\and CNES Centre Spatial de Toulouse, 18 avenue Edouard Belin, 31401 Toulouse Cedex 9, France\relax\label{inst:0029}\vfill
\and Centre for Astrophysics Research, University of Hertfordshire, College Lane, AL10 9AB, Hatfield, United Kingdom\relax\label{inst:0032}\vfill
\and APAVE SUDEUROPE SAS for CNES Centre Spatial de Toulouse, 18 avenue Edouard Belin, 31401 Toulouse Cedex 9, France\relax\label{inst:0037}\vfill
\and Theoretical Astrophysics, Division of Astronomy and Space Physics, Department of Physics and Astronomy, Uppsala University, Box 516, 751 20 Uppsala, Sweden\relax\label{inst:0039}\vfill
\and Royal Observatory of Belgium, Ringlaan 3, 1180 Brussels, Belgium\relax\label{inst:0040}\vfill
\and European Space Agency (ESA), European Space Astronomy Centre (ESAC), Camino bajo del Castillo, s/n, Urbanizacion Villafranca del Castillo, Villanueva de la Ca\~{n}ada, 28692 Madrid, Spain\relax\label{inst:0041}\vfill
\and Data Science and Big Data Lab, Pablo de Olavide University, 41013, Seville, Spain\relax\label{inst:0042}\vfill
\and Observational Astrophysics, Division of Astronomy and Space Physics, Department of Physics and Astronomy, Uppsala University, Box 516, 751 20 Uppsala, Sweden\relax\label{inst:0043}\vfill
\and Dipartimento di Fisica e Astronomia ""Ettore Majorana"", Universit\`{a} di Catania, Via S. Sofia 64, 95123 Catania, Italy\relax\label{inst:0052}\vfill
\and LESIA, Observatoire de Paris, Universit\'{e} PSL, CNRS, Sorbonne Universit\'{e}, Universit\'{e} de Paris, 5 Place Jules Janssen, 92190 Meudon, France\relax\label{inst:0053}\vfill
\and Universit\'{e} Rennes, CNRS, IPR (Institut de Physique de Rennes) - UMR 6251, 35000 Rennes, France\relax\label{inst:0054}\vfill
\and Aurora Technology for European Space Agency (ESA), Camino bajo del Castillo, s/n, Urbanizacion Villafranca del Castillo, Villanueva de la Ca\~{n}ada, 28692 Madrid, Spain\relax\label{inst:0056}\vfill
\and IPAC, Mail Code 100-22, California Institute of Technology, 1200 E. California Blvd., Pasadena, CA 91125, USA\relax\label{inst:0058}\vfill
\and Dpto. de Inteligencia Artificial, UNED, c/ Juan del Rosal 16, 28040 Madrid, Spain\relax\label{inst:0073}\vfill
\and INAF - Osservatorio astronomico di Padova, Vicolo Osservatorio 5, 35122 Padova, Italy\relax\label{inst:0078}\vfill
\and Institute of Global Health, University of Geneva\relax\label{inst:0080}\vfill
\and Applied Physics Department, Universidade de Vigo, 36310 Vigo, Spain\relax\label{inst:0083}\vfill
\and Sorbonne Universit\'{e}, CNRS, UMR7095, Institut d'Astrophysique de Paris, 98bis bd. Arago, 75014 Paris, France\relax\label{inst:0087}\vfill
}

\date{Received 25 February 2022 / Accepted 30 May 2022}

 
  \abstract
   {As part of the third \gaia data release, we present the contributions of the non-stellar and classification modules from the eighth coordination unit (CU8) of the Data Processing and Analysis Consortium, which is responsible for the determination of source astrophysical parameters using \gaia\ data. This is the third in a series of three papers describing the work done within CU8 for this release.}
   {For each of the five relevant modules from CU8, we summarise their objectives, the methods they employ, their performance, and the results they produce for \gdr{3}. We further advise how to use these data products and highlight some limitations.} 
   {The Discrete Source Classifier (\dsc) module provides classification probabilities associated with five types of sources: quasars, galaxies, stars, white dwarfs, and physical binary stars. A subset of these sources are processed by the Outlier Analysis (OA) module, which performs an unsupervised clustering analysis, and then associates labels with the clusters to complement the \dsc classification. The Quasi Stellar Object Classifier (\qsoc) and the Unresolved Galaxy Classifier (\ugc)  determine the redshifts of the sources classified as quasar and galaxy by the \dsc module. Finally, the Total Galactic Extinction (TGE) module uses the extinctions of individual stars determined by another CU8 module to determine the asymptotic extinction along all lines of sight for Galactic latitudes $|b| > 5\deg$.}
   {\gdr{3} includes 1591 million sources with \dsc classifications; 56 million sources to which the \oa clustering is applied; 1.4 million sources with redshift estimates from \ugc; 6.4 million sources with \qsoc redshift; and 3.1 million level 9 HEALPixes of size $0.013$ \sqdeg \,where the extinction is evaluated by \tge. 
   }
  {Validation shows that results are in good agreement with values from external catalogues; for example   $90\%$ of the \qsoc redshifts have absolute error lower than $0.1$ for sources with empty warning flags, while \ugc redshifts have a mean error of $0.008 \pm 0.037$ if evaluated on a clean set of spectra.
  An internal validation of the \oa results further shows that $30$ million sources are located in high confidence regions of the clustering map.
  }


\maketitle




\section{Introduction}
\label{sec:introduction}

The ESA \gaia mission was designed to create the most precise three dimensional map of the Milky way, along with its kinematics, through the repeated observation of about two billion stars. \gaia\ observes all objects in the sky down to an apparent $G$ magnitude of about $21$ mag, which includes millions of galaxies and quasars. \citep{DR1-DPACP-18}. The data collected between 25 July 2014 and 28 May 2017 (34 months) have been processed by the \gaia Data Processing and Analysis Consortium (DPAC) to provide the third data release of the \gaia catalogue,  \gdr{3}. 

For sources with $\gmag \leq 17$ mag, typical positional uncertainties are on the order of $80$ $\mu$as; parallax uncertainties on the order of $100$ $\mu$as; proper motion uncertainties on the order of $100$ $\mu$as yr$^{-1}$; and $G$ magnitude uncertainties on the order of $1$ mmag. In addition to this exquisite astrometric and photometric performance, \gaia provides high-resolution spectroscopy ($R = \lambda/\Delta \lambda \approx 11700$) centred around the calcium triplet ($845$--$872$ nm), hence its name radial velocity spectrometer (RVS), as well as low-resolution spectrophotometry from two instruments: the blue photometer (BP) covering the wavelength range $330$--$680$ nm with $30 \leq R \leq 100$, and the red photometer (RP) covering the wavelength range $640$--$1050$ nm with $70 \leq R \leq 100$ \citep{2021A&A...652A..86C}.

Eight coordination units (CUs) were set up within the DPAC, each focusing on a particular aspect of the \gaia processing: CU1 for managing the computer architecture; CU2 for the data simulations; CU3 for the core astrometric processing; CU4 for the analysis of non-single stars, Solar System objects, and extended objects; CU5 for the photometric \bporrp processing; CU6 for the spectroscopic RVS processing; CU7 for the variability analysis; and CU8 for the determination of the astrophysical parameters (APs) of the observed sources. Finally, a ninth CU is responsible for the catalogue validation, access, and publication.

This paper is the third in a series of three papers describing the processing done within CU8. The first of these, \cite{DR3-DPACP-157}, summarises the work done in CU8 and the various APs it produces. The second, \cite{DR3-DPACP-160}, describes stellar APs. The present paper discusses the object classification and the non-stellar APs produced by CU8, namely the redshifts of extragalactic sources and total Galactic extinction map. We describe the results and methods of the relevant modules, as they have evolved since their description given prior to launch \citep{Apsis2013}, while focusing on technical details. A thorough scientific analysis of these results, seen from a cross-CU perspective, can be found in performance verification papers like in \cite{DR3-DPACP-101}, where the classification and characterisation of the extragalactic sources are discussed in more details.


We provide an overview of the data products from the classification and non-stellar modules in Section \ref{sec:overview}. The Discrete Source Classifier (\dsc), which classifies sources probabilistically into five classes that are known a priori from its training set (quasar, galaxy, star, white dwarf, and physical binary star), is described in Section \ref{sec:dsc}. The Outlier Analysis (\oa), which complements the DSC classification through a clustering algorithm applied to \bporrp spectra of sources with low \dsc probability, is described in Section \ref{sec:oa}. The quasar classifier (QSOC) and Unresolved Galaxy Classifier (UGC), both based on \bporrp spectra, make use of the DSC probabilities in order to identify quasars and galaxies and subsequently determine their redshifts; these are described in Sections \ref{sec:qsoc} and \ref{sec:ugc}, respectively. Finally, the global stellar parameters of giant stars, as inferred from \bporrp spectra, allow the Total Galactic Extinction (TGE) module to derive the Galactic extinction seen along a given line-of-sight as described in Section \ref{sec:tge}. Finally, we summarise the improvements that are currently foreseen for \gdr{4} in \secref{sec:beyond_gdr3}.
Additional information on the design and performance of the modules can be found in the \gaia online documentation\footnote{\href{\linktodoc/index.html}{\linktodoc/index.html}}.
\section{Overview of the non-stellar astrophysical parameters from CU8 in Gaia DR3}
\label{sec:overview}

\begin{figure*}[t!]
    \centering
    \includegraphics[width=0.9\textwidth]{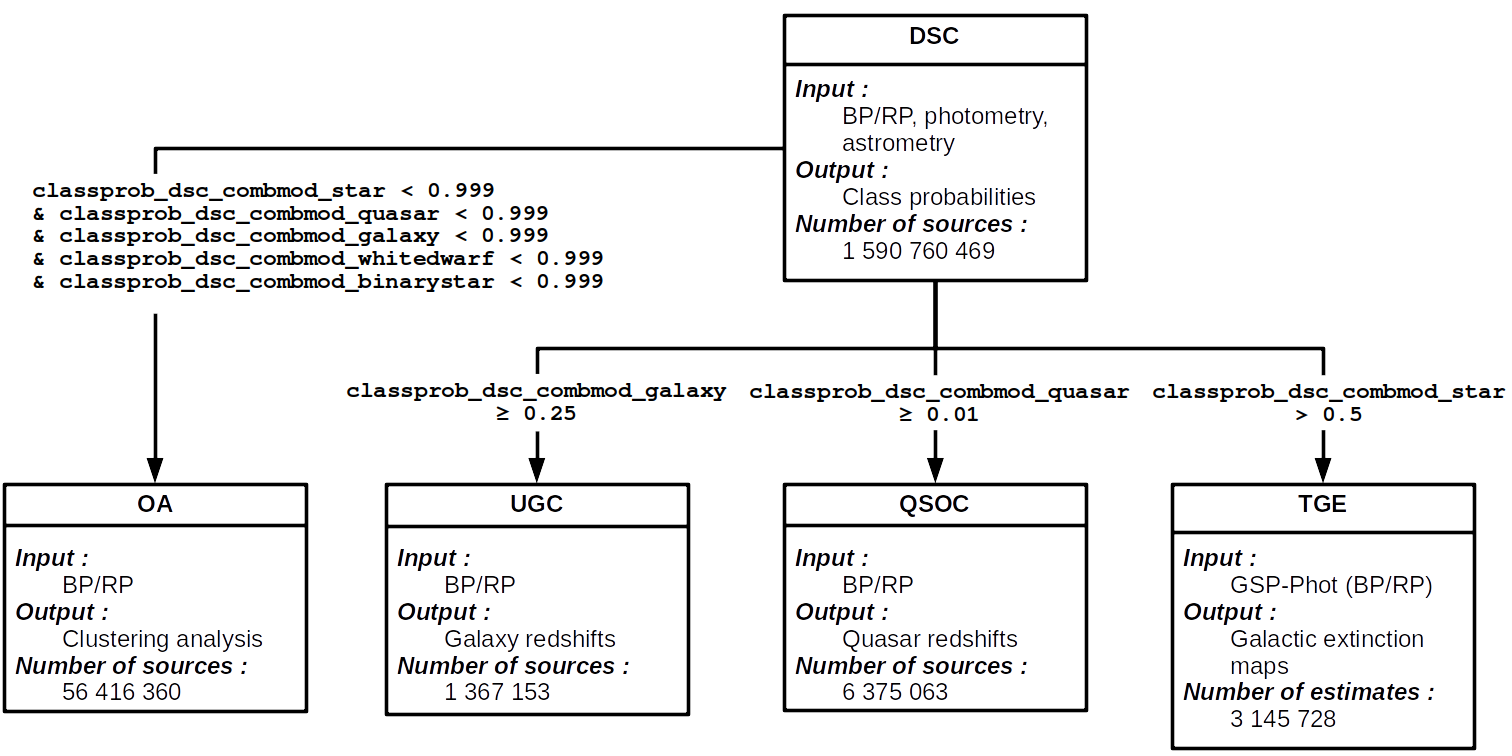}
    \caption{Dependency of the \oa, \ugc, \qsoc, and \tge modules on the \dsc combined probabilities for the selection of the sources to be processed (\texttt{classprob\_dsc\_combod}, see \secref{sec:dsc} for a definition). For each module, we provide a synthetic view of their input and output, and the number of sources for which the module produces results in \gdr{3}. In the case of \tge, we provide the number of extinction estimates that were computed in level 9 HEALPixes (see \secref{sec:tge}). Unlike the other modules described here, \tge additionally relies on the General Stellar Parametrizer from Photometry (\gspphot) for its source selection and processing, which is described in \cite{DR3-DPACP-156}.}
    \label{fig:overview}
\end{figure*}

The five non-stellar modules together contribute to $110$ unique fields in the \gdr{3}. Table \ref{tab:aps_overview} provides an overview  of the tables and fields that each of the modules contributes to, including the resulting number of entries in each table. These fields are spread over eight different tables and concern about $1.6$ billion unique sources. Figure \ref{fig:overview} sketches the inter-dependency between these modules, the selection they apply on the \dsc probabilities, their input, output, and the number of sources for which they produce results in \gdr{3}. The different selection policies from each module are clearly seen in this plot; each leads to a different associated completeness and purity. The filtering applied by each module on the results they produced is not mentioned here, although we should generally not expect the number of sources satisfying the provided \dsc selection criteria to be equal to the number of sources for which there are results in \gdr{3} for each module.

\begin{table*}
    \centering
    \caption{Individual contributions of the non-stellar CU8 modules to the \gdr{3}. See the sections dedicated to each module for a complete description of the fields and tables listed herein. Fields from module-specific tables (i.e. \oa and \tge) are not listed here. \label{tab:aps_overview} }
    \footnotesize
    \begin{tabular}{p{2cm}|p{8.5cm}p{4cm}}
    \hline
    Module & Table and field names & Number of non-empty rows\\
    \hline \hline
  \multirow{11}{2cm}{\dsc (source classification)}
    & -~\linktoAPTable{astrophysical_parameters} \\
      & \hspace{1cm}{\tt classprob\_dsc\_allosmod}$^a$ & 1\,370\,759\,105  \\ 
      & \hspace{1cm}{\tt classprob\_dsc\_specmod}$^b$, {\tt classprob\_dsc\_combmod}$^c$ & 1\,590\,760\,469 \\
    & -~\linktoMainTable{gaia_source} \\
      & \hspace{1cm} {\tt classprob\_dsc\_combmod}$^c$ & 1\,590\,760\,469 \\
    & -~\linktoEGTable{galaxy_candidates} \\
       & \hspace{1cm}{\tt classprob\_dsc\_combmod}$^c$, \linktoEGParam{galaxy_candidates}{classlabel_dsc}, & 4\,841\,799 \\ 
       & \hspace{1cm}\linktoEGParam{galaxy_candidates}{classlabel_dsc_joint} \\
    & -~\linktoEGTable{qso_candidates} \\ 
       & \hspace{1cm}{\tt classprob\_dsc\_combmod}$^c$, \linktoEGParam{qso_candidates}{classlabel_dsc}, & 6\,647\,511 \\
       & \hspace{1cm}\linktoEGParam{qso_candidates}{classlabel_dsc_joint} \\   
  \hline 
  \multirow{9}{2cm}{\oa (source classification based on self-organising map)} 
    & -~\linktoAPTable{oa_neuron_information} (78 fields) & 900 (1 per neuron) \\
    & -~\linktoAPTable{oa_neuron_xp_spectra} (7 fields) & 78\,300 (900 neurons $\times$ 87 samples per spectrum) \\
    & -~\linktoAPTable{astrophysical_parameters} \\
      & \hspace{1cm}\linktoAPParam{astrophysical_parameters}{neuron_oa_id}, \linktoAPParam{astrophysical_parameters}{neuron_oa_dist} & 56\,416\,360 \\
      & \hspace{1cm}\linktoAPParam{astrophysical_parameters}{neuron_oa_dist_percentile_rank}, \linktoAPParam{astrophysical_parameters}{flags_oa} \\
    & -~\linktoEGTable{galaxy_candidates} \\ 
      & \hspace{1cm}\linktoEGParam{galaxy_candidates}{classlabel_oa} & 1\,901\,026 \\
    & -~\linktoEGTable{qso_candidates} \\ 
      & \hspace{1cm} \linktoEGParam{qso_candidates}{classlabel_oa} & 2\,803\,225 \\
  \hline
  \multirow{4}{2cm}{\qsoc (quasar redshift determination)} 
    & -~\linktoEGTable{qso_candidates} \\ 
      & \hspace{1cm}\linktoEGParam{qso_candidates}{redshift_qsoc}, \linktoEGParam{qso_candidates}{redshift_qsoc_lower} & 6\,375\,063 \\
      & \hspace{1cm}\linktoEGParam{qso_candidates}{redshift_qsoc_upper}, \linktoEGParam{qso_candidates}{ccfratio_qsoc}, \\
      & \hspace{1cm} \linktoEGParam{qso_candidates}{zscore_qsoc}, \linktoEGParam{qso_candidates}{flags_qsoc}\\
  \hline
  \multirow{3}{2cm}{\ugc (galaxy redshift determination)} 
    & -~\linktoEGTable{galaxy_candidates} \\ 
      & \hspace{1cm}\linktoEGParam{galaxy_candidates}{redshift_ugc}, \linktoEGParam{galaxy_candidates}{redshift_ugc_lower}, & 1\,367\,153 \\
      & \hspace{1cm} \linktoEGParam{galaxy_candidates}{redshift_ugc_upper} \\
  \hline
  \multirow{2}{2cm}{\tge (Galactic extinction)}  
    & -~\linktoAPTable{total_galactic_extinction_map} (10 fields) & 4\,177\,920 (49\,152 in HEALPix level 6, 196\,608 in level 7, 786\,432 in level 8, 3\,145\,728 in level 9) \\
    & -~\linktoAPTable{total_galactic_extinction_map_opt} (7 fields) & 3\,145\,728 (HEALPix level 9) \\
  \hline
    \multicolumn{3}{p{15cm}}{$^a$ Corresponding to \linktoAPParam{astrophysical_parameters}{classprob_dsc_allosmod_quasar},  \linktoAPParam{astrophysical_parameters}{classprob_dsc_allosmod_galaxy} and \linktoAPParam{astrophysical_parameters}{classprob_dsc_allosmod_star}} \\
    \multicolumn{3}{p{15cm}}{$^b$ Corresponding to \linktoAPParam{astrophysical_parameters}{classprob_dsc_specmod_quasar},  \linktoAPParam{astrophysical_parameters}{classprob_dsc_specmod_galaxy},  \linktoAPParam{astrophysical_parameters}{classprob_dsc_specmod_star},  \linktoAPParam{astrophysical_parameters}{classprob_dsc_specmod_whitedwarf} and \linktoAPParam{astrophysical_parameters}{classprob_dsc_specmod_binarystar}}\\
    \multicolumn{3}{p{15cm}}{$^c$ Corresponding to \linktoAPParam{astrophysical_parameters}{classprob_dsc_combmod_quasar},  \linktoAPParam{astrophysical_parameters}{classprob_dsc_combmod_galaxy},  \linktoAPParam{astrophysical_parameters}{classprob_dsc_combmod_star},  \linktoAPParam{astrophysical_parameters}{classprob_dsc_combmod_whitedwarf} and \linktoAPParam{astrophysical_parameters}{classprob_dsc_combmod_binarystar}}
    \end{tabular}
\end{table*}

\section{Source classification (DSC)}
\label{sec:dsc}

\subsection{Objectives} \label{subsec:dsc_objective}

\dsc classifies \gaia sources probabilistically into five classes: quasar, galaxy, star, white dwarf, and physical binary star. These classes are defined by the training data, which are \gaia\ data, with labels provided by external catalogues. \dsc comprises three classifiers: Specmod uses \bporrp spectra to classify into all five classes; Allosmod uses various other features to classify into just the first three classes; Combmod takes the output class probabilities of the other two classifiers and combines them to give combined probabilities in all five classes.

\subsection{Method} \label{subsec:dsc_method}

\subsubsection{Algorithms and I/O}

Specmod uses an ExtraTrees classifier, which is an ensemble of classification trees. Each tree maps the 100-dimensional input space of the \bporrp spectrum ---60 samples each, minus 5 samples that are rejected at the edges of each spectrum--- into regions that are then identified with each of the five classes. By using an ensemble of hundreds of trees, these individual discrete classifications are turned into class probabilities.

Allosmod uses a Gaussian Mixture Model (GMM). For each class, the distribution of the training data in an eight-dimensional feature space is modelled by a mixture of 25 Gaussians. This is done independently for all three classes (quasar, galaxy, star).
Once appropriately normalised and a suitable prior applied, each GMM gives the probability that a feature vector (i.e.\ a new source) is of that class. The eight features are as follows; they are fields in the Gaia source table or are computed from these fields: 
\begin{itemize}[leftmargin=0.5cm]
    \item sine of the Galactic latitude, $\sin \linktoMainParam{gaia_source}{b}$,
    \item parallax, \linktoMainParam{gaia_source}{parallax},
    \item total proper motion, 
    \linktoMainParam{gaia_source}{pm},
    \item unit weight error (uwe), \\
    $=\sqrt{\frac{\linktoMainParam{gaia_source}{astrometric_chi2_al}}
    {\linktoMainParam{gaia_source}{astrometric_n_good_obs_al} - 5} }$,
    \item \gmag\ band magnitude, \linktoMainParam{gaia_source}{phot_g_mean_mag},
    \item colour \bming, \linktoMainParam{gaia_source}{bp_g},
    \item colour \gminr, \linktoMainParam{gaia_source}{g_rp},
    \item The relative variability in the \gmag\ band (relvarg), \\
    $=\sqrt{ \linktoMainParam{gaia_source}{phot_g_n_obs} / \linktoMainParam{gaia_source}{phot_g_mean_flux_over_error} }$.
\end{itemize}
All eight features must exist for a given source for Allosmod to provide a probability. 
As explained below, we exploit some of the `failures' of these features to help identify objects. For example, galaxies should have true proper motions (and parallaxes) very close to zero. Yet they sometimes have larger measured proper motions in \gdr{3} on account of their physical extent combined with the variability in the calculation of the centroid during each scan made by Gaia (obtained at different position angles). This can give rise to spuriously large proper motions (although the uncertainties are also larger). In many cases, these solutions are rejected by the astrometric solutions (to give the so-called 2p solutions; see
\citealt{2021A&A...649A...2L} for the definitions), meaning that many galaxies lack parallaxes and proper motions and are therefore not processed by Allosmod.

Allosmod models the distribution of the data, and so it provides likelihoods. When combined with the class prior, this gives posterior class probabilities, which are the output from Allosmod. Specmod, in contrast, is a tree-based model that does not strictly provide posterior probabilities. Moreover, its output is influenced by the distribution in the training data (see below). However, by using the simple method described in the \linktosec{cu8par}{apsis}{dsc} we can adjust the outputs from Specmod so that they are analogous to posterior probabilities that incorporate our desired class prior.
Allosmod is described in more detail in~\cite{2019MNRAS.490.5615B}, where it is applied to \gdr{2} data.

The third DSC classifier, Combmod, takes the probabilities from Specmod and Allosmod for a source and combines them into a new posterior probability over all five classes. This is not entirely trivial, because it has to ensure that the global prior is not counted twice, and it has to allow for the fact that Specmod has more classes than Allosmod. The combination algorithm is described in Appendix~\ref{app:combmod_definition}.

\subsubsection{Class prior}

\begin{table*}
\begin{center}
  \caption[DSC class prior]{DSC class prior. The first row gives these as fractions relative to the stars, and the second row gives their decimal values summing to 1.0. This is the class prior for Specmod. The prior for the star class in Allosmod is the sum of star, white dwarf, and physical binary star.
  \label{tab:cu8par_apsis_dsc_classprior}
  }
\begin{tabular}{rrrrrr}
\hline
  &  {\tt quasar} & {\tt galaxy} & {\tt star} & {\tt white dwarf} & {\tt physical binary star} \\
  \hline
$\propto$ &  1/1000 & 1/5000 & 1 & 1/5000 & 1/100 \\
$=$           &   0.000989 & 0.000198 & 0.988728 & 0.000198 & 0.009887 \\
\hline
\end{tabular}
\end{center}
\end{table*}

Single stars hugely outnumber extragalactic sources in \gaia, and failing to take this into account would give erroneous probabilities and classifications. 
Specifically, if we were to assume equal priors for all classes, 
then when the attributes of a given source do not provide a strong discrimination between the classes, the source would be classified as any class with near equal probabilities. However, in reality, the source is far more likely to be a star, because extragalactic sources are so rare.
We must therefore set appropriate priors for the classes. Failing to do so
corresponds to the well-known base rate fallacy. 
We choose here to adopt a global prior that reflects the expected fraction of each class (as we define them) in the entire \gdr{3} data set.
This prior is given in Table~\ref{tab:cu8par_apsis_dsc_classprior}. As the relative fraction of extragalactic to Galactic objects that Gaia observes varies with quantities such as magnitude and Galactic latitude, we could make the prior a function of these (and potentially other) quantities; but we have not introduced this in \gdr{3}.

Using the correct prior is important. A classifier with equal priors would perform worse on the rare objects than a classifier with appropriate priors, because the former would tend to misclassify many stars as being extragalactic. However, we would not notice this if we erroneously validated the classifier on a balanced set (equal numbers in each class), because such a validation set has an artificially low fraction of stars, and hence far too few potential contaminants. The classifier would perform worse but would appear to be performing better. This is demonstrated in Table 1 of \cite{2019MNRAS.490.5615B}.
We address this issue in the context of our validation data in section~\ref{subsec:dsc_performances}.

\subsubsection{Training data}\label{sec:dsc_training_data}

\dsc is trained empirically, meaning it is trained on a labelled subset of the actual Gaia data it will be applied to (except for binary stars). The classes were defined by selecting sources of each class from an external database and cross-matching them to \gdr{3}. 
The sources used to construct the training sets ---and which therefore define the classes--- are as follows ( see the \linktosec{cu8par}{apsis}{dsc} and~\cite{LL:CBJ-094} for more details):
\begin{itemize}[leftmargin=0.5cm]
    \item Quasars: 300\,000 spectroscopically confirmed quasars from the fourteenth release of the Sloan Digital Sky Survey (SDSS) catalogue, SDSS-DR14 \citep{2018A&A...613A..51P}. 
    \item Galaxies: 50\,000 spectroscopically confirmed galaxies from SDSS-DR15 \citep{2019ApJS..240...23A}. 
    \item Stars: 720\,000 objects drawn at random from \gdr{3} that are not in the quasar or galaxy training sets. Strictly speaking, this is therefore an `anonymous' class. But as the vast majority of sources in \gaia are stars, and the majority of those will appear in (spectro)photometry and astrometry as single stars, we call this class `stars'.
    \item White dwarfs: 40\,000 white dwarfs from the Montreal White Dwarf Database\footnote{\href{http://www.montrealwhitedwarfdatabase.org}{http://www.montrealwhitedwarfdatabase.org}} that have coordinates and that are not known to be binaries using the flag provided in that table. This class is not in Allosmod.
    \item Physical binary stars: 280\,000 \bporrp spectra formed by summing the two separate components in spatially-resolved binaries in \gdr{3} (see the \linktosec{cu8par}{apsis}{dsc}). This is only done for the \bporrp spectra, not for astrometry or photometry, so physical binaries are not a class in Allosmod.
\end{itemize}
The quasar, galaxy, and star class definitions are more or less the same as in~\cite{2019MNRAS.490.5615B}. 

The selected sources were filtered in order to remove obvious contaminants or problematic measurements (as described in the \linktosec{cu8par}{apsis}{dsc}).
The numbers above refer to what remains after this filtering.
The remaining set was then split into roughly equally sized training and validation sets (per class).
Generally speaking, the relative number of objects of each class ---the {\em class fraction}--- in the training data affects the output probabilities of  a classifier, because it acts as an implicit prior in the classifier. However, for both Specmod and Allosmod, we remove this influence to ensure that their priors correspond to our class prior.
We are therefore free to choose as many training examples in each class as we need, or can obtain, in order to learn the data distributions.

We note that for the common classes between Specmod and Allosmod, that is,\ quasars, galaxies, and stars, a common sample with complete input data was used to train both modules.
In particular, this means that even though Specmod does not require parallaxes and proper motions as inputs, its training sample is restricted to those sources that do have parallaxes and proper motions. This is important because Specmod is also applied to sources that lack parallaxes and proper motions, meaning that some of its results are on types of objects that are not represented in its training set.
This is particularly important for galaxies. 

\begin{figure*}
\centering
\includegraphics[width=0.8\textwidth,angle=0]{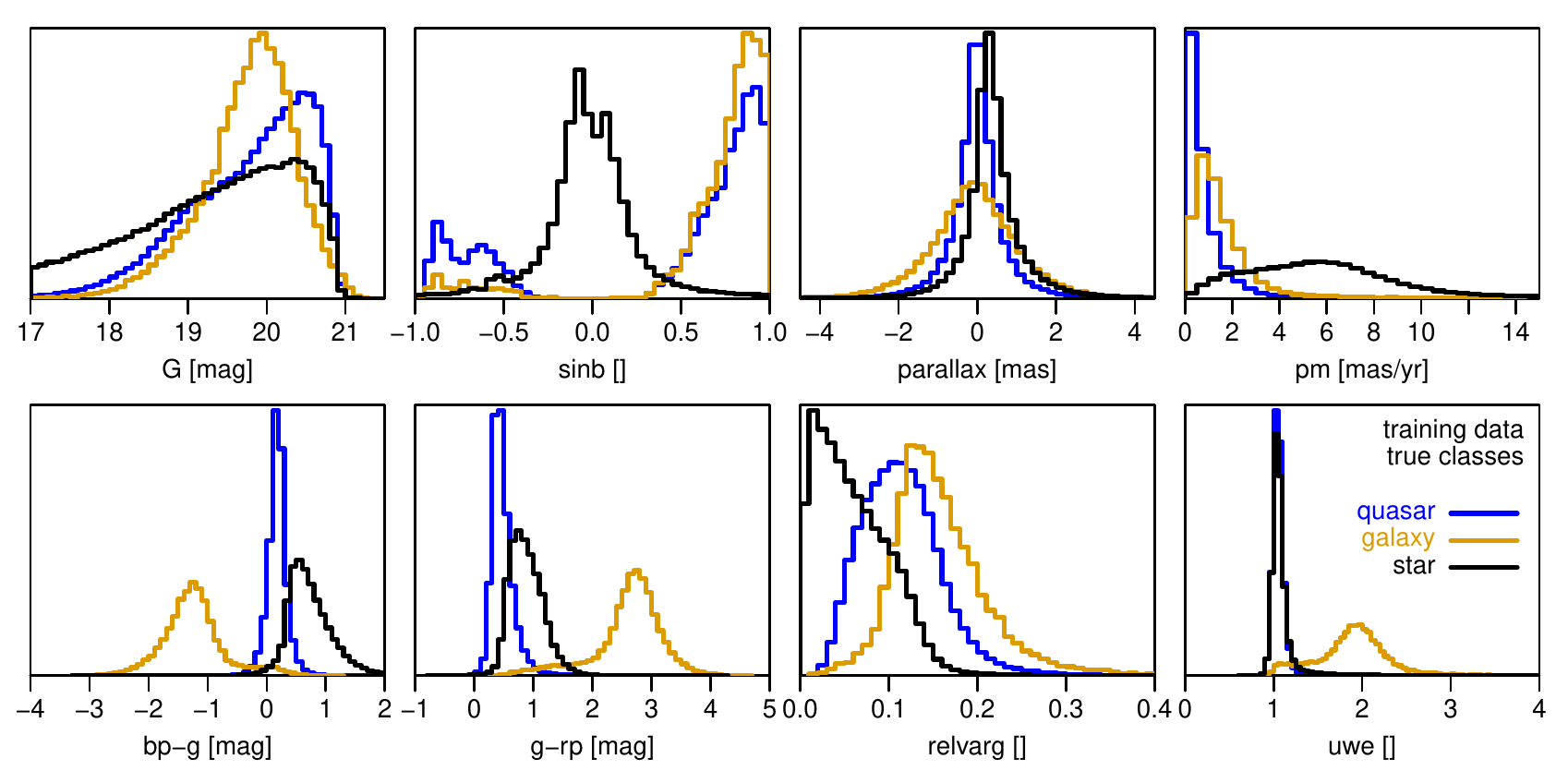}
\rule{0.85\textwidth}{0.25pt}
\includegraphics[width=0.8\textwidth,angle=0]{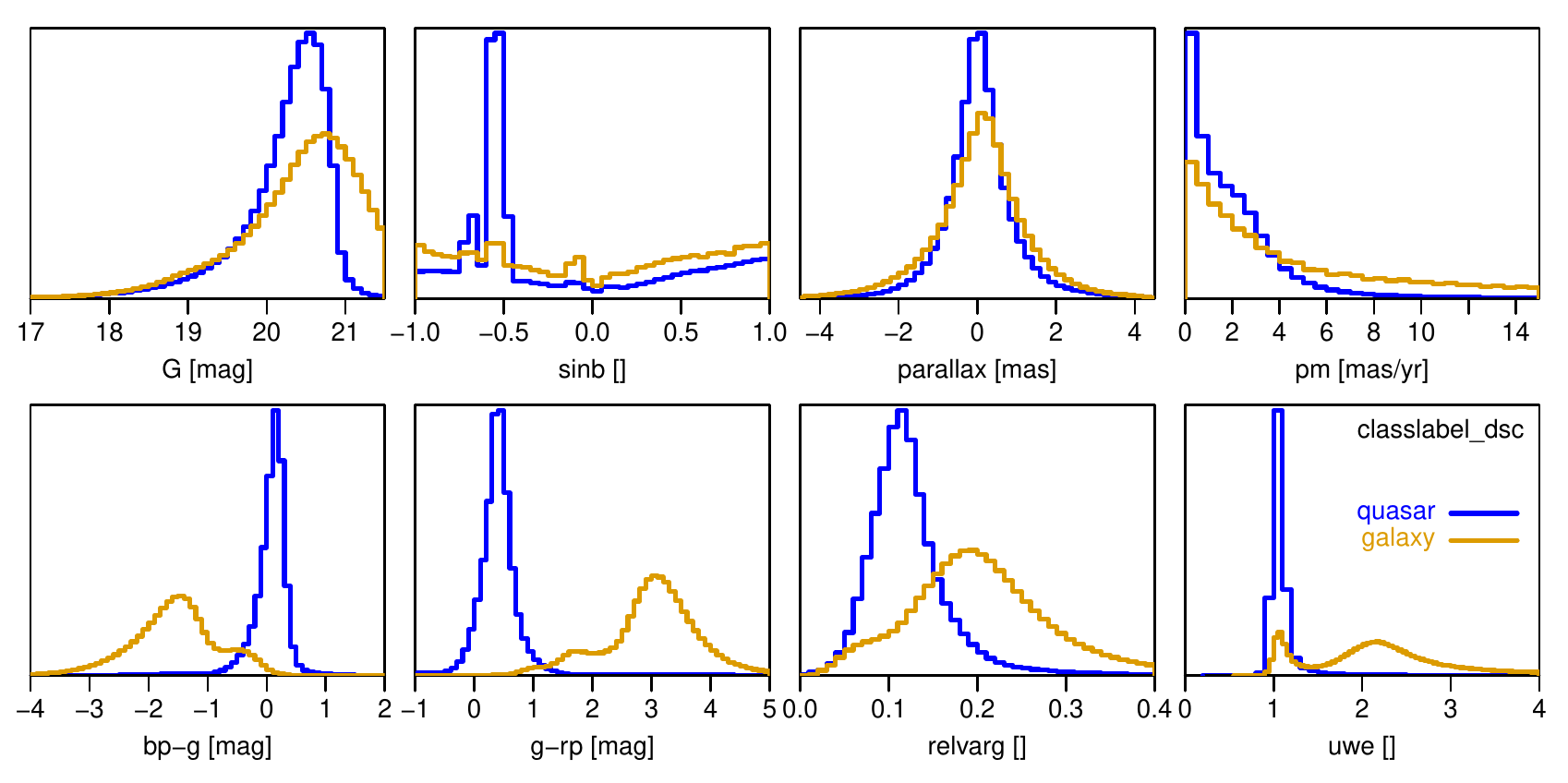}
\rule{0.85\textwidth}{0.25pt}
\includegraphics[width=0.8\textwidth,angle=0]{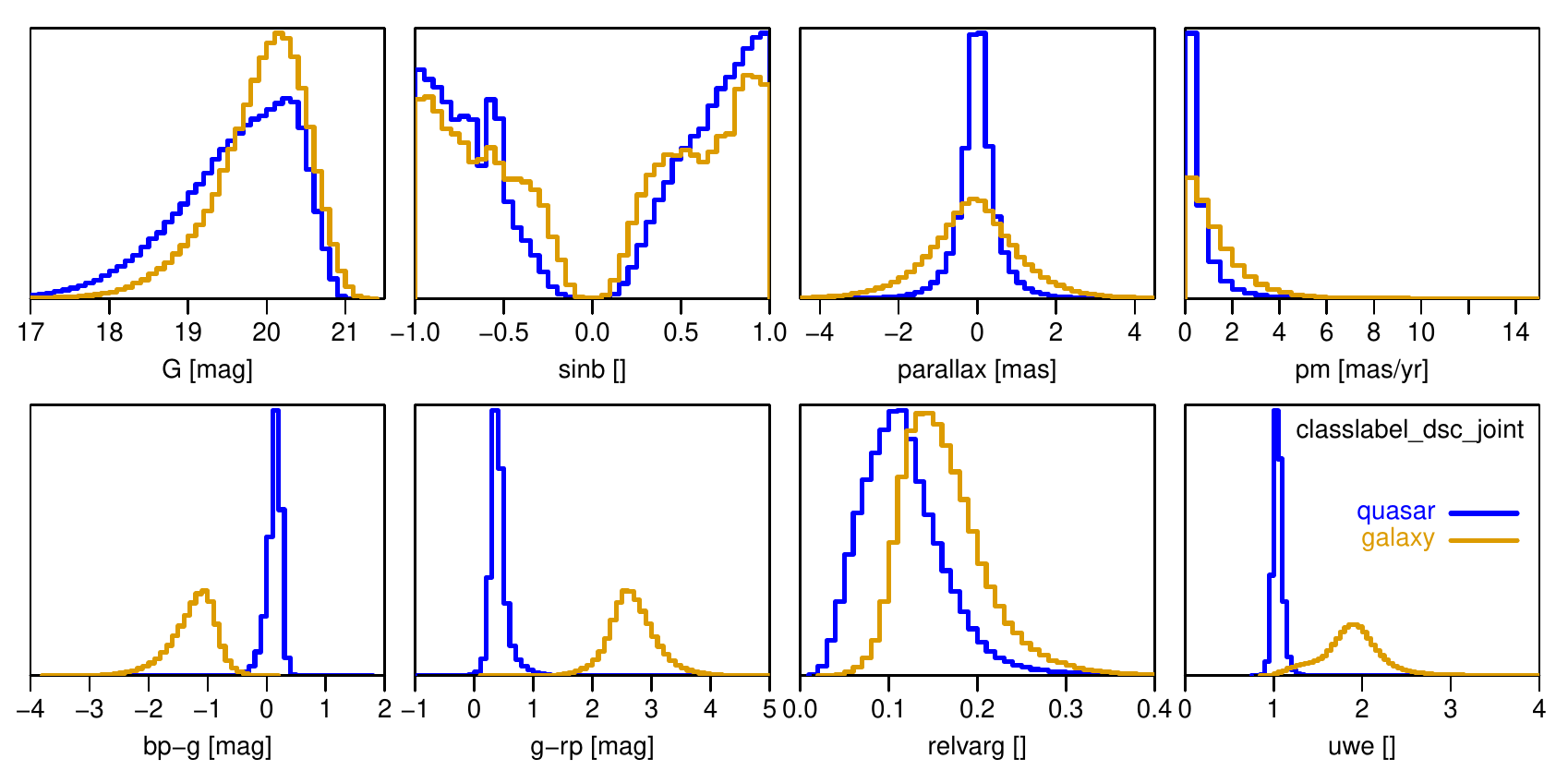}
\caption{Distribution (linear scale) of \gaia\ features for various samples used in DSC. 
Top: Training data for quasars (blue), galaxies (orange), and stars (black). When training Allosmod, the sin\,b distributions for quasars and galaxies are replaced with uniform ones.
Middle: \gaia sources assigned {\tt classlabel\_dsc='quasar'} (blue) and  {\tt classlabel\_dsc='galaxy'} (orange).
Bottom: \gaia sources assigned {\tt classlabel\_dsc\_joint='quasar'} (blue) and  {\tt classlabel\_dsc\_joint='galaxy'} (orange).
\label{fig:dsc_featurehist}
}
\end{figure*}


\begin{figure}[t!]
\begin{center}
\includegraphics[width=0.40\textwidth,angle=0]{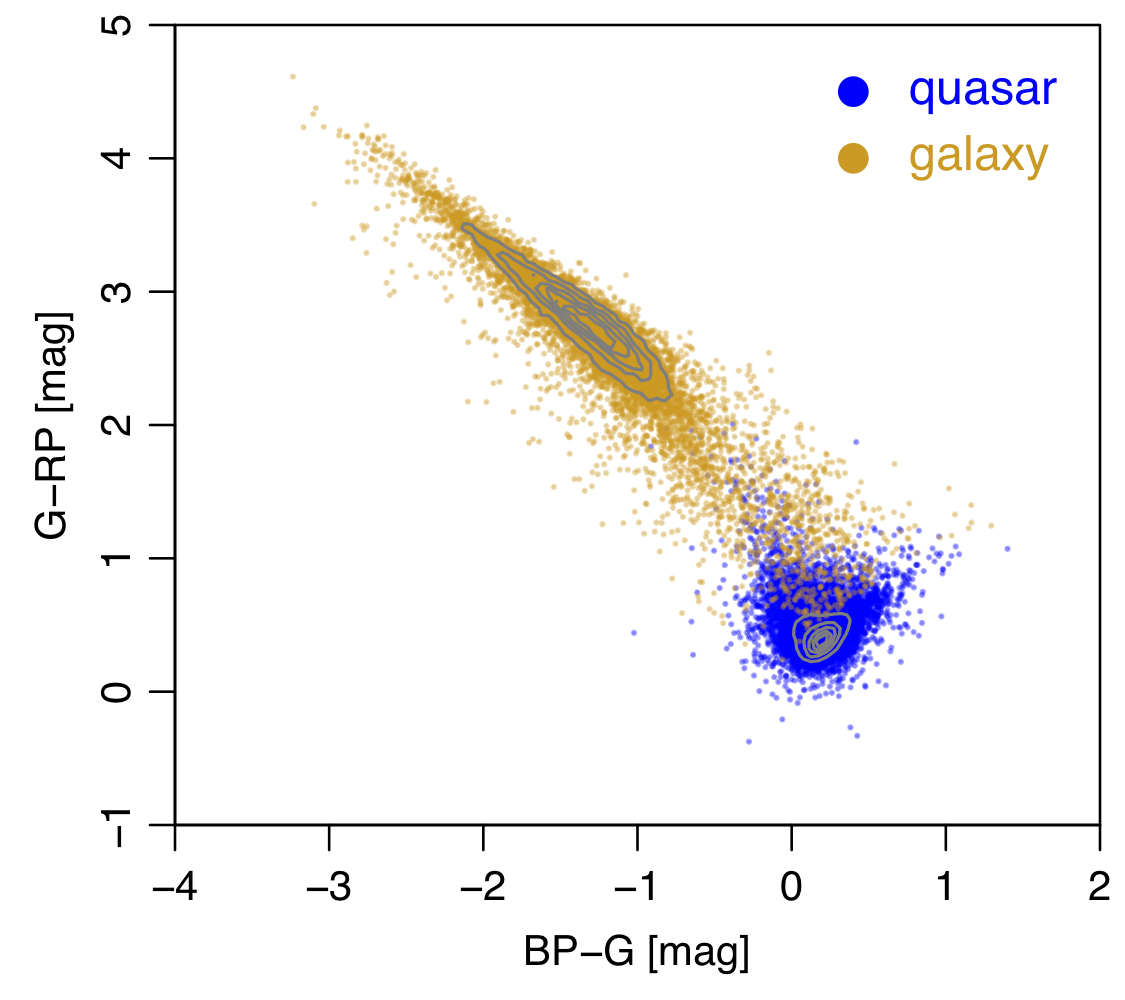}
\includegraphics[width=0.40\textwidth,angle=0]{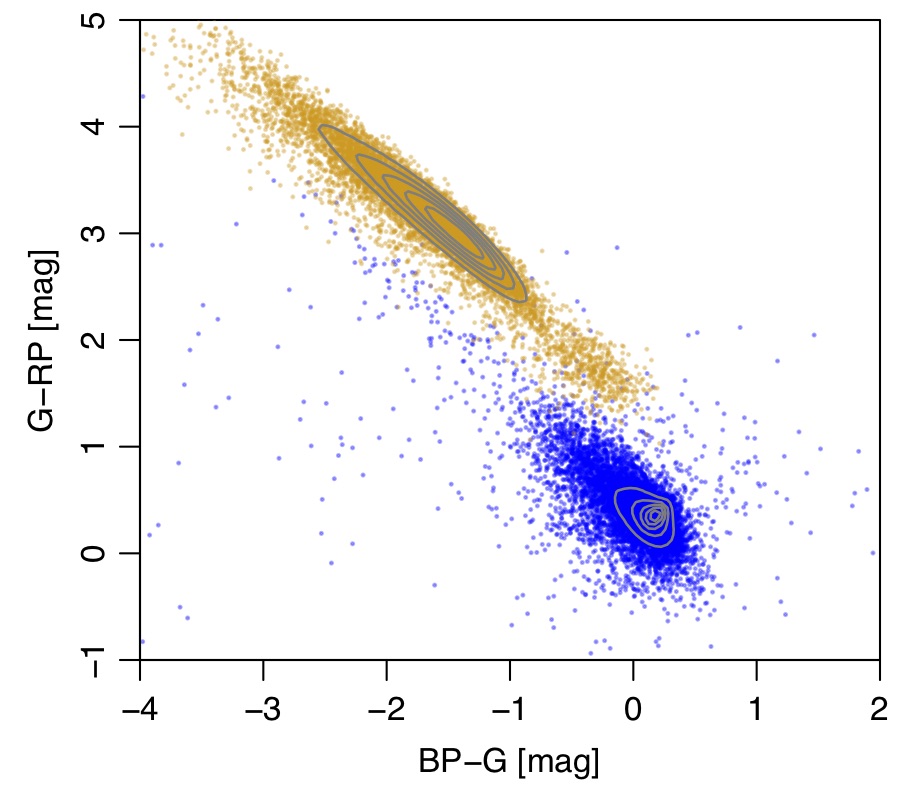}
\includegraphics[width=0.40\textwidth,angle=0]{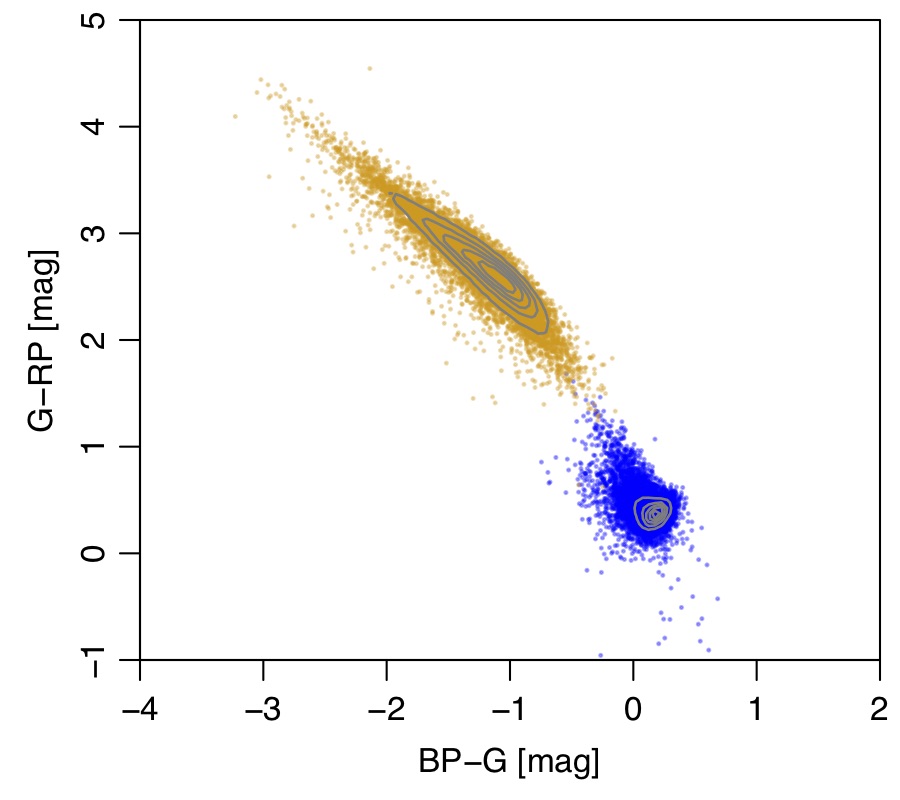}
\caption{Colour--colour diagrams for various samples used in DSC.
Top: Training data for quasars (blue) and galaxies (orange). 
Middle: \gaia sources assigned {\tt classlabel\_dsc='quasar'} (blue) and  {\tt classlabel\_dsc='galaxy'} (orange).
Bottom: \gaia sources assigned {\tt classlabel\_dsc\_joint='quasar'} (blue) and  {\tt classlabel\_dsc\_joint='galaxy'} (orange). The differences in the distributions are due to the various levels of completeness and purity in the two types of class label.
\label{fig:dsc_ccd}
}
\end{center}
\end{figure}

Figure~\ref{fig:dsc_featurehist} (top) shows the distribution of the eight Allosmod features in the training data for the quasar and galaxy classes. As we do not want the model to learn the $\sin b$ distribution of extragalactic objects, which is just the SDSS footprint (shown in the plot), we replace this with a random value drawn from a uniform distribution in $\sin b$ (i.e.\ uniform sky density) when training Allosmod. This plot also shows, for comparison, the distribution of the features for the star class in the training data.
Figure~\ref{fig:dsc_ccd} (top) shows the distribution of the two colours of the quasars and galaxies in a colour--colour diagram.

\subsubsection{Class labels}\label{sec:dsc_class_labels}

The main output from \dsc is the class probabilities from all three classifiers. For convenience, we also compute two class labels from the probabilities, which appear only for sources in the \linktoEGTable{qso_candidates} and \linktoEGTable{galaxy_candidates} tables in the data release.
The first label, \fieldName{classlabel_dsc}, is set to the class that gets the highest posterior probability in Combmod that is greater than 0.5.
If none of the output probabilities are above 0.5, this class label is {\tt unclassified}.
This gives a sample that is fairly complete for quasars and galaxies, but not very pure.

The second class label, \fieldName{classlabel_dsc_joint}, identifies a purer set of quasars and galaxies. It  is set to the class that achieves a probability above 0.5 in both Specmod and Allosmod.
This produces purer samples because the Specmod and Allosmod probabilities are not perfectly correlated. This lack of correlation may be unexpected, but is what we want, because it means the classifiers are providing non-redundant information.

Because DSC is not the only contributor to the \linktoEGTable{qso_candidates} and \linktoEGTable{galaxy_candidates} tables, sources in the {\tt qso\_candidates} table can have either classlabel set to {\tt galaxy}, and vice versa. 


\subsection{Performance: Purity and completeness}\label{subsec:dsc_performances}

By assigning each source to the class with the largest probability, it is uniquely classified. An alternative is to additionally adopt a minimum probability threshold, in which case we can get multiple classifications if the threshold is low enough, or no classification if it is high enough.
Doing this on sources with known classes (assumed to be correct), we can then compute the confusion matrix, which tells us how many sources of each true class are assigned to each DSC class. From this, we then compute, for each class, the completeness --the fraction of true positives among all trues--- and the purity ---the fraction of true positives among all positives.

Here we use the largest probabilities to compute the completenesses and purities on the validation sets.\footnote{The validation data for the binaries is not the one mentioned in section~\ref{sec:dsc_training_data}, namely synthetically-combined single stars,
but instead a set of unresolved binaries directly from \gaia. See the \linktosec{cu8par}{apsis}{dsc} for more details.}
As the class fractions in this validation set are not representative of what they are in \gaia, the raw purities are meaningless. Specifically, stars are far less common in the validation data than they are in a random sample of \gaia data, and so there are too few potential contaminants of the other classes in the validation data, resulting in significantly overestimated purities.
This fact is sometimes overlooked in the validation of classification results in the literature.
Fortunately, we can easily correct for this. As explained in section 3.4 (especially\ equation 4) of \cite{2019MNRAS.490.5615B}, we can modify the confusion matrix to correspond to a validation set that has class fractions equal to the class prior. The purity computed from this modified confusion matrix is then appropriate for any randomly selected sample of \gaia sources. (This modification does not affect the completeness.) We note that this modification is independent of the fact that \dsc\ probabilities are already posterior probabilities that take into account this class prior (i.e.\ both modifications must be done).
This should also serve as a warning when assessing any classifier: if the validation data set does not have a representative fraction of contamination, or if this is not adjusted, the predicted purities will be erroneous.

\begin{table*}[t]
\begin{center}
  \caption[DSC performance]{DSC performance evaluated on the validation data set.
  Classification is done by assigning the class with the largest posterior probability. Performance is given in terms of completeness (compl.) and purity, for each classifier and for each class. Purities have been adjusted to reflect the class prior (given in Table~\ref{tab:cu8par_apsis_dsc_classprior}).
  Results on the `binary' class are largely meaningless due to the incongruity of the class definitions in the training and validation data sets.
   These results reflect performance for sources drawn at random from the entire \gaia\ data set, in particular for all magnitudes and latitudes.
  The final two columns labelled `Spec\&Allos' refer to samples obtained by requiring a probability larger than 0.5 from both Specmod and Allosmod for a given class: this is identical to \fieldName{classlabel_dsc_joint} in the \linktoEGTable{qso_candidates} and \linktoEGTable{galaxy_candidates} tables.
  The bottom two rows refer to extragalactic sources at higher Galactic latitudes ($|b|>11.54\deg$), where the prior is more favourable for detecting quasars and galaxies. These are conservative estimates, accounting only for reduced numbers of stars, not the better visibility of extragalactic objects on account of less interstellar extinction and source confusion.
  \label{tab:cu8par_apsis_dsc_resvst_defset_cp}
  }
  \begin{tabular}{rrrrrrrrr}
    \hline
    & \multicolumn{2}{c}{Specmod}  & \multicolumn{2}{c}{Allosmod}   & \multicolumn{2}{c}{Combmod}    & \multicolumn{2}{c}{Spec\&Allos} \\
  & compl.\ & purity &  compl.\ & purity   & compl.\ & purity    & compl.\ & purity  \\
  \hline
  quasar           & 0.409 & 0.248 & 0.838 & 0.408 & 0.916 & 0.240 & 0.384 & 0.621 \\
  galaxy           & 0.831 & 0.402 & 0.924 & 0.298 & 0.936 & 0.219  & 0.826 & 0.638 \\
  star   & 0.998 & 0.989 & 0.998 & 1.000 & 0.996 & 0.990 & --  & -- \\
  white dwarf  & 0.491 & 0.158 & --       & --       & 0.432 & 0.250  & --  & --  \\
  physical binary star           & 0.002 & 0.096 & --       & --      & 0.002 & 0.075  & --  & --  \\
  \hline
  quasar, $|\sin\,b\,|>0.2$ & 0.409 & 0.442 & 0.881 & 0.603 & 0.935 & 0.412 & 0.393 & 0.786 \\
  galaxy, $|\sin\,b\,|>0.2$ & 0.830 & 0.648 & 0.928 & 0.461 & 0.938 & 0.409 & 0.827 & 0.817 \\
\hline
\end{tabular}
\end{center}
\end{table*}

\tabref{tab:cu8par_apsis_dsc_resvst_defset_cp} shows the completenesses and purities for the DSC classes and classifiers. This is the performance we expect for a sample selected at random from the entire Gaia dataset that has complete input data for both Specmod and Allosmod. 
It accommodates the rareness of all these classes, as specified by the global class prior (Table~\ref{tab:cu8par_apsis_dsc_classprior}), both in the probabilities and the application data set.
It is important to bear in mind that these purity and completeness measures only
refer to the types of objects in the validation set. For extragalactic objects, this means objects classified as such by SDSS using the SDSS spectra. The overall population of extragalactic objects classified by \dsc is of course broader than this, and so the completeness and purity evaluated on other subsets of extragalactic objects could differ. 

Due to the dominance of single stars in \gaia, we are not really interested in the performance on this class. Indeed, it is trivial to get an excellent single-star classifier: simply call everything a single star and your classifier has 99.9\% completeness and 99.9\% purity.

The performance is modest overall, for reasons that are further discussed in section~\ref{subsec:dsc_use}.
Results on binaries are very poor, partly because the validation set we used to compute the confusion matrix is not representative of the training set. This is because the validation set comprises only real Gaia objects, and so known unresolved binaries, whereas the training set was made by combining single star spectra. However, the internal performance on binaries was also poor. This suggests an intrinsic difficulty in separating binaries (as we define them) from single stars.

The performance in Table~\ref{tab:cu8par_apsis_dsc_resvst_defset_cp} refers to objects covering the full \gaia\ parameter space, in particular all magnitudes and Galactic latitudes. The purities tend to increase for brighter magnitudes, as can be seen from the plots in the
\linktosec{cu8par}{apsis}{dsc} and in~\cite{LL:CBJ-094}. There we see, for example, that for $\gmag \leq 18$\,mag, the purities for quasars and galaxies when using Allosmod alone is 80\% or higher. 
However, when looking at the performance in a specific part of the parameter space, one should adopt a new prior that is appropriate for that part of the parameter space, for example\ fewer extragalactic objects visible at low latitudes.
We then recompute the posterior probabilities (Appendix~\ref{sec:cu8par_apsis_dsc_adjusting_probabilities}) and the completenesses and purities (remembering that the adjustment of the confusion matrix must use the class fractions in this subset of the validation set). This we have done for sources outside of the Galactic plane, with results reported in the bottom two lines of Table~\ref{tab:cu8par_apsis_dsc_resvst_defset_cp}.
For $|b|>11.54\deg$, we adopt a prior probability for quasars of $2.64 \times 10^{-3}$ ($9.9 \times 10^{-4}$ globally), and a prior probability for galaxies of $5.3 \times 10^{-4}$ ($2 \times 10^{-4}$ globally).
The purities of the quasar and galaxy samples are significantly higher, as expected because
there are fewer contaminating stars per square degree. Using a probability threshold increases the purities even further, albeit at the expense of completeness (see \linktosec{cu8par}{apsis}{dsc}  for more plots).
Clearly, if we were willing and able to push the prior for extragalactic objects higher, we would obtain higher purities.

\subsection{Results}
\label{subsec:dsc_results}

\dsc was applied to all \gaia sources that have the required input data. Its results were not filtered in any way. In particular, we did not remove sources with lower quality input data, or that have input data lying outside the range of the training data. By including all results, we allow the user to apply their own filters according to their own goals and needs.

\begin{figure*}
\begin{center}
\includegraphics[width=0.49\textwidth,angle=]{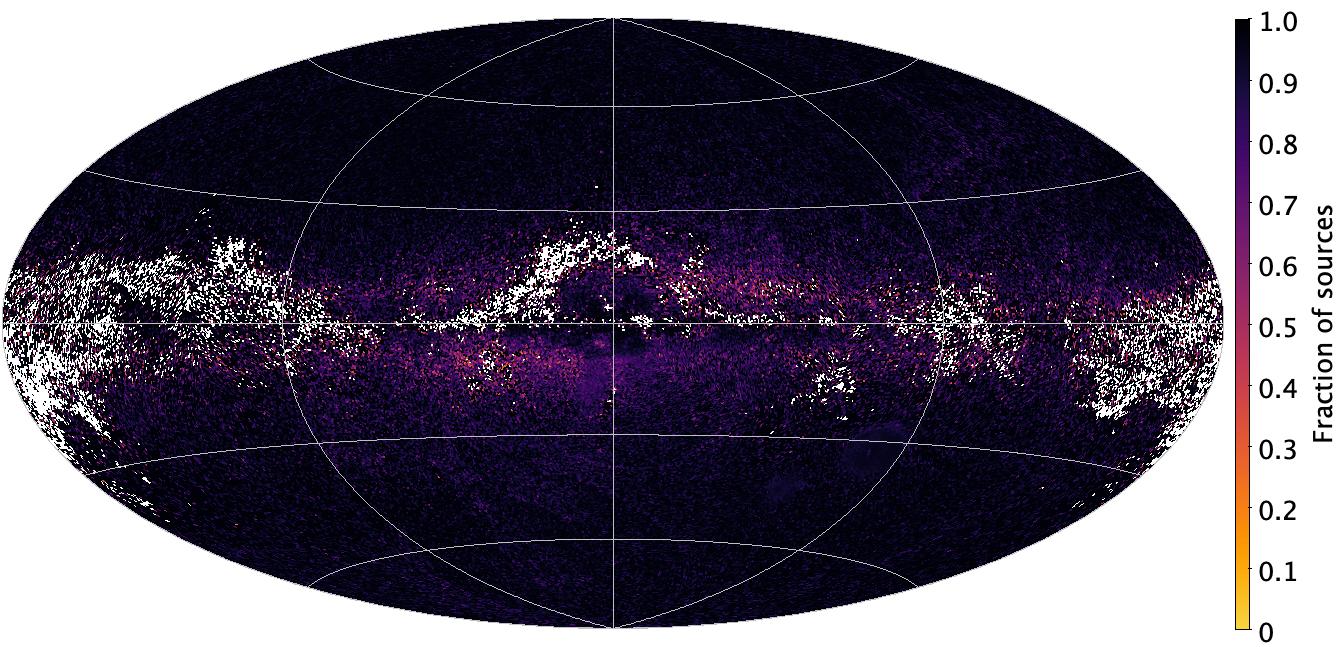}
\includegraphics[width=0.49\textwidth,angle=]{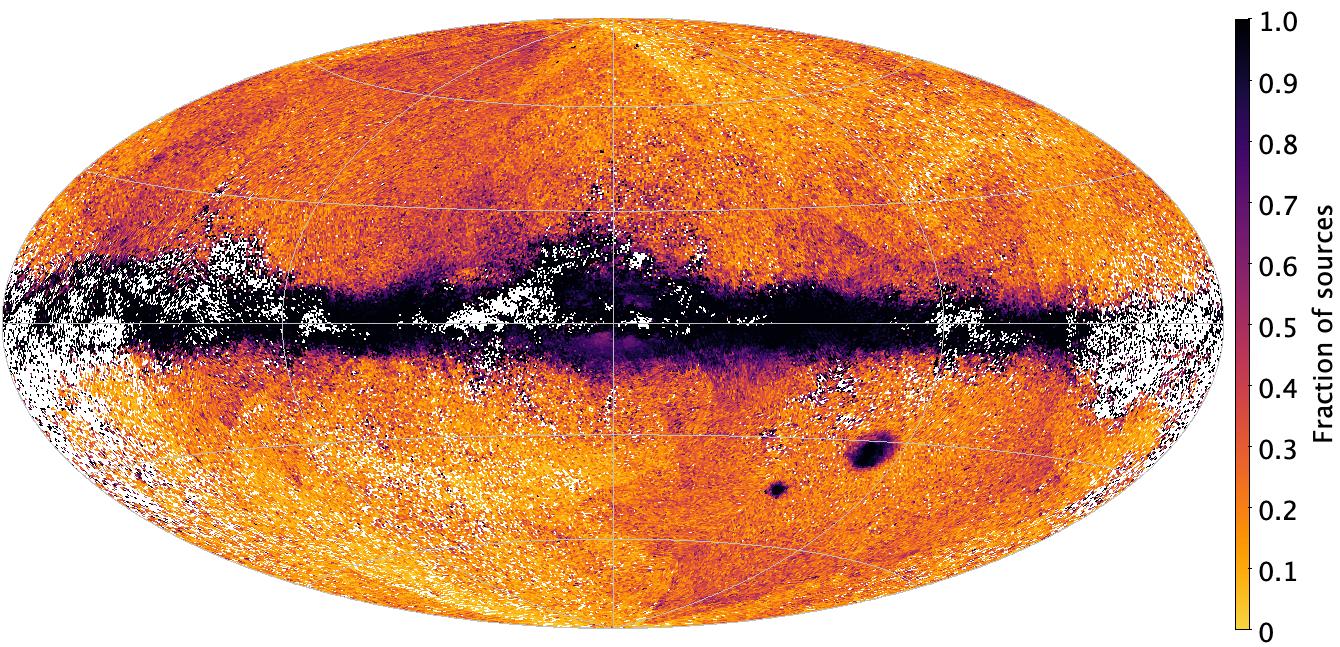}
\caption{Galactic sky distribution of the fraction of sources that have 5p/6p astrometric solutions (i.e.\ have parallaxes and proper motions) for sources that also have
{\tt dsc\_classlabel='quasar'} (left) and {\tt dsc\_classlabel='galaxy'} (right).
The plot is shown at HEALPix level 7 (0.210 \sqdeg) in a Hammer--Aitoff equal area projection with the Galactic centre in the middle, north up, and longitude increasing to the left.
White indicates no sources.
\label{fig:dsc_parallaxfrac_skyplot}}
\end{center}
\end{figure*}

\dsc produces outputs for
1\,590\,760\,469 sources. All of these have probabilities from Combmod and Specmod, whereas 
1\,370\,759\,105 (86.2\%) have probabilities from Allosmod.\footnote{It so happens that all sources which have Allosmod results also have Specmod results, but not vice versa.}
This lower number from Allosmod is due to missing input data, usually missing parallaxes and proper motions (or missing colours in a few cases). That is, sources must have 5p or 6p astrometric solutions from the \gaia Astrometric Global Iterative Solution (AGIS) in order to have Allosmod results. 
This can be seen in Figure~\ref{fig:dsc_parallaxfrac_skyplot}, which shows the fraction of sources (per HEALPix) that have 5p/6p solutions, for those with
{\tt dsc\_classlabel='quasar'} (left) and {\tt dsc\_classlabel='galaxy'} (right). While most objects classified as quasars have measured parallaxes (i.e.\ 5p or 6p solutions), most sources outside of the Galactic plane classified as galaxies do not. Those objects that lack parallaxes and proper motions (the 2p solutions) also lack Allosmod results, and so their Combmod results (and hence {\tt dsc\_classlabel}) are determined only by Specmod. We explore the differences between the 5p/6p and 2p solutions at the end of this section.

The vast majority of sources have high probabilities of being stars, and because the purities of the white dwarf and physical binary classes are low (see the online documentation), we focus here on the results for the quasar and galaxy classes.

\begin{figure*} 
\begin{center}
\includegraphics[width=1.0\textwidth,angle=0]{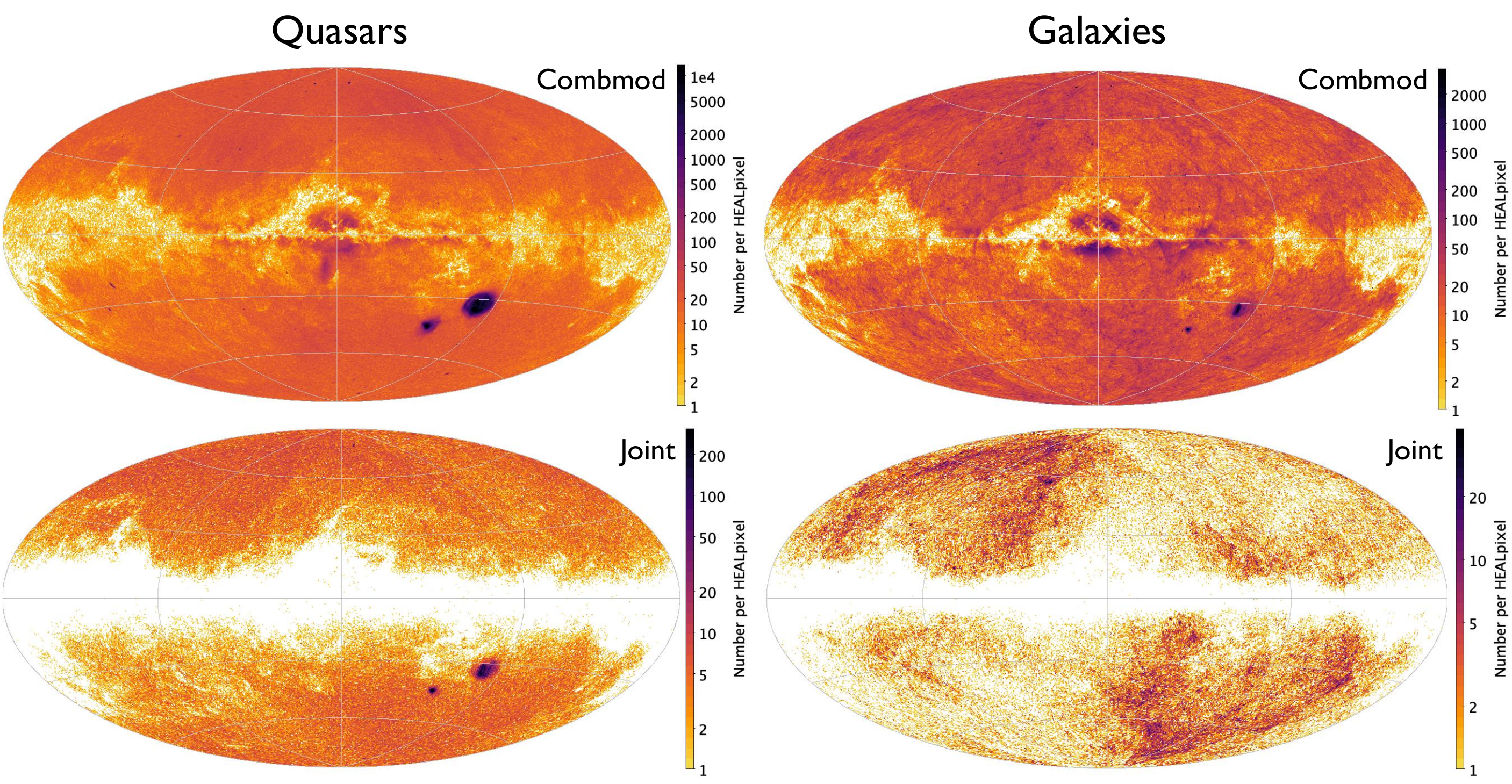}
\caption{Galactic sky distribution of the number of DSC sources classified as quasars (left) and galaxies (right) according to {\tt classlabel\_dsc} (top) and {\tt classlabel\_dsc\_joint} (bottom)  (see Section~\ref{sec:dsc_class_labels} for the label definition). The plot is shown at HEALPix level 7 (0.210 \sqdeg). The logarithmic colour scale covers the full range for each panel, and is
therefore different for each panel. 
\label{fig:dsc_number_skyplots}}
\end{center}
\end{figure*}


The label {\tt classlabel\_dsc} (defined in section~\ref{sec:dsc_class_labels}) classifies 5\,243\,012 sources as quasars and 3\,566\,085 as galaxies.
Their sky distributions are shown in the top two panels of Figure~\ref{fig:dsc_number_skyplots}.
The analysis in section~\ref{subsec:dsc_performances} suggests that these samples are not very pure (see Table~\ref{tab:cu8par_apsis_dsc_resvst_defset_cp}).
In these sky plots, we see large overdensities of supposed quasars in 
several regions, in particular the LMC and SMC, 
suggesting that this sample is not very pure. However, such 
overdensities are expected when we have a constant misclassification rate over the whole sky, because any high-density region will have a high density of both correctly and incorrectly classified objects. 
However, it turns out that the {fraction} of sources classified as quasars is also higher than average in these regions (see below).
The LMC and SMC are so dense that 38\% of all the quasar identifications using 
{\tt classlabel\_dsc}
are in the LMC, and 6.4\% are in the SMC.\footnote{For this purpose, the LMC is defined as a circle of 9\deg\ radius centred on RA=81.3\deg, Dec.=-68.7\deg, and the SMC as a circle of 6\deg\ radius centred on RA=16.0\deg, Dec.=-72.8\deg.} These percentages are much smaller for galaxies: just
3\% for the LMC and 1\% for the SMC.

The bottom row of Figure~\ref{fig:dsc_number_skyplots} shows the distribution of the 547\,201 sources classified as quasars and the 251\,063 sources classified as galaxies by the purer class label
{\tt classlabel\_dsc\_joint}. The overdensities of quasars in the LMC and SMC regions are now greatly reduced, to 4\% and 1\% of all sources respectively. 

\begin{figure*}
\begin{center}
\includegraphics[width=1.0\textwidth,angle=0]{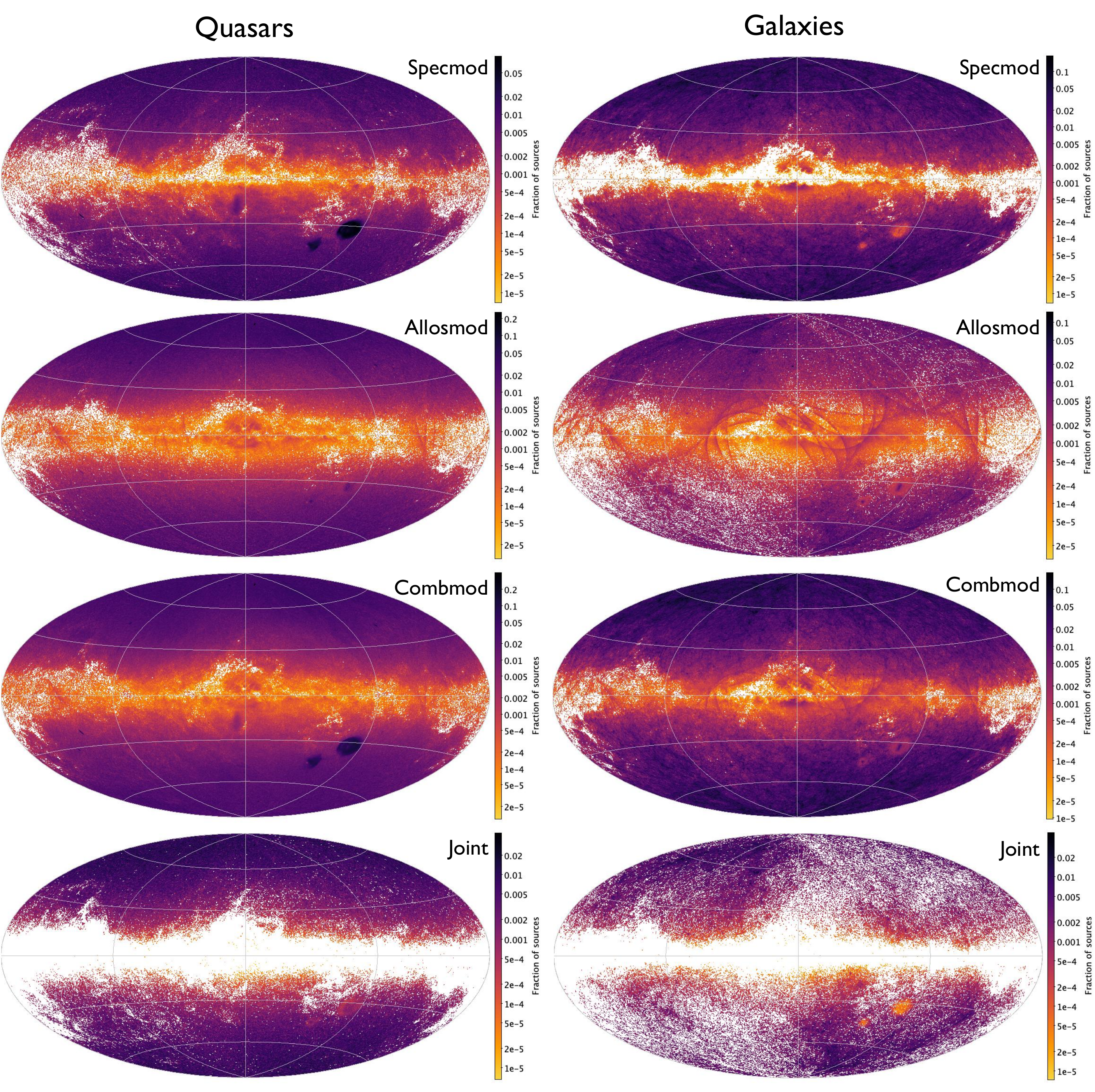}
\caption{Galactic sky distribution of the fraction of DSC sources classified as quasars (left) and galaxies (right) according to Specmod (top), Allosmod (second), Combmod (third), and Specmod and Allosmod (bottom) probabilities being greater than 0.5 for that class.
The bottom two rows are identical to {\tt classlabel\_dsc} and {\tt classlabel\_dsc\_joint} (respectively) being set to the appropriate class (see section~\ref{sec:dsc_class_labels}).
The plot is shown at HEALPix level 7 (0.210 \sqdeg) with each cell showing the ratio of the sources classified to the total number of sources with DSC results (1.59 billion over the whole sky).
The logarithmic colour scale covers the full range for each panel, and is therefore different for each panel.
\label{fig:dsc_fraction_skyplots}}
\end{center}
\end{figure*}

Figure~\ref{fig:dsc_fraction_skyplots} shows the same sky distribution as before, but now expressing the numbers as a fraction of the total number of sources in that HEALPix\footnote{For details on the HEALPix scheme used by Gaia, see \citet{LL:BAS-020}} (classified by DSC as anything).
As most of the sources are stars, these plots essentially show the ratio of extragalactic to Galactic objects per HEALPix, albeit with varying degrees of contamination. The four rows of the plot correspond to four possible ways of classifying extragalactic sources: the top three rows are for probabilities above 0.5 for Specmod, Allosmod, and Combmod, respectively, whereby the latter is identical to {\tt classlabel\_dsc}. The bottom row is {\tt classlabel\_dsc\_joint}.
Looking at the third row ---for {\tt classlabel\_dsc}--- we see a higher fraction of extragalactic sources (plus contamination) has been discovered outside of the Galactic plane than at lower latitudes. This we expect, as high extinction from Galactic dust obscures extragalactic objects, and also there are far more stars in the Galactic plane. However, we also see a higher fraction of supposed quasars (left) in the LMC and SMC ---clear misclassifications--- indicating a higher contamination in these regions. Looking at the top two left panels in 
Figure~\ref{fig:dsc_fraction_skyplots} for Specmod and Allosmod, respectively, we see that this contamination comes from Specmod, that is,\ misclassification of the \bporrp spectra, but not from Allosmod, which uses photometry and astrometry.
It is probably not due to  crowding in the LMC/SMC corrupting the \bporrp spectra, because we do not see such high contamination in the crowded Galactic plane; it is more likely due to faint blue sources in the LMC/SMC being confused with quasars, something which does not occur as much in the Galactic plane due to the higher reddening there.

The top three rows of the right column of Figure~\ref{fig:dsc_fraction_skyplots} show the corresponding plots for galaxies. The stripes are artefacts of the \gaia scanning law. They are much more prominent in Allosmod than in Specmod, and we see in   Table~\ref{tab:cu8par_apsis_dsc_resvst_defset_cp} that Allosmod is expected to have a lower purity for galaxies than Specmod (the opposite is true for quasars).

When we use {\tt classlabel\_dsc\_joint} for classification, we get smaller but purer samples (see~\cite{DR3-DPACP-101}). 
The sky distributions for these samples (bottom row of Fig.~\ref{fig:dsc_fraction_skyplots})
show that low-latitude regions are excluded. In other words, only sources at higher latitudes were classified with probabilities above 0.5 by both Specmod and Allosmod. We also note that the overdensities in the LMC and SMC are greatly reduced with {\tt classlabel\_dsc\_joint}.

The middle panels of Figure~\ref{fig:dsc_featurehist} show the distributions of various \gaia features for the sources classified as quasar (in blue) and galaxy (in orange) by  {\tt classlabel\_dsc}.  The middle panel of Figure~\ref{fig:dsc_ccd} shows the two colours as a colour--colour diagram. These may be compared to the distributions of the training data in the upper panels in both cases.
There are some noticeable differences.
The most obvious is the spike in the latitude distribution for (apparent) quasars at the LMC. Recall that, when training Allosmod, we used a flat $\sin{b}$ distribution (see section~\ref{subsec:dsc_method}). We also see that the objects classified ---galaxies in particular--- extend to fainter magnitudes than the training data. This is not surprising given that the training sample had to have SDSS spectroscopic classifications, whereas we apply DSC to all \gaia sources, which extend to fainter magnitudes, where  misclassifications are more frequent.
The observed galaxies also show larger (anomalous) proper motions, plus more (anomalous) photometric variability according to the relative variability, {\tt relvarg}, parameter.
Finally, we also see differences in the colour distributions compared to the training data for both classes (Figure~\ref{fig:dsc_ccd}). Some of this is due to the different populations being sampled (the training objects are brighter), as well as contamination.

The bottom panels of Figures~\ref{fig:dsc_featurehist} and~\ref{fig:dsc_ccd} show the features and colour--colour diagrams for objects classified using the purer {\tt classlabel\_dsc\_joint} label. These show tighter distributions that are more similar to the training data. We note in particular the reduction of faint galaxies.

\begin{figure*}[t]
\begin{center}
\includegraphics[width=0.40\textwidth,angle=0]{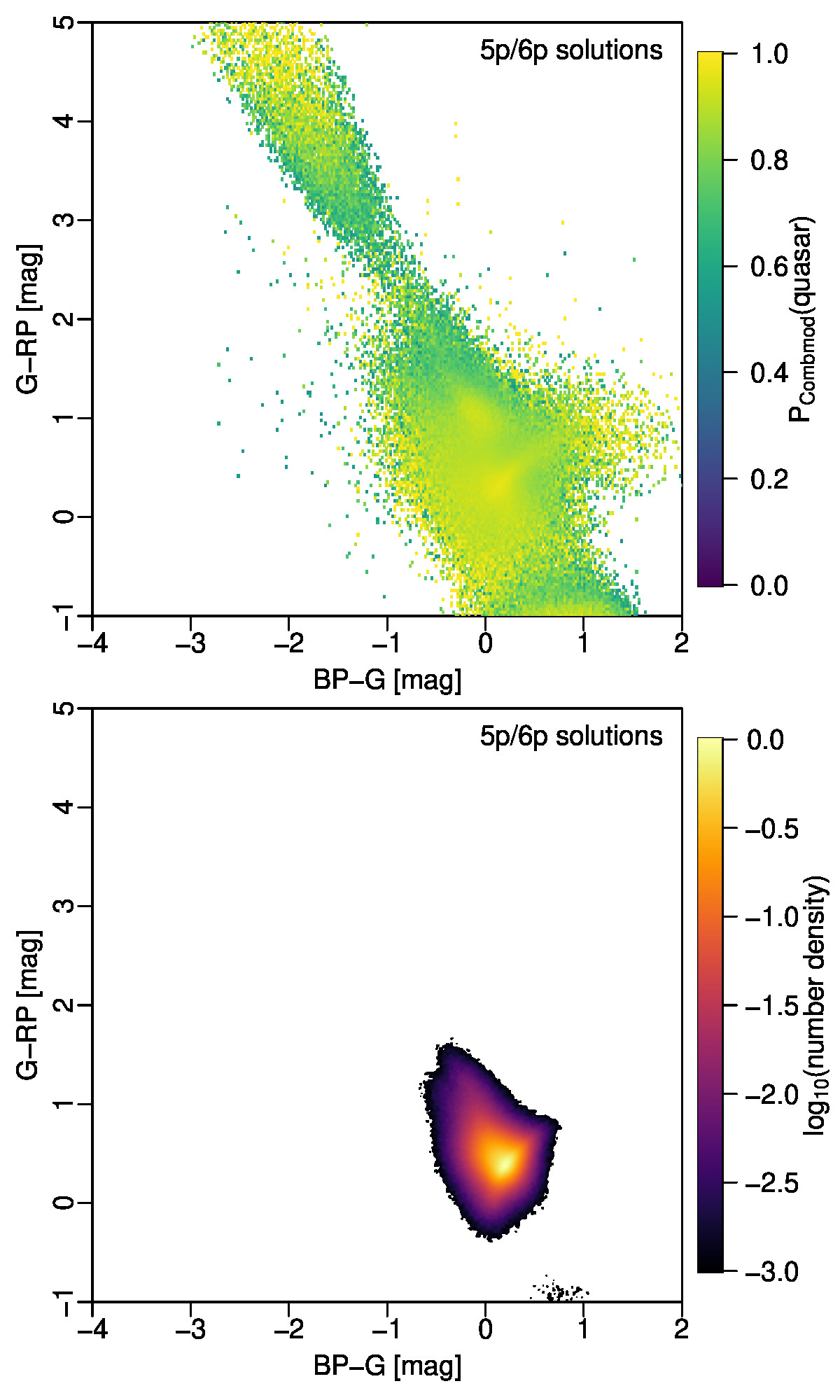}
\includegraphics[width=0.40\textwidth,angle=0]{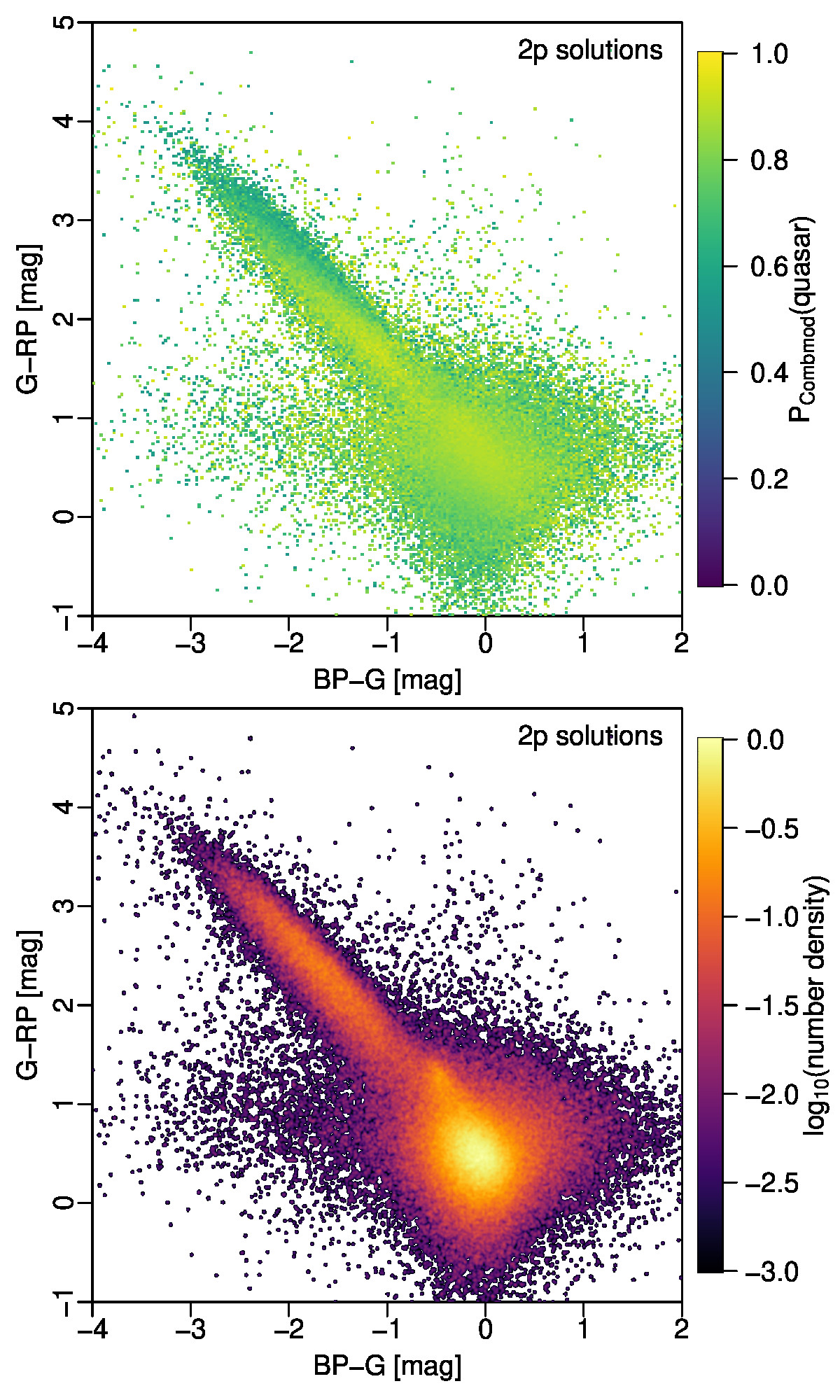}
\caption{Colour--colour diagram for sources in the \linktoEGTable{qso_candidates} table with \linktoEGParam{qso_candidates}{classlabel_dsc}{\tt ='quasar'}, excluding regions around the LMC and SMC.
The left column shows sources with 5p/6p solutions (2.64 million sources), the right column shows sources with 2p solutions (0.14 million sources).
These numbers refer to plotted sources, i.e.\ that have all \gaia bands.
The colour coding in the upper panel shows the mean DSC-Combmod probability for the quasar class (the field \linktoAPParam{astrophysical_parameters}{classprob_dsc_combmod_quasar}). The colour coding in the lower panel shows the density of sources on a log scale relative to the peak density in that panel.
\label{fig:dr3int5_qsotable_astrometry_classlabel_dsc_ccd_grp_bpg}
}
\end{center}
\end{figure*}

\begin{figure*}[t]
\begin{center}
\includegraphics[width=0.40\textwidth,angle=0]{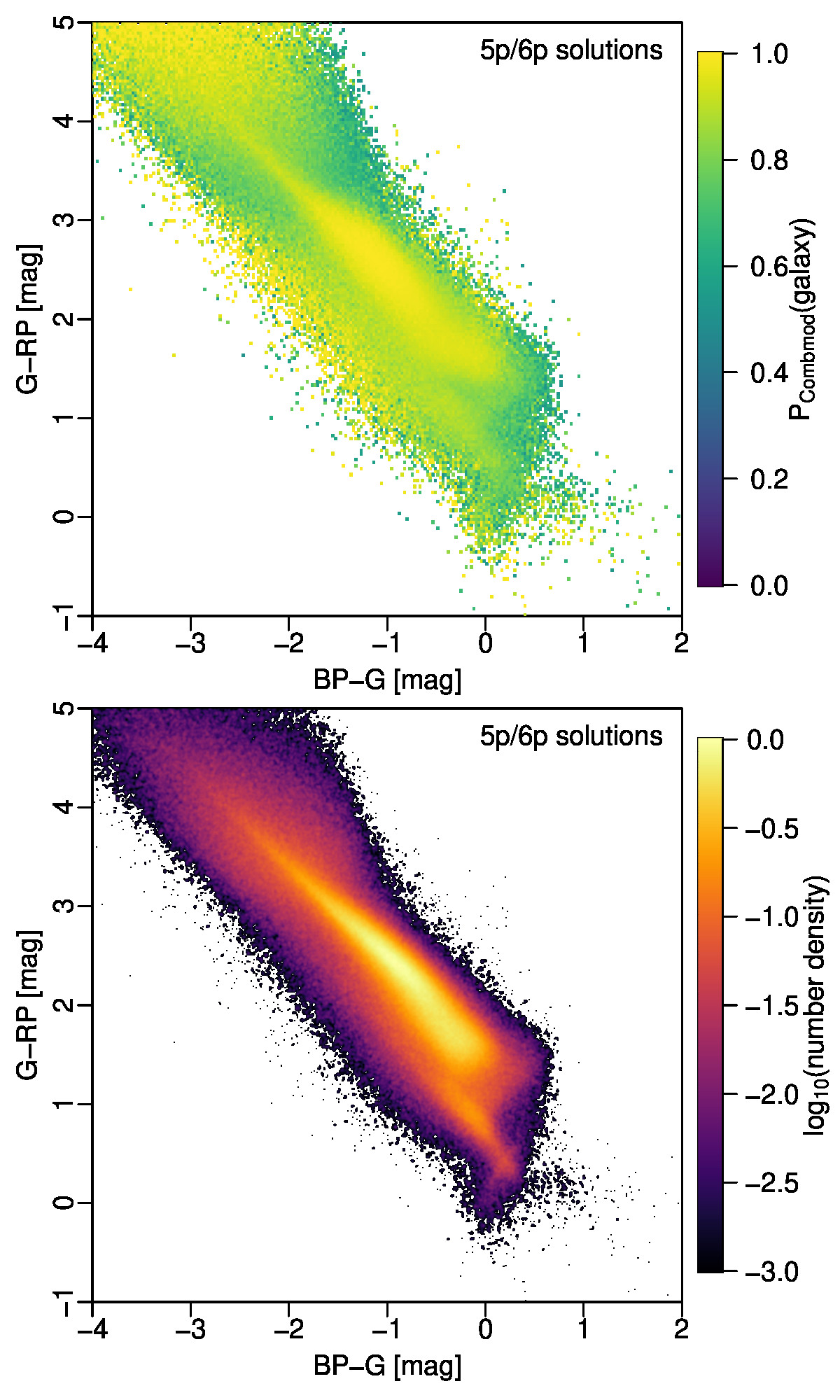}
\includegraphics[width=0.40\textwidth,angle=0]{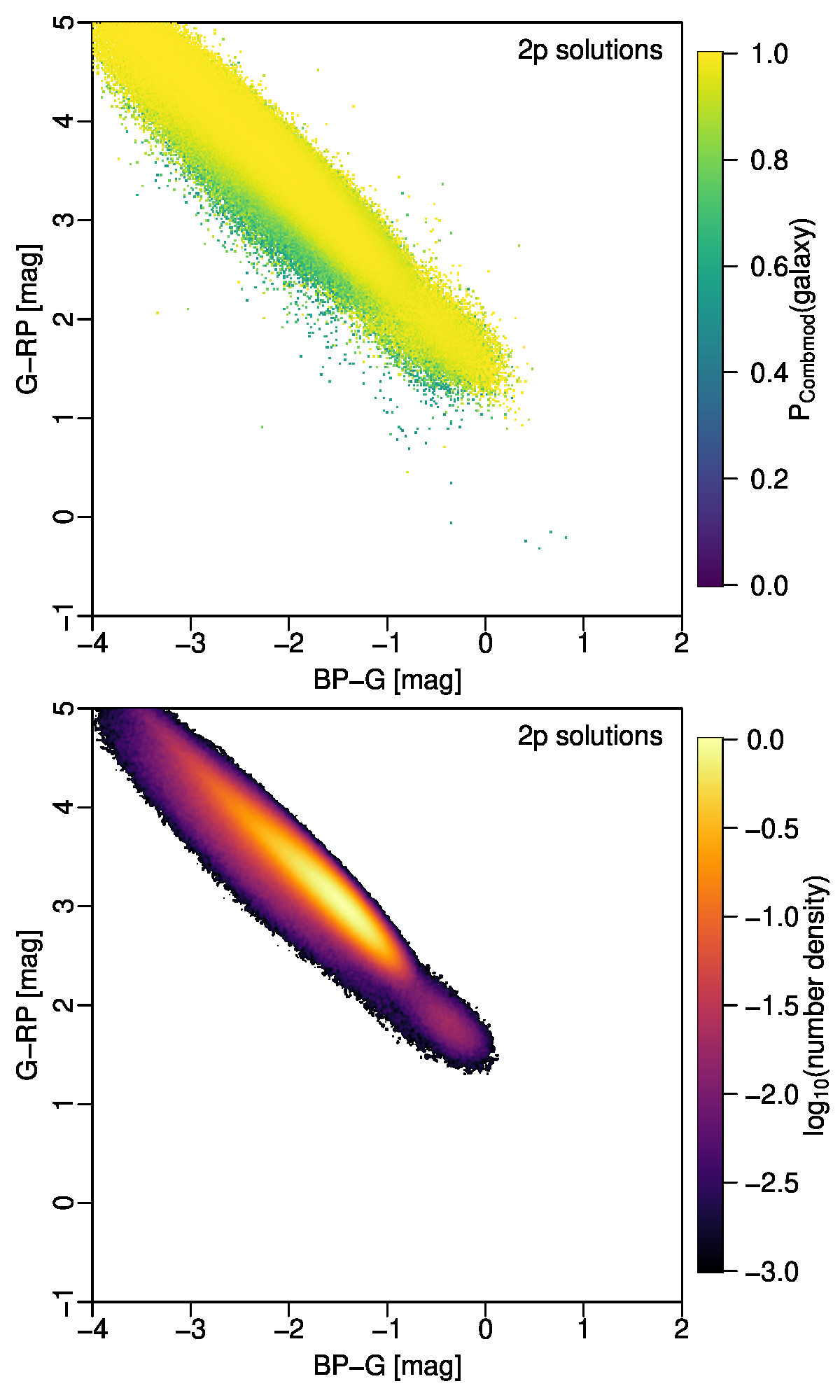}
\caption{
As in Figure~\ref{fig:dr3int5_qsotable_astrometry_classlabel_dsc_ccd_grp_bpg}
but for sources in the \linktoEGTable{galaxy_candidates} table
with \linktoEGParam{galaxy_candidates}{classlabel_dsc}{\tt ='galaxy'}, excluding regions around the LMC and SMC.
The left column shows sources with 5p/6p solutions (0.91 million sources), and the right column shows sources with 2p solutions (2.32 million sources).
These numbers refer to plotted sources, i.e.\ that have all \gaia bands.
\label{fig:dr3int5_galaxytable_astrometry_classlabel_dsc_ccd_grp_bpg}
}
\end{center}
\end{figure*}

We now return to the issue of the 5p/6p and 2p solutions. Figure~\ref{fig:dr3int5_qsotable_astrometry_classlabel_dsc_ccd_grp_bpg} shows the colour--colour diagram for all sources with {\tt classlabel\_dsc='quasar'}, excluding those in the regions around the LMC and SMC, for sources with (5p/6p) and without (2p) parallaxes and proper motions. 
The DSC-Comdmod probabilities for 5p/6p solutions 
come from both Specmod and Allosmod, whereas for the 2p solutions they only come from Specmod. Of the objects classified here as quasars, 95\%  have 5p/6p solutions.
We see that the 5p/6p solutions are confined to a smaller range of colours than are the 2p solutions. That is, demanding the existence of parallaxes and proper motions yields a slightly different population of objects in colour space. We reiterate the fact that there is significant stellar contamination in the 
{\tt classlabel\_dsc='quasar'} sample as a whole.
The (purer) subset defined by {\tt classlabel\_dsc\_joint='quasar'} has a distribution (not shown) similar to that of the 5p/6p solutions in the bottom left panel of Figure~\ref{fig:dr3int5_qsotable_astrometry_classlabel_dsc_ccd_grp_bpg}.

Figure~\ref{fig:dr3int5_galaxytable_astrometry_classlabel_dsc_ccd_grp_bpg} shows the colour--colour diagram for the galaxies. 
Again we see a difference in the colour distribution of the two types of astrometric solution, but now it is the 2p solutions that cover a narrower range of colours. Galaxies are partially resolved by \gaia, and their structure can induce a spurious parallax and proper motion in AGIS (which DSC-Allosmod tries to exploit). Many of these astrometric solutions are rejected by AGIS, turning them into 2p solutions, and these sources can only be classified by Specmod. Of the objects classified here as galaxies, 72\%  have 2p solutions, compared to 5\% for the quasars.
Thus, the Specmod and Allosmod results reported in \gdr{3} are not for identical populations of objects, because of the different input data requirements of these classifiers.

\begin{figure*}[t]
\begin{center}
\includegraphics[width=0.40\textwidth,angle=0]{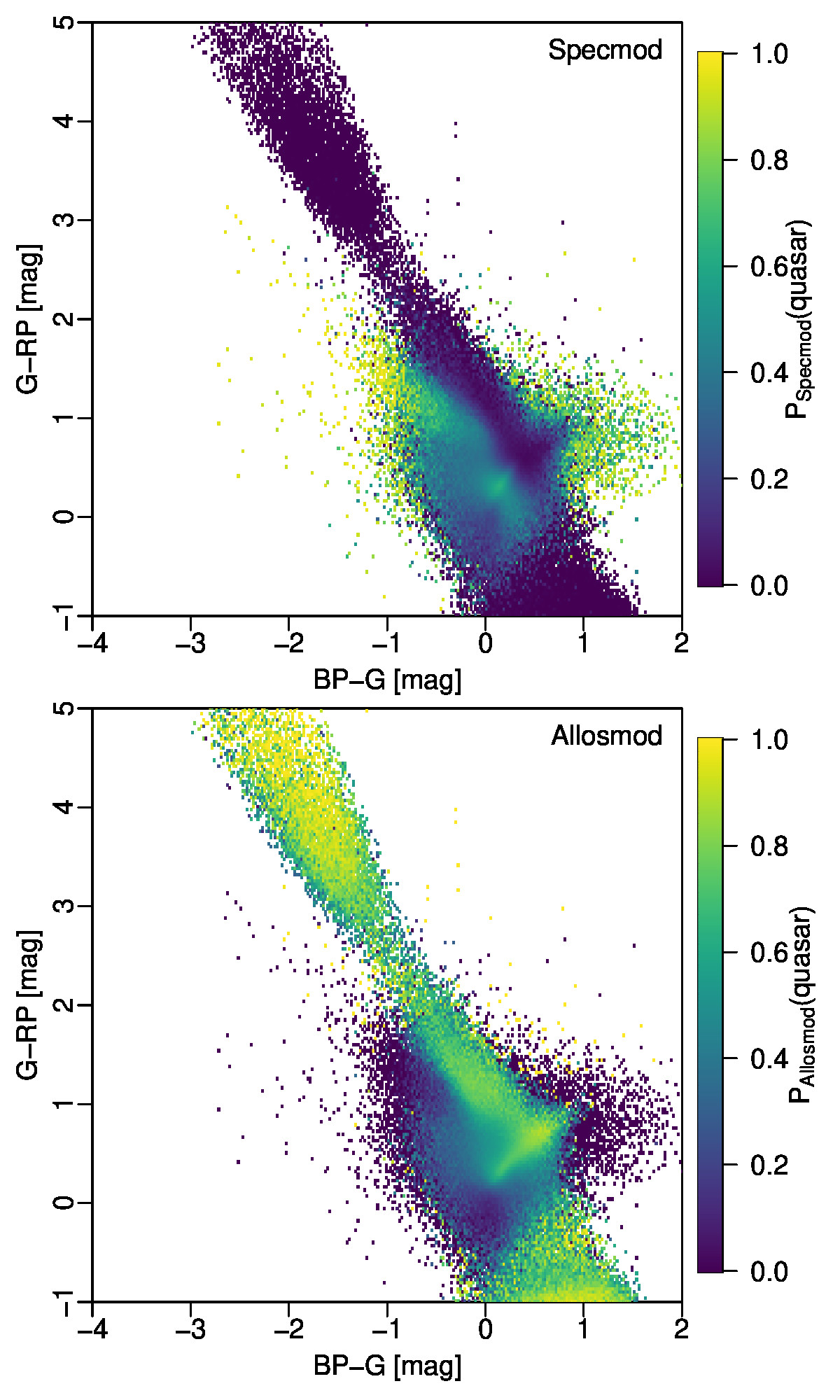}
\includegraphics[width=0.40\textwidth,angle=0]{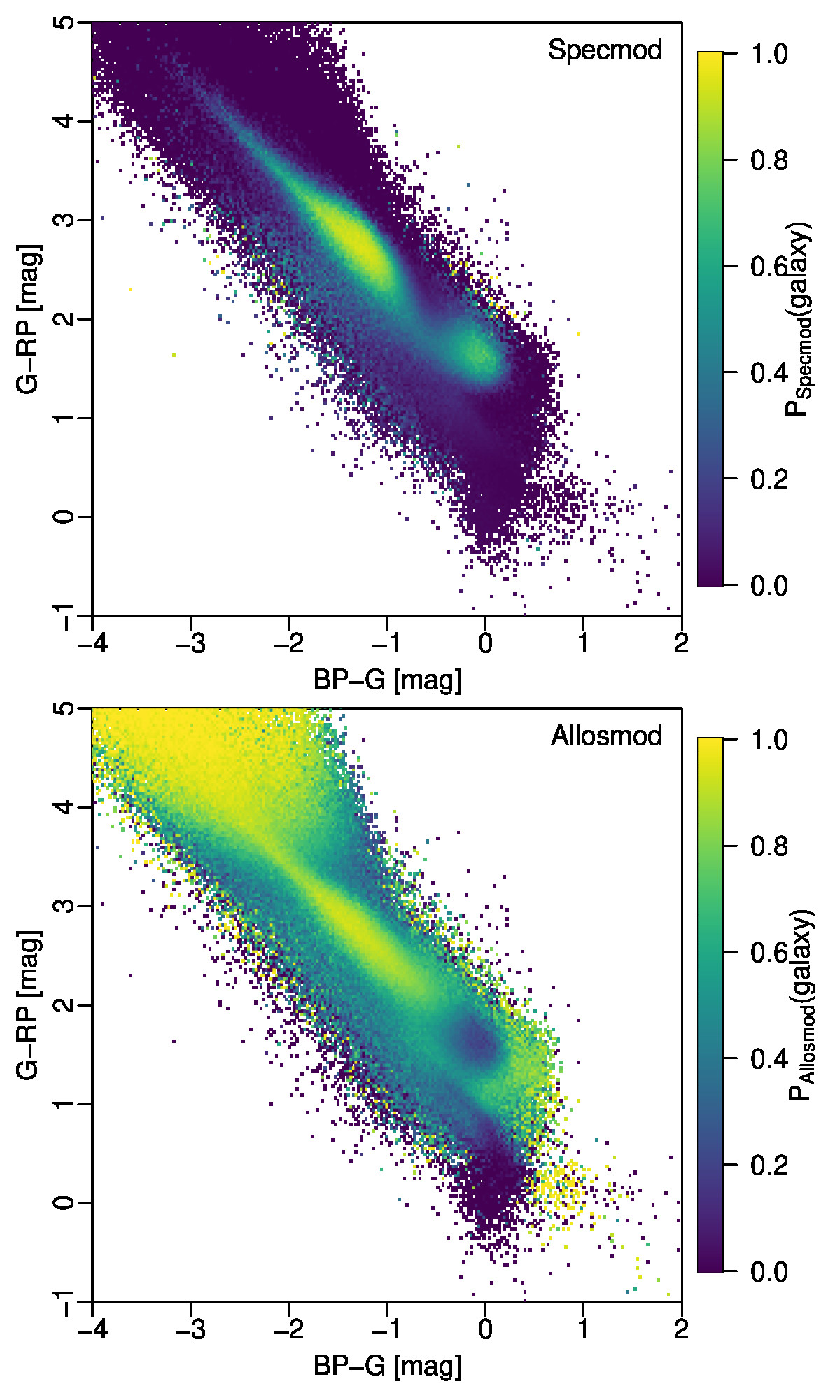}
\caption{Colour--colour diagram for sources in the \linktoEGTable{qso_candidates} table with \linktoEGParam{qso_candidates}{classlabel_dsc}{\tt ='quasar'} (left) and in the \linktoEGTable{galaxy_candidates} table with \linktoEGParam{galaxy_candidates}{classlabel_dsc}{\tt ='galaxy'} (right), in both cases excluding regions around the LMC/SMC, that have both Specmod and Allosmod results.
The upper and lower panels show the mean DSC-Specmod probability and the mean DSC-Allosmod probability, respectively, for a common sample.
\label{fig:dr3int5_bothtables_specmod_and_allosmod_classlabel_dsc_ccd_grp_bpg}
}
\end{center}
\end{figure*}

As Specmod and Allosmod use different data, it is interesting to see how their classification probabilities differ for a common set of sources. We investigate this by selecting sources that have results from both Specmod and Allosmod, and have {\tt classlabel\_dsc} set. This is shown for the quasar candidates in the left column of Fig.~\ref{fig:dr3int5_bothtables_specmod_and_allosmod_classlabel_dsc_ccd_grp_bpg}.
These plots do not convey the number of sources in each part of the diagram, and should therefore be interpreted with that in mind.
Nonetheless, although we see regions where Specmod and Allosmod have similar probabilities, there are also regions where their probabilities are quite different. Because {\tt classlabel\_dsc\_joint} is only set to `quasar' when both Specmod and Allosmod probabilities are above 0.5, these figures explain why that set is comparatively small.
The right column of Figure~\ref{fig:dr3int5_bothtables_specmod_and_allosmod_classlabel_dsc_ccd_grp_bpg} shows the same for the galaxy candidates, and again we see a significant lack of correlation between Specmod and Allosmod. This shows that the different data used by these two classifiers convey rather different information. 


\subsection{Use of DSC results}\label{subsec:dsc_use}

The \dsc class probabilities exist primarily to help users identify quasars and galaxies. 
The performance on white dwarfs and binaries is rather poor. These probabilities will be of limited use to the general user and we do not recommend their use to build samples. One could add these probabilities to the star probability for each source, and thereby end up with a three-class classifier. 

Classification can be done by selecting sources with class probabilities above a given threshold. A threshold of 0.5 gives a selection (and performance) very similar to what would be obtained when taking the maximum probability. A threshold of 0.5 applied to the Combmod outputs is identical to the \fieldName{classlabel_dsc} label (section~\ref{sec:dsc_class_labels}).
With this choice of threshold, the purities for galaxies and quasars are rather modest, as we can see from Table~\ref{tab:cu8par_apsis_dsc_resvst_defset_cp}.
This is unsurprising, because with a threshold of 0.5 we expect up to half of the objects to be incorrectly classified even with a perfect classifier. Increasing the threshold does increase the purity at the cost of decreased completeness, but because the 
DSC probabilities tend to be rather extreme (see plots in~\citealt{LL:CBJ-094}), this does not help as much as one might hope.
The fact that the purities are often lower than the limit expected from the threshold may be due not only to an imperfect classifier, but also to an imperfect calibration of the probabilities in Specmod and Combmod (although not Allosmod).\footnote{The issue of expected sample purity is discussed in section 5.2 of \cite{2008MNRAS.391.1838B}. Even with an imperfect classifier, it is possible to infer the expected number of true sources from the inferred numbers by inverting the confusion matrix, as shown by \cite{2019MNRAS.490.5615B}.}

The DSC completenesses, especially with Combmod, are quite good, but the purities are rather modest, as discussed earlier. This is a consequence of primarily two factors.

The first factor is the intrinsic rareness of the quasars and galaxies. If only one in every thousand sources were extragalactic, then even if our classifier had 99.9\% accuracy, the resulting sample would only be around 50\% pure. This is the situation we have: the intrinsic ability of \dsc to separate the classes is actually very good, with purities of the order of 99\% on balanced test sets. 
However, when it is then applied to a randomly selected set of Gaia data 
there are so many stars that even though a small {\em fraction} of these are misclassified, this is still a large {\em number}. We cannot overcome this problem by adopting a different prior. If we used uniform priors, for example, this would classify many more sources ---both true and false---- as extragalactic. This would increase the completeness of this class. It is not immediately obvious what happens to the purity, 
but \cite{2019MNRAS.490.5615B} found that for Allosmod in \gdr{2}, the purities for quasars and galaxies were actually significantly reduced. 
 
The extreme rareness of the extragalactic objects places high demands on the classifiers, and the performance may be limited by the second factor, namely the ability of the data to distinguish between the classes. We experimented with using different or additional Gaia features (e.g.\ colour excess factor) as inputs to Allosmod, but this did not help. 
Performance might improve if we define synthetic filters from the \bporrp spectra instead of using the entire spectrum, or by generating other features from the Gaia data, but this has not been explored\footnote{One obvious example is to compute the absolute magnitude, because this together with colour -- i.e.\ the HRD -- clearly separates out white dwarfs when the parallax uncertainties are not too large.}. The inclusion of non-Gaia data, such as infrared photometry, should help but was beyond the scope of the activities for \gdr{3}.

A third potential limiting factor is the 
set of training examples we use. Although the SDSS spectroscopic classifications are believed to be very good, they may have errors, and they may also not provide the clearest distinction between galaxies and quasars.

The fact remains that the classification performance depends unavoidably on the intrinsic rareness, that is,\ on the prior. Users may want to adopt a different prior from ours (Table~\ref{tab:cu8par_apsis_dsc_classprior}), which would be particularly appropriate if they focus on a subset of parameter space. To recompute the DSC  probabilities with a new prior we do not need to re-train or re-apply DSC. The fact that DSC provides posterior probabilities as outputs makes it simple to strip off our prior and apply a new one, as shown in appendix~\ref{sec:cu8par_apsis_dsc_adjusting_probabilities}.

It is important to realise that the performances in Table~\ref{tab:cu8par_apsis_dsc_resvst_defset_cp} are (a) only for the classes as defined by the training data and (b) an average over the entire Gaia sample, and are therefore dominated by faint sources with lower quality data.
Our galaxy class in particular is a peculiar subset of all galaxies, because \gaia tends not to observe extended objects, and even then may not measure them correctly (see section~\ref{subsec:dsc_method}).

\dsc misclassifies some very bright sources that are obviously not extragalactic, for example. 
As these are easily removed by the user, we chose not to filter the DSC results in any way.
One may likewise wonder why there are some objects classified as quasars with statistically significant proper motions . We do use proper motion as a classification feature, but in a continuous fashion, not as a hard cut. A more conservative approach to classification is to apply a series of necessary conditions, that is,\ a simple decision tree. This could increase the purity ---and could be tuned to guarantee that certain known objects come out correctly--- but at the expense of completeness. 
We do nevertheless provide 
the class label \fieldName{classlabel_dsc_joint}
as a means to select a purer subsample of extragalactic sources (section~\ref{sec:dsc_class_labels}), as can be seen from the last two columns of Table \ref{tab:cu8par_apsis_dsc_resvst_defset_cp}.

\section{Outlier analysis (OA)}
\label{sec:oa}

\subsection{Objectives}
\label{subsec:oa_objectives}

The Outlier Analysis (\oa) module aims to complement the overall classification performed by the \dsc module, by processing those objects with lower classification probability from \dsc (see \secref{sec:dsc}). \oa is intended to analyse abnormal or infrequent objects, or artefacts, and was applied to all sources that received \dsc Combmod probabilities below 0.999 in all of its five classes. This threshold was chosen so as to process a limited number of 134 million sources, corresponding to about $10\%$ of the total number of sources for which \dsc produced probabilities. Subsequently, a selection of the sources to be processed is carried out based on several quality criteria, the most restrictive being that the mean spectra correspond to at least five transits (see details in the \linktosec{cu8par}{apsis}{oa}). The resulting filtering leads us to process a total of 56\,416\,360 sources. Such sources tend to be fainter and/or have noisier data. For these objects, \oa provides an unsupervised classification ---where the true object types are not known---  that complements the one produced by \dsc, which follows a supervised approach based on a set of fixed classes.

\subsection{Method} \label{subsec:oa_method}
The method used by \oa to analyse the physical nature of classification outliers is based on a self-organising map \citep[SOM,][]{Kohonen1982}, which groups objects with similar \bporrp spectra (see \secref{subsubsec:oa_method_preprocessing}) according to a Euclidean distance measure. The SOM performs a projection of the multidimensional input space of \bporrp into a two-dimensional grid of size $30\times 30$, which facilitates the visual interpretation of clustering results. Such a projection is characterised by its preservation of the topological order, in the sense that, for a given distance metric, similar data in the input space will belong to the same or to neighbouring neurons in the output space. Each one of these neurons has a prototype, which is adjusted during the training phase and that best represents the input spectra that are closest to this neuron. In \gdr{3,} each prototype is the average spectrum of the pre-processed\footnote{The \oa pre-processing of \bporrp spectra is later described in \secref{subsubsec:oa_method_preprocessing}.} \bporrp spectra of the sources assigned to that particular neuron, which correspond to those closest to the neuron according to the Euclidean distance between the neuron prototype and the pre-processed \bporrp spectrum of the source. Neuron prototypes are reported in the \linktoAPTable{oa_neuron_xp_spectra} table. A centroid is also identified for each neuron, which is the source whose pre-processed \bporrp spectrum is the closest to the prototype of the neuron, according to the Euclidean distance. Centroids can be found in the \linktoAPParam{oa_neuron_information}{centroid_id} field of the \linktoAPTable{oa_neuron_information} table along with statistics of the main Gaia observables for the sources belonging to this neuron: \gmag, \gbp, and \grp magnitudes, proper motions, Galactic latitude, parallax, number of \bporrp transits, renormalised unit weight error (\linktoMainParam{gaia_source}{ruwe}), \bporrp flux excess factor, and \bpminrp colour.

\subsubsection{\bporrp spectra preprocessing}\label{subsubsec:oa_method_preprocessing}

The sampled mean \bporrp spectra produced by \smsgen are transformed in order to remove artefacts, and to improve the clustering produced by the SOMs: (a) Pixels with negative or zero flux values are linearly interpolated, provided that they do not affect more than 10\% of the effective wavelength in a consecutive manner or more than 25\% of the entire effective wavelength. Such a filtering was imposed because most of the spectra that did not meet such criteria were usually of low quality and had a low number of transits. These filtered spectra are not analysed; (b) BP and RP spectra are downsampled to 60 pixels each; (c) both spectra are trimmed to avoid the low transmission regions of the CCD, so that \oa uses the effective wavelength ranges $375$--$644$nm for BP and $644$--$1050$nm for RP; (d) spectra are concatenated to obtain a single spectrum; and, (e) the joint spectrum is normalised so that the sum of its flux is equal to one.

\begin{figure*}[t]
    \centering
    \includegraphics[width=0.75\textwidth]{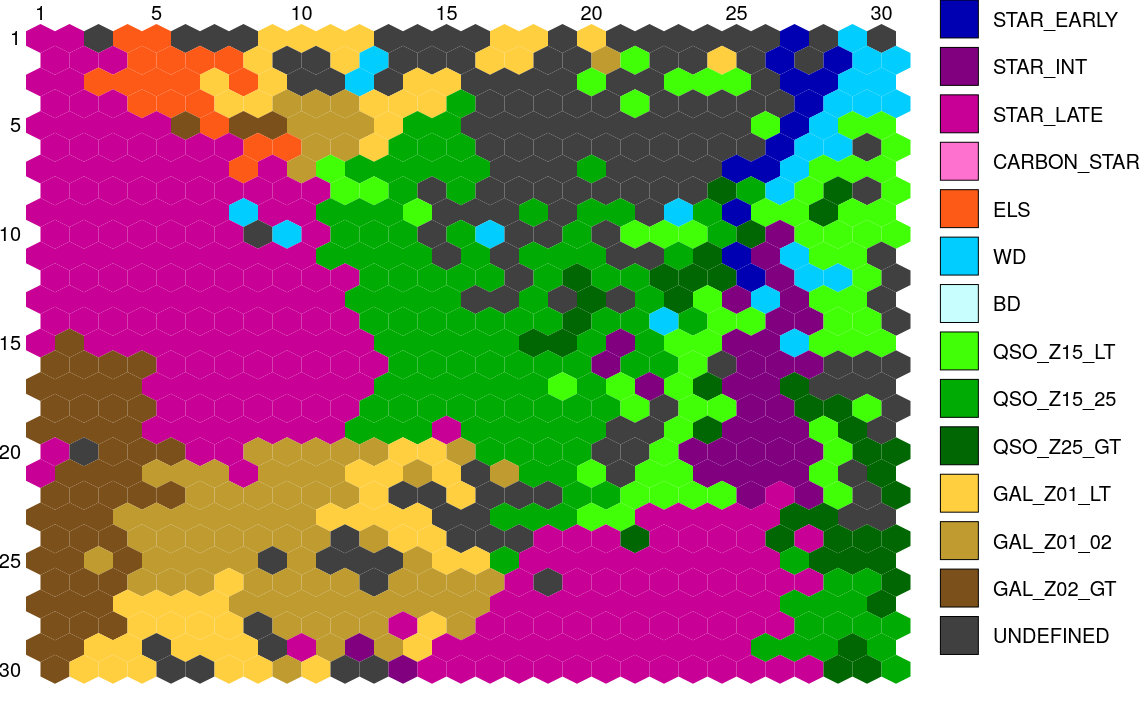}
    \caption{SOM grid from the \oa module visualised through the GUASOM tool~\citep{Alvarez2021}. Each cell corresponds to a neuron from the SOM, most of which were assigned a class label. Those neurons that did not meet the quality criteria defined to establish a class label remain `undefined', as explained in Section~\ref{subsubsec:oa_method_labelling}}
    \label{fig:oa_method_labelling_combined_map}
\end{figure*}

\subsubsection{Quality assessment}\label{subsubsec:oa_method_quality}

The performance of \oa cannot be measured through metrics such as completeness and purity because of the unsupervised nature of the technique. Therefore, a descriptive approach based on the intra-neuron and inter-neuron distances~\citep{Alvarez2021} was followed in order to analyse the quality of the clustering. We decided to use the squared Euclidean distance as a proxy for distance because the SOM algorithm uses it as a measurement of mean quantisation error for processing elements. The intra-neuron distance of each source is then computed as the squared value of the Euclidean distance between the source and the prototype of the neuron it belongs to, whereas the inter-neuron distance is computed as the squared Euclidean distance between two different neuron prototypes. 
In order to assess the quality of the clustering, we selected the three parameters that we thought best describe the distribution of the intra-neuron distances: (a) the width of the distribution according to the value of the full width at half maximum ($FWHM$); (b) the skewness ($S$), which measures its asymmetry; and, (c) the kurtosis excess ($K$), which measures the level of concentration of distances. 
A high-quality clustering will result from neurons with low values of the $FWHM$ parameter, and large positive values of both skewness and kurtosis. 
Finally, in order to facilitate the interpretation of such quality measurements, a categorical index named $QC$ was derived based on the values obtained for $S$, $K$, and a normalised version of $FWHM$ (which is reversed in order for the higher quality neurons to take larger values). 
To this purpose, seven quality categories were established, according to the values taken by such parameters with respect to six arbitrarily chosen percentiles ($95^{\rm th}$, $90^{\rm th}$, $75^{\rm th}$, $50^{\rm th}$, $32^{\rm th}$, and $10^{\rm th}$), which are computed independently for each one of the parameters listed above over the entire map. 
For each neuron, we determine the lowest percentile in which the three parameters are above their respective percentile values. 
Thus, if a value is above the $95^{\rm th}$ percentile, then $QC$ will take the value of zero; if it is in the $90^{\rm th}$ percentile, then $QC$ will correspond to category one, and so on up to category six, which will correspond to those neurons whose poorest quality indicator is outside the lowest percentile that has been considered, $10^{\rm th}$. Accordingly, the best-quality neurons will have $QC=0$ and the worst ones $QC=6$.  It should be emphasised here that $QC$ only assesses the quality of the clustering (i.e. how closely the pre-processed \bporrp spectra in a neuron match their prototype) compared to the overall intra-neuron distances, such that no assumption should be made on the quality of the spectra they contain, nor on the labelling of the individual neurons described below.

\subsubsection{Neuron labelling}\label{subsubsec:oa_method_labelling}

Unsupervised methods do not directly provide any label to the samples that are being analysed. For this reason, a set of reference \bporrp spectra templates for prototypical astronomical objects was built by taking into account validation sources from the various \apsis modules (see the \linktosec{cu8par}{apsis}{oa}). These reference templates are used to label the neurons in \gdr{3} by identifying the closest template to the neuron prototype according to the Euclidean distance. In addition, to guarantee the suitability of the assigned templates (and class labels), two conditions were imposed: (a) the squared Euclidean distance between a template and the neuron prototype must not exceed a threshold of $3.58 \times 10^{-2}$; and, (b) the neuron must have $QC < 6$. \figref{fig:oa_method_labelling_combined_map} shows the SOM built by \oa for \gdr{3}, where around 80\% of the neurons were assigned a template, and hence a class label. The limit of $3.58 \times 10^{-2}$ on the squared distance was set during the template-building process and is detailed in the \linktosec{cu8par}{apsis}{oa}.

\subsubsection{GUASOM visualisation tool}\label{subsubsec:oa_guasom}

To help the user to analyse and visualise the clustering results, we designed an application called Gaia Utility for the Analysis of Self-Organising Maps (GUASOM)~\citep{Alvarez2021}. It can be run over the internet, and contains several visualisation utilities that allow an interactive analysis of the information present on the map. The tool provides both classical and specific domain representations such as U-matrix, hits, parameter distributions, template labels, colour distribution, and category distribution.

\subsection{Performance and results}\label{subsec:oa_results}

\begin{figure}[t]
    \centering
    \includegraphics[width=0.9\columnwidth]{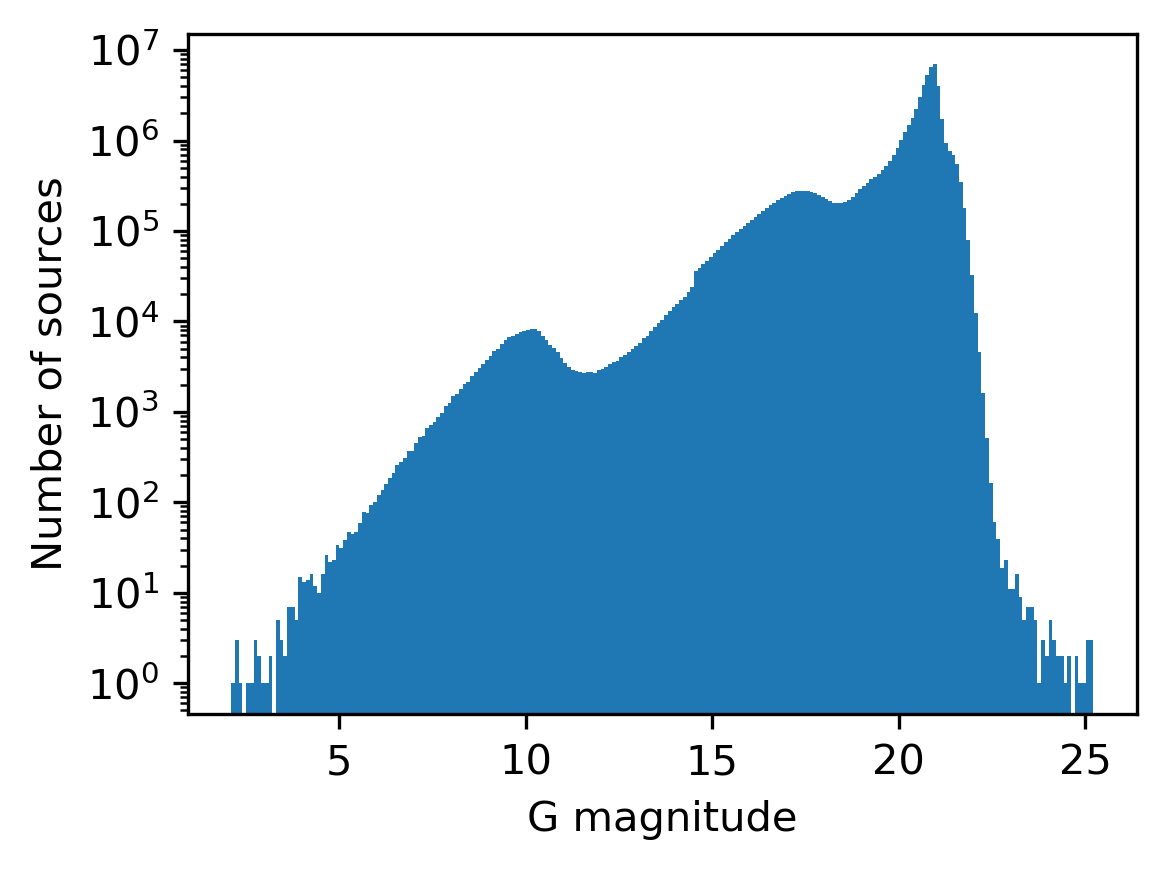}
    \caption{\gmag mag distribution of the $56\,416\,360$ sources processed by the \oa module in \gdr{3} (bin width of $0.1$).}
    \label{fig:oa_results_sources_g_mag_distribution}
\end{figure}

\oa processed $56\,416\,360$ objects in \gdr{3}. \figref{fig:oa_results_sources_g_mag_distribution} displays their \gmag magnitude distribution, demonstrating that \oa covers a wide range of \gmag magnitudes with a significant fraction of faint objects.

\begin{figure}[t]
    \centering
    \includegraphics[width=0.9\columnwidth]{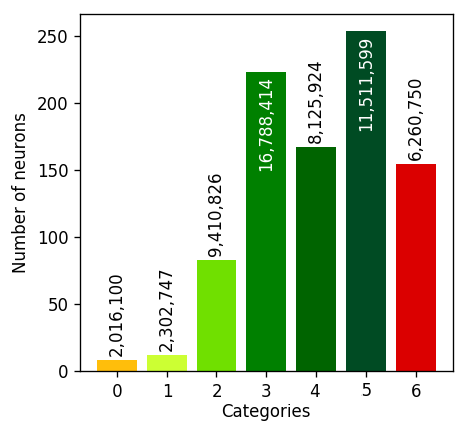}
    \caption{Histogram of
neuron quality categories for the sources processed by the \oa in \gdr{3}. The number of sources per category is superimposed along with the bars. Those neurons with $0 \leq QC \leq 3$ are considered high-quality neurons.}
    \label{fig:oa_method_labelling_quality_histogram}
\end{figure}

\begin{figure}[t]
    \centering
    \includegraphics[width=\columnwidth]{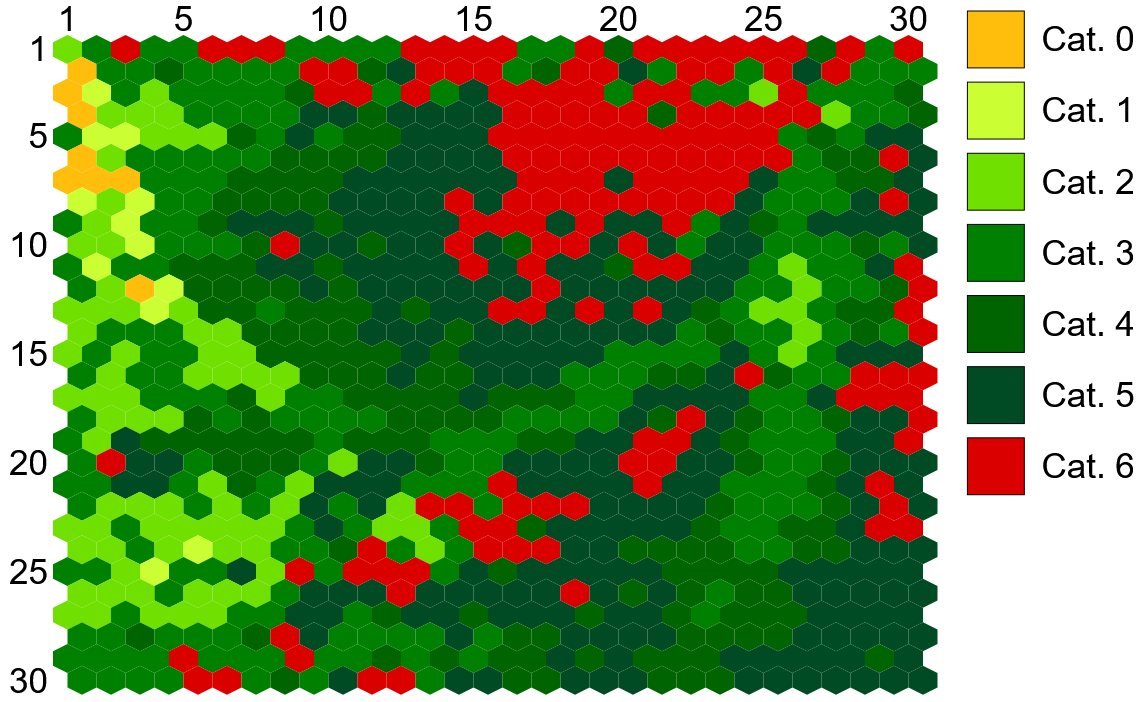}
    \caption{SOM grid visualised through the GUASOM tool~\citep{Alvarez2021} to represent the quality category~($QC$) assigned to each neuron.}
    \label{fig:oa_method_results_quality_map}
\end{figure}

\figref{fig:oa_method_labelling_quality_histogram} shows the histogram of neuron quality categories, $QC$, where the total number of sources belonging to such neurons is superimposed. Approximately 35\% of the neurons have $0 \leq QC \leq 3$ and are hence referred to as `high-quality neuron': these comprise around 55\% of the sources processed. The rest of the neurons can be considered as low-quality neurons. 
\figref{fig:oa_method_results_quality_map} shows how the quality categories are distributed over the SOM. 

It is worth mentioning that the SOM does not directly label neurons, nor does it provide quality measurements on the clustering they produce, which means that we have to apply the procedures described in Sections~\ref{subsubsec:oa_method_quality} and~\ref{subsubsec:oa_method_labelling} after we build the map. As a result, \figref{fig:oa_method_results_quality_map} shows the quality category associated with each neuron in our grid of $30 \times 30$ neurons. These quality categories assess how well the sources fit to the prototype of the neuron they belong to: neurons with the lowest quality category are composed of sources whose spectra are the most homogeneous (i.e. neurons of highest quality). Similarly, in \figref{fig:oa_method_labelling_combined_map}, the label assigned to each neuron provides a hint as to the astronomical type of the sources they contain. Comparing Figures \ref{fig:oa_method_labelling_combined_map} and \ref{fig:oa_method_results_quality_map}, we can see that high-quality neurons mostly correspond to stars and galaxies, while quasars are usually associated with low-quality neurons. The reason for this mostly stands in the wide range of cosmological redshifts that is observed amongst those objects, in their different continuum shapes and emission-line equivalent widths.

\tabref{tab:oa_results_confusion_matrix} represents the contingency table between \dsc Combmod and \oa class labels. \dsc labels are determined according to the class with the highest \dsc Combmod probability, except for those that take a probability below $0.5$, which are labelled as `unknown'. Sources with \dsc `binary star' class are considered as `star' as the former class is not present in \oa. Similarly, \oa class labels are aggregated into more generic ones in order to enable comparison with the \dsc class labels. Recalling that \oa only processes sources with all \dsc Combmod probabilities below $0.999$, the \oa results can be summarised as follows. 
\begin{itemize}[leftmargin=0.5cm]
    \item {Galaxies: There is close agreement for galaxies, as around 80\% of the galaxies identified by \dsc are also confirmed by \oa.}
    \item {Quasars: The agreement with \dsc decreases to 35\%. A large fraction of those quasars identified by \dsc are considered as stars or white dwarfs by \oa.}
    \item {Stars: Around 40\% of those identified by \dsc were also confirmed by \oa. However, a large fraction of them were considered as extragalactic objects by \oa.}
    \item {White dwarfs: In this case, the agreement between both modules is around 50\%. Most of the remaining objects are considered as stars by \oa.}
\end{itemize}
Around 11\% of the sources are assigned to a neuron that was not labelled by \oa because of their poor quality (category six). In particular, approximately 2\,510 sources could not be classified by \oa and have {\tt classlabel\_dsc = 'unclassified'}, meaning that studying their nature may require a deeper analysis.


\begin{table*}[t]
\centering
\begin{tabular}{|ll|rrrrr|r|}
        \cline{3-7}
        \multicolumn{2}{c|}{} & \multicolumn{5}{c|}{\textbf{\oa class label}}\\
        \multicolumn{2}{c|}{} & \textbf{STAR} & \textbf{WD} & \textbf{QSO} & \textbf{GAL} & \textbf{UNDEFINED} & \multicolumn{1}{c}{\textbf{Total}}\\
        \hline
        \multirow{6}{*}{\rotatebox[origin=c]{90}{\textbf{\dsc}}} & \textbf{STAR} & $40$\% & $3$\% & $22$\% & $24$\% & $11$\% & $53\,295\,527$\\
        & \textbf{WD} & $42$\% & $51$\% & $3$\% & $0$\% & $4$\% & $92\,186$\\
        & \textbf{QSO} &  $29$\% &  $21$\% &  $35$\% &  $2$\% &  $13$\% & $2\,158\,916$\\
        & \textbf{GAL} &  $4$\% &  $0$\% &  $9$\% &  $83$\% &  $4$\% & $851\,127$\\
        & \textbf{UNKNOWN} &  $22$\% &  $7$\% &  $35$\% &  $22$\% &  $13$\% & $18\,604$\\
        \hline
        \multicolumn{2}{r|}{\textbf{Total}} & $21\,763\,876$ & $2\,240\,195$ & $12\,680\,763$ & $13\,470\,776$ & $6\,260\,750$\\
        \cline{3-7}
\end{tabular}

\caption{Contingency table between DSC taken from predominant probabilities produced by \dsc Combmod and OA classifications, grouped into generic types. Unknown means that the DSC predominant probability was below $0.5$, whereas for OA it means that no template was assigned due to quality constraints. Fractions are computed with respect to the total number of sources in each \dsc class.}
\label{tab:oa_results_confusion_matrix}
\end{table*}

\subsection{Use of \oa clustering}\label{subsec:oa_use}

The analysis performed by the \oa module can be useful for different purposes. For instance, high-quality neurons can help to assess the physical nature of some sources with \dsc combmod probabilities below the chosen threshold ($0.999$) in all classes or to identify objects that were potentially misclassified. As \oa provides an unsupervised classification based on a normalised SED comparison, for a given neuron there are sources with different degrees of similarity to the prototype. For that reason, we encourage the user to isolate clean samples for each neuron through the quality measurements provided in the \linktosec{cu8par}{apsis}{oa}. In particular, we suggest combining both the categorical quality index~($QC$) and the classification distance in order to retrieve the best classified sources from \oa. \tabref{tab:oa_use_reliable} shows the number of sources per class that are assigned to a high-quality neuron (from category zero to three), and whose classification distance between the pre-processed \bporrp spectrum of the source and the neuron prototype is below $0.001$ (i.e. what we consider here as reliable predicted classes). As can be seen, around 13 million stars, 9 million galaxies, 2 million quasars, and $1.5$ million white dwarfs meet these criteria.


\begin{table}[t]
\centering
\begin{tabular}{lr}
\hline
\textbf{Class label} &  \textbf{Number of sources}\\
\hline
STAR\_LATE  & $8\,966\,955$ \\
GAL\_Z01\_02 & $3\,917\,749$ \\
STAR\_INT   & $3\,158\,041$ \\
GAL\_Z02\_GT & $2\,952\,297$ \\
GAL\_Z01\_LT & $2\,355\,895$ \\
WD & $1\,561\,204$ \\
QSO\_Z15\_LT & $1\,138\,832$ \\
QSO\_Z15\_25 & $1\,020\,337$ \\
STAR\_EARLY & $914\,470$ \\
ELS & $489\,551$ \\
QSO\_Z25\_GT & $92\,460$ \\
\hline
\end{tabular}
\caption{Number of sources in each \oa class that belong to a high-quality neuron while having a classification squared Euclidean distance below $0.001$ (i.e. what we consider here as reliable). We note that there may be considerable contamination in these class assignments.}
\label{tab:oa_use_reliable}
\end{table}

\section{Quasar classifier (QSOC)}
\label{sec:qsoc}

\subsection{Objectives} \label{subsec:qsoc_objective}

The quasar classifier (\qsoc) module is designed to determine the redshift, $z$, of the sources that are classified as quasars by the \dsc module (see Section \ref{sec:dsc} for more details). In order to produce redshift estimates for the most complete set of sources, we considered a very low threshold on the \dsc quasar probability of \linktoEGParam{qso_candidates}{classprob_dsc_combmod_quasar} $\geq 0.01$, meaning that we expect a significant fraction of the processed sources to be stars or galaxies. Users interested in purer sub-samples may then require that \linktoEGParam{qso_candidates}{classlabel_dsc_joint} {\tt = 'quasar'}, as explained in Section \ref{sec:dsc_class_labels}, or may use more sophisticated filtering, as explained in \cite[Section 8]{DR3-DPACP-101}.

\subsection{Method}

\subsubsection{Overview}
\label{subsubsec:qsoc_overview}

\qsoc is based on a $\chi^2$ approach that compares the observed \bporrp spectra sampled by \smsgen \citep[see][and the \linktosec{cu8par}{apsis}{smsgen}]{DR3-DPACP-157} to quasar rest-frame templates in order to infer their redshift.  The predicted redshifts take values in the range $0.0826 < z < 6.12295$. As the effective redshift is not necessarily the one associated with the minimal $\chi^2$ (see Section \ref{subsubsec:qsoc_quasar_algorithm}), it is complemented by an indicator of the presence of quasar emission lines ($Z_{\rm score}$ from Equation \ref{eq:qsoc_zscore}) and these are converted into a redshift score, $S$, from Equation \ref{eq:qsoc_redshift_score}. For a given source, the redshift with the highest score is then the one that is selected by the algorithm. Quasar templates are described in Section \ref{subsubsec:qsoc_quasar_templates} while the redshift determination algorithm is described in Section  \ref{subsubsec:qsoc_quasar_algorithm}.

\subsubsection{Quasar templates} \label{subsubsec:qsoc_quasar_templates}

\begin{figure*}
    \centering
    \includegraphics[width=\textwidth]{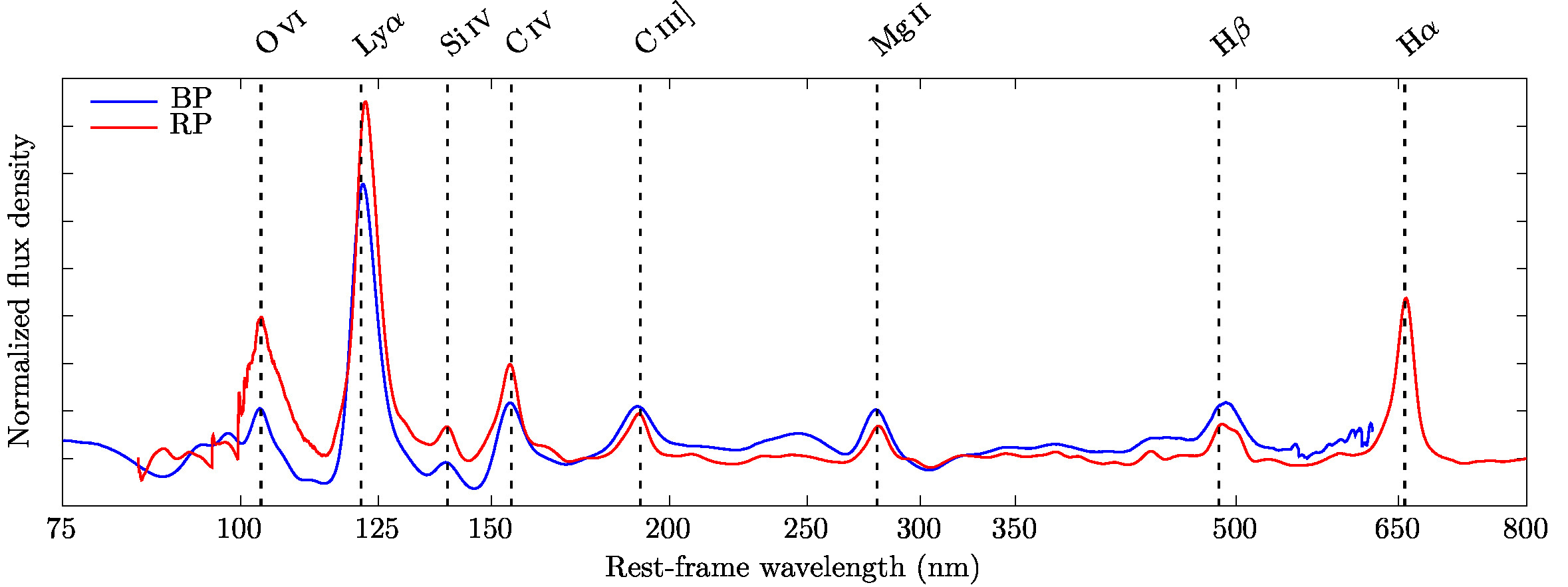}
    \caption{Rest-frame quasar templates used by \qsoc. These correspond to the dominant templates taken over the 32 templates that are computed based on the method described in \cite{2015MNRAS.446.3545D} and applied to 297\,264 quasars from the DR12Q catalogue that are converted into \bporrp spectra through the use of the \bporrp spectrum simulator provided by CU5.}
    \label{fig:qsoc_templates}
\end{figure*}

The quasar templates used by \qsoc were built based on the method described in \cite{2015MNRAS.446.3545D} and applied to 297\,264 quasars\footnote{We note that for 37 of the 297\,301 quasars originally contained in the DR12Q catalogue, the $\ell$-1 norm fit of the continuum to the observed spectrum (later described) did not converge and these were accordingly not included in the final sample we used.} from the twelfth release of the Sloan Digital Sky Survey Quasar catalogue of \cite[DR12Q]{2017A&A...597A..79P}. These spectra are first extrapolated to the wavelength range of the \gaia \bporrp spectro-photometer (i.e. $300$--$1100$ nm) with a linear wavelength sampling of $0.1$ nm using a procedure similar to the one used by \cite{2018MNRAS.473.1785D}. They are subsequently converted into \bporrp spectra through the use of the \bporrp spectrum simulator provided by CU5 and described in \cite{EDR3-DPACP-120}. An artificial spectrum with a uniform SED (i.e. of constant flux density per wavelength) was also converted through the \bporrp spectrum simulator in order to produce the so-called `flat \bporrp spectrum'. We then divided each simulated \bporrp spectrum by its flat counterpart before subtracting a quadratic polynomial that is fitted to the observations in a least absolute deviation sense (i.e. $\ell$-1 norm minimisation), leaving pure emission line spectra. We note that, in order to avoid fitting emission lines, a second-order derivative of the flux density was estimated around each sampled point, $d^2 f_i / d \lambda_i^2$, and later used to scale the associated uncertainties by a factor of $\operatorname{max}(\left| d^2 f_i / d \lambda_i^2 \right| / M, 0.01),$ where $M$ is a normalisation factor equal to the maximal absolute value of the second-order derivatives evaluated over all the sampled points. As the continuum regions often have very low curvatures compared to the emission lines, they are usually overweighted by a factor of up to 100 in the $\ell$-1 norm minimisation. A logarithmic wavelength sampling of $\log L = 0.001$ was then used for both the BP and RP templates, ensuring that the resolution of the \bporrp spectra, as sampled by \smsgen, is preserved. We extracted 32 \bporrp templates based on these 297\,264 simulated spectra using the weighted principal component analysis method described in \cite{2015MNRAS.446.3545D}; nevertheless, only the dominant \bporrp templates ---corresponding to the mean of the weighted principal component analysis method--- were used because cross-validation tests performed on the simulated spectra show that a larger number of templates
significantly increases the degeneracy between redshift predictions.

The resulting templates, illustrated in Figure \ref{fig:qsoc_templates}, closely match the typical composite spectra of quasar emission lines \citep[see e.g.][Section 7]{DR3-DPACP-101}, although they are convolved by the \gaia line spread function which is averaged over the entire set of rest-frame wavelengths. The templates cover the rest-frame wavelength range from $45.7$ nm to $623.3$ nm in BP and from $84.6$ nm to $992.3$ nm in RP. These limits, along with the observed wavelength coverage imposed by \smsgen of $325$--$680$ nm in BP and $610$--$1050$ nm in RP allow \qsoc to predict redshifts in the range $0.0826 < z < 6.1295$\footnote{As the cross correlation function computed by \qsoc is extrapolated by $\pm \log L$ at its border, the range of the \qsoc redshift predictions is slightly wider than one would expect from a straight comparison of the observed and rest-frame wavelengths.}. 

\subsubsection{Algorithm}
\label{subsubsec:qsoc_quasar_algorithm}

The determination of the redshift of quasars by \qsoc is based on the fact that the redshift, $z$, turns into a simple offset once considered on a logarithmic wavelength scale:
\begin{equation}
Z =  \log (z + 1 ) = \log \lambda_{\rm obs} - \log \lambda_{\rm rest},
\label{eq:qsoc_log_redshift}
\end{equation}
where we assume that a given spectral feature located at rest-frame wavelength $\lambda_{\rm rest}$ is observed at wavelength $\lambda_{\rm obs}$. Consider such a logarithmic sampling $\lambda_i = \lambda_0\, L^i$, where $\lambda_0$ is a reference wavelength and $L$ is the logarithmic wavelength sampling we use, here $\log L = 0.001$ (or $L \approx 1.001$). Then for a given set of $n$ rest-frame templates, $\mat{T}$, and an observation vector, $\vec{s}$, which are both logarithmically sampled with $L$, the derivation of the optimal shift, $k$, between $\mat{T}$ and $\vec{s}$ can be formulated as a $\chi^2$ minimisation problem through
\begin{equation}
\chi^2(k) = \sum_i \frac{1}{\sigma_i^2} \left( s_i - \sum_{j=1}^{n} a_{j,k} T_{i+k,j} \right)^2
\label{eq:qsoc_chi2}
,\end{equation}
where $\sigma_i$ is the uncertainty on $s_i$ and $a_{j,k}$ are the coefficients that enable the fit of $\mat{T}$ to $\vec{s}$ in a weighted least squares sense while considering a shift $k$ that is applied to the templates. The redshift that is associated with the shift $k$ is then given by $z = L^k - 1$. A continuous estimation of the redshift is then obtained by fitting a quadratic polynomial to $\chi^2(k)$ in the vicinity of the most probable shift.

Despite its appealing simplicity, Equation \ref{eq:qsoc_chi2} is known to have a cubic time complexity on $N$, as shown in \cite{2016MNRAS.460.2811D}, where $N$ is the number of samples contained in each template. In the same manuscript, it is shown that the computation of the \emph{cross-correlation function} (CCF), defined as
\begin{equation}
\operatorname{ccf}(k) = \left(\sum_i \frac{s_i^2}{\sigma_i^2}\right) - \chi^2(k) = C - \chi^2(k),
\label{eq:qsoc_ccf}
\end{equation}
requires only $\mathcal{O}\left(N \log N\right)$ floating point operations. Furthermore, given that $C$ is independent of the explored shift, $k$, maximising $\operatorname{ccf}(k)$ is equivalent to minimising $\chi^2(k)$. 

However, some features of the \bporrp spectra complicate the computation of the CCF. First, the BP and RP spectra are distinct such that the effective CCF is actually composed of the sum of two CCFs associated with the BP and RP spectra and templates, $\operatorname{ccf}_{\rm bp}(k)$ and $\operatorname{ccf}_{\rm rp}(k)$, respectively:
\begin{equation}
\operatorname{ccf}(k) = \operatorname{ccf}_{\rm bp}(k) + \operatorname{ccf}_{\rm rp}(k).
\label{eq:qsoc_ccf_xp}
\end{equation}
Secondly, the \bporrp spectra have bell shapes (i.e.\ their flux smoothly goes to zero at the borders of the spectra), and have spectral flux densities that are integrated over wavelength bins of different sizes, as explained in \cite{DR3-DPACP-157}. Equation \ref{eq:qsoc_ccf} is therefore not directly applicable to these spectra. In order to overcome these difficulties, we divided each \bporrp spectrum by the previously mentioned flat \bporrp spectrum (i.e. \bporrp spectrum coming from a constant flux density and converted through the \bporrp spectrum simulator) and updated their uncertainties accordingly. This solution enables us to solve both the bell shape issue and the varying wavelength size of each pixel, passing from units of flux to units of flux density. Finally, most of the quasar flux resides in its continuum, which we model here as a second-order polynomial, concatenated to the set of templates, $\mat{T}$, and subsequently fitted to the observations in Equation \ref{eq:qsoc_ccf}.

\begin{table*}
\caption{The QSOC parameters used to compute the redshift score of quasars from Equation \ref{eq:qsoc_redshift_score} and the $Z_{\rm score}$ from Equation \ref{eq:qsoc_zscore}. The rest-frame wavelengths, $\lambda$, of each emission line were  retrieved from the quasar templates described in Section \ref{subsubsec:qsoc_quasar_templates}. Theoretical emission line intensities, $I_\lambda$, and score parameters, $w_0$, $w_1$, and $p$, were computed based on a global optimisation procedure that is designed to maximise the score of the redshift predictions with $|\Delta z| < 0.1$ amongst 88\,196 randomly selected sources with a redshift estimate from DR12Q. We note that another set of 89\,839 observations was then kept as a test set, though the two sets provide a similar distribution of scores.}
\begin{center}
\footnotesize
\begin{tabular}{rccccccccc}
\hline
\multicolumn{10}{c}{\bf Parameters of the redshift score} \\
\hline
& \multicolumn{3}{c}{$w_0 = 0.71413$} & \multicolumn{3}{c}{$w_1 = 0.28587$} & \multicolumn{3}{c}{$p = 0.24365$} \\ \\
\hline
\multicolumn{10}{c}{\bf Parameters of the $\boldsymbol{Z_{\rm score}}$ for BP spectra} \\
\hline
& \ion{O}{iv} & Ly$\alpha$ & \ion{Si}{iv} & \ion{C}{iv} & \ion{C}{iii}] & \ion{Mg}{ii} & H$\gamma$ & H$\beta$ &  \\
$\boldsymbol{\lambda}$ {\bf[nm]} & 103.202 & 121.896 & 139.349 & 154.658 & 189.957 & 279.259 & 437.904 & 491.899 & \\
$\boldsymbol{I_\lambda}$ & 0.017 & 1.0039 & 0.01 & 0.13202 & 0.31359 & 0.94396 & 0.23848 & 0.93124 & \\ \\
\hline
\multicolumn{10}{c}{\bf Parameters of the $\boldsymbol{Z_{\rm score}}$ for RP spectra} \\
\hline
& \ion{O}{iv} & Ly$\alpha$ & \ion{Si}{iv} & \ion{C}{iv} & \ion{C}{iii}] & \ion{Mg}{ii} & H$\gamma$ & H$\beta$ &  H$\alpha$ \\
$\boldsymbol{\lambda}$ {\bf[nm]} & 103.353 & 122.388 & 139.563 & 154.588 & 190.398 & 280.470 & 435.600 & 488.952 & 657.736 \\
$\boldsymbol{I_\lambda}$ & 0.062484 & 0.10984 & 0.18982 & 0.07023 &  0.1409 & 0.22011 &  0.4101 & 0.25137 & 0.59948 \\ \\
\hline
\end{tabular}
\end{center}
\label{tbl:qsoc_shift_score_parameters}
\end{table*}

\begin{table*}
\caption{Binary warning flags used in the QSOC redshift selection procedure and reported in the \linktoEGParam{qso_candidates}{flags_qsoc} field. Sources with \linktoEGParam{qso_candidates}{flags_qsoc} $ = 0$ encountered no issues during their processing and are based on reliable spectra which means that they are more likely to contain reliable predictions.}
\begin{center}
\begin{tabular}{p{2.5cm}|p{0.75cm}|p{0.75cm}|p{10cm}}
\hline
Warning flag & Bit & Value  & Condition(s) for rising \\
\hline
\verb+Z_AMBIGUOUS+ & 1 & 1 & The CCF has more than one maximum with $\chi_r^2(k) > 0.85$, meaning that at least two redshifts lead to a similar $\chi^2$ and the solution is ambiguous. \\
\verb+Z_LOWCHI2R+ & 2 & 2 & $\chi_r^2(k) < 0.9$ \\
\verb+Z_LOWZSCORE+ & 3 & 4 & $Z_{\rm score}(k) < 0.9$ \\
\verb+Z_NOTOPTIMAL+ & 4 & 8 & The selected solution did not correspond to the global maximum (i.e. $\chi_r^2(k) < 1$) \\
\verb+Z_BADSPEC+ & 5 & 16 & The \bporrp spectra upon which this prediction is based are considered as unreliable. An unreliable spectrum has a number of spectral transits in BP, $N_{\rm bp}$ or RP, $N_{\rm rp}$ that is lower than or equal to ten transits or $G \geq 20.5$ mag or $G \geq 19 + 0.03 \times (N_{\rm bp} - 10)$ mag or $G \geq 19 + 0.03 \times (N_{\rm rp} - 10)$ mag (see the \linktosec{cu8par}{apsis}{qsoc} for more information on the derivation of these limits). \\
\hline
\end{tabular}
\end{center}
\label{tbl:qsoc_zwarning}
\end{table*}

As highlighted in \cite{2018MNRAS.473.1785D}, the global maximum of the CCF may not always lead to a physical solution as, for example, some characteristic emission lines of quasars (e.g. Ly$\alpha$, \ion{Mg}{ii,} or H$\alpha$) may be omitted from the fit while some emission lines can be falsely fitted to absorption features. This global maximum may also result from the fit of noise in the case of very low signal-to-noise-ratio (S/N) spectra. In order to identify these sources of error, we define a score, $0 \leq S(k) \leq 1$, that is associated with each shift; the shift associated with the highest score is the one that is selected by the algorithm. This score is computed as a weighted $p$-norm of the chi-square ratio defined as the value of the CCF evaluated at $k$ over the maximum of the CCF,
\begin{equation}
\chi_r^2(k) = \frac{\operatorname{ccf}(k)}{\operatorname{max}_k(\operatorname{ccf})} \hspace{0.5cm} \mbox{where} \hspace{0.5cm} 0 \leq \chi_r^2(k) \leq 1,
\label{eq:qsoc_chi2r}
\end{equation}
and of an indicator of the presence of quasar emission lines,
\begin{equation}
Z_{\rm score}(k) =  \prod_\lambda \left[ \frac{1}{2} \left(1 + \operatorname{erf} \frac{e_\lambda}{\sigma(e_\lambda) \sqrt{2}}\right) \right]^{I_\lambda},
\label{eq:qsoc_zscore}
\end{equation}
where $e_\lambda$ is the value of the \bporrp flux of the continuum-subtracted emission line at rest-frame wavelength $\lambda$ if we consider the observed spectrum to be at redshift $z = L^k - 1$; $\sigma(e_\lambda)$ is the associated uncertainty and $I_\lambda$ is the theoretical intensity\footnote{Theoretical emission line intensities should be regarded as weights. They do not refer to a particular theoretical model of the emission lines of quasars but to the values inferred in Table \ref{tbl:qsoc_shift_score_parameters}.} of the emission line located at $\lambda$, which is normalised so that the total intensity of all emission lines in the observed wavelength range is equal to one. Equation \ref{eq:qsoc_zscore} can then be viewed as a weighted geometric mean of a set of normal cumulative distribution functions of mean zero and standard deviations $\sigma(e_\lambda)$ evaluated at $e_\lambda$. A $Z_{\rm score}$ close to one indicates that all the emission lines that we expect at redshift $z$ are found in the spectra while missing a single emission line often leads to a very low $Z_{\rm score}$. The final formulation of the score is then given by 
\begin{equation}
S(k) =  \sqrt[p]{w_0 \, \left[\,\chi_r^2(k)\,\right]^p + w_1 \, \left[\,Z_{\rm score}(k) \,\right]^{p}},
\label{eq:qsoc_redshift_score}
\end{equation}
where $w_0$, $w_1$, and $p$ are parameters of the weighted $p$-norm, as listed in Table \ref{tbl:qsoc_shift_score_parameters}.

Table \ref{tbl:qsoc_shift_score_parameters} summarises the various parameters used in the computation of the redshift score, $S(k)$. Also, in order to facilitate the filtering of these potentially erroneous redshifts by the final user, we define binary processing flags, \linktoEGParam{qso_candidates}{flags_qsoc}, which are listed in Table \ref{tbl:qsoc_zwarning}. As later highlighted in \secref{subsec:qsoc_filtering}, most secure predictions often have bits 1--4 unset (i.e. \linktoEGParam{qso_candidates}{flags_qsoc} = 0 or \linktoEGParam{qso_candidates}{flags_qsoc} = 16).

Finally, the uncertainty on the selected redshift, $\sigma_z$, is derived from the uncertainty on the associated shift, $\sigma_k$, using the asymptotic normality property of the $\chi^2$ estimator, which states that $k$ is asymptotically normally distributed with a variance that is inversely proportional to the curvature of the CCF around the optimum. In particular, the variance on $k$ is asymptotically given by $\sigma_k^2 = -2\, dk^2 / d^2 \operatorname{ccf}(k),$ and as $Z = k \log\left(L\right)$,  the logarithmic redshift, $Z = \log(z+1)$, is also normally distributed with a variance of
\begin{equation}
\sigma_Z^2 = 2 \, \left|\frac{d^2 \operatorname{ccf}(k)}{dk^2}\right|^{-1} \log^2 \left(L\right).
\label{eq:qsoc_log_redshift_variance}
\end{equation}
Furthermore, as $z = \exp Z - 1$, the redshift that is reported by \qsoc is distributed as a log-normal distribution of mean $Z$ and variance $\sigma_Z^2$, although this distribution is shifted by $-1$. Accordingly, the squared uncertainty on the computed redshift is given by
\begin{equation}
\sigma_z^2 = (z + 1)^2 \left(\exp \sigma_Z^2 - 1.0\right) \exp \sigma_Z^2,
\label{eq:qsoc_redshift_variance}
\end{equation}
while its lower and upper confidence intervals, taken as its $0.15866$ and $0.84134$ quantiles, respectively, are given by
\begin{equation}
z_{\rm low} = \exp(Z - \sigma_Z) - 1 \hspace{0.5cm} \mbox{and} \hspace{0.5cm} z_{\rm up} = \exp(Z + \sigma_Z) - 1.
\label{eq:qsoc_redshift_confidence_interval}
\end{equation}

\subsection{Performance and results} \label{subsec:qsoc_performances}

The \qsoc contributions to \gdr{3} can be found in the \linktoEGTable{qso_candidates} table and consist of: \linktoEGParam{qso_candidates}{redshift_qsoc}, the quasar redshift, $z$; \linktoEGParam{qso_candidates}{redshift_qsoc_lower}/\linktoEGParam{qso_candidates}{redshift_qsoc_upper}, the lower and upper confidence intervals, $z_{\rm low}$ and $z_{\rm up}$, corresponding to the 16\% and 84\% quantiles of $z$, respectively, as given by Equation \ref{eq:qsoc_redshift_confidence_interval}; \linktoEGParam{qso_candidates}{ccfratio_qsoc}, the chi-square ratio, $\chi_r^2$, from Equation \ref{eq:qsoc_chi2r}; \linktoEGParam{qso_candidates}{zscore_qsoc}, the $Z_{\rm score}$ from Equation \ref{eq:qsoc_zscore}, and \linktoEGParam{qso_candidates}{flags_qsoc}, the \qsoc processing flags, $z_{\rm warn}$, from Table \ref{tbl:qsoc_zwarning}.

\begin{figure}
    \centering
    \includegraphics[width=0.49\textwidth]{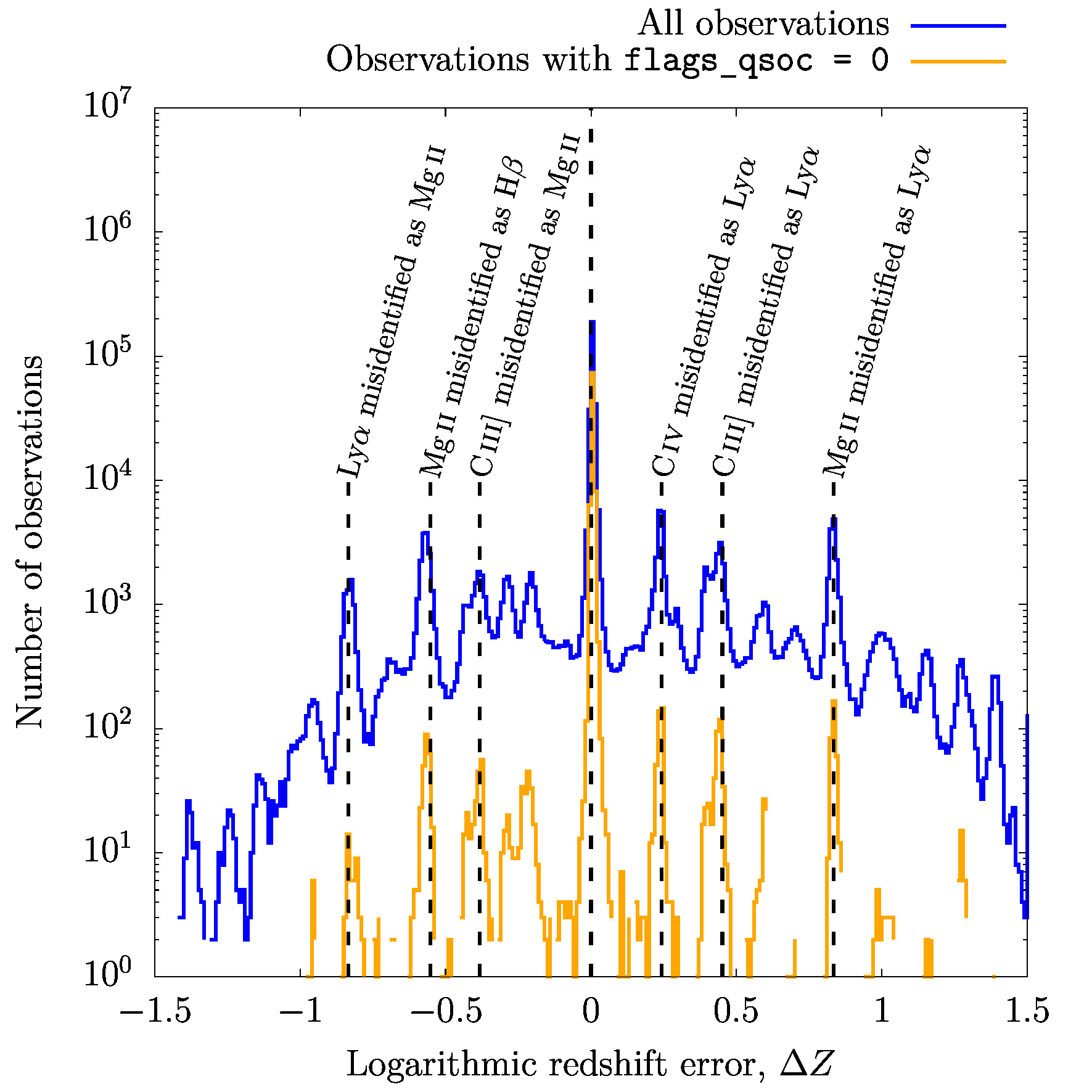}
    \caption{Histogram of the logarithmic redshift error, $\Delta Z = \log(z + 1) - \log(z_{\rm true} + 1)$ between \qsoc redshift, $z,$ and literature redshift, $z_{\rm true}$, for 439\,127 sources contained in the Milliquas 7.2 catalogue. A bin width of $0.01$ was used for both curves.}
    \label{fig:qsoc_Zerr}
\end{figure}

We quantitatively assess the quality of the \qsoc outputs by comparing the predicted reshifts against values from the literature. For this purpose, we cross-matched 6\,375\,063 sources with redshift estimates from \qsoc with 790\,776 quasars that have spectroscopically confirmed redshifts in the Milliquas 7.2 catalogue of \cite{2021arXiv210512985F} (i.e. {\tt type = 'Q'} in Milliquas). Using a 1$\arcsec$ search radius, we found 439\,127 sources in common between the two catalogues. It should be emphasised here that the distributions of the redshifts and $G$ magnitudes of the cross-matched sources are not representative of the intrinsic quasar population as they inherit the selection and observational biases that are present in both the Milliquas catalogue and in \gaia. The numbers reported here should therefore be interpreted with that in mind. A straight comparison between the \qsoc prediction and the Milliquas spectroscopic redshifts, illustrated in Figure \ref{fig:qsoc_Zerr} on a logarithmic scale, shows that $63.7\%$ of the sources have an absolute error on the predicted redshift, $|\Delta z|$ , that is lower than 0.1. This ratio increases to $97.6\%$ if only {\tt flags\_qsoc = 0} sources are considered. 

As most of the DR12Q quasars we use for building our templates are also contained in the Milliquas catalogue (161\,278 \qsoc predictions are contained in both the DR12Q and Milliquas catalogue), one may  wonder whether these induce a positive bias on the fraction of sources with $| \Delta z | < 0.1$. In order to answer this question, we note that the \qsoc templates were built based on a statistically significant number of 297\,264 sources, and so we expect the computed templates to be representative of the whole quasar population under study while not being too specific to the particular set of spectra we used (i.e.\ any other set of spectra of the same size would have provided us with very similar templates). Nevertheless, $71\%$ of the sources in the DR12Q catalogue have $|\Delta z| < 0.1$. This compares to $59.5\%$ of the sources with $|\Delta z| < 0.1$ that are not in the DR12Q catalogue.
If we consider only sources with \linktoEGParam{qso_candidates}{flags_qsoc}$ = 0$, then these numbers are $97\%$ and $98.8\%$, respectively. The observed differences can be explained primarily by the fact that, due to the selection made in the SDSS-III/BOSS survey, $31.7\%$ of the DR12Q sources that are found among the \qsoc predictions have $2 < z < 2.6,$ where the presence of the Ly$\alpha$+\ion{Si}{iv}+\ion{C}{iv}+\ion{C}{iii} emission lines allows secure determination of the redshift ($81.4\%$ of the sources in this range have $| \Delta z | < 0.1$). In contrast, the redshift distribution of the sources that are found only in Milliquas peaks in the range $1.2 < z < 1.4$ where only $50.5\%$ of the sources have $| \Delta z | < 0.1$, owing to the sole presence of the \ion{Mg}{ii} emission line in this redshift range (see Section \ref{subsec:qsoc_filtering} for more information on these specific redshift ranges). However, both subsets have a comparable fraction of predictions with $| \Delta z | < 0.1$ once these are computed over narrower redshift ranges, as expected.

We further investigate the distribution of the logarithmic redshift error, defined as 
\begin{equation}
\Delta Z = \log(z + 1) - \log(z_{\rm true} + 1),
\label{eq:qsoc_log_redshift_error}
\end{equation}
between \qsoc redshift, $z$, and the literature redshift, $z_{\rm true}$, in Figure \ref{fig:qsoc_Zerr}. If we assume that a spectral feature at rest-frame wavelength $\lambda_{\rm true}$ is falsely identified by \qsoc as another spectral feature  at $\lambda_{\rm false}$, then the resulting logarithmic redshift error will be equal to $\Delta Z = \log \lambda_{\rm true} - \log \lambda_{\rm false}$, such that $\Delta Z$, besides its ability to identify good predictions, can also be used to highlight common mismatches between emission lines. In Figure \ref{fig:qsoc_Zerr}, we can see that most of the predicted (logarithmic) redshifts are in good agreement with their literature values while emission line mismatches mainly occur with respect to two specific emission lines: \ion{C}{iii]} and \ion{Mg}{ii}. In the most frequent case, the \ion{C}{iv} emission line is misidentified as Ly$\alpha$, because the separation between these two lines is comparable to the separation between \ion{C}{iv} and \ion{C}{iii]} when considered on a logarithmic wavelength scale. The Ly$\alpha$ and \ion{C}{iii]} lines are subsequently fitted to noise or wiggles in the very blue part of BP and in RP, respectively. By requiring that {\tt flags\_qsoc = 0}, we can mitigate the effect of these emission-line mismatches without affecting the central peak of correct predictions too much.

Finally, we note that the distribution of $\Delta Z / \sigma_Z$, where $\sigma_Z = [ \log (z_{\rm up}+1) - \log (z_{\rm low}+1) ] / 2$ is defined in Equation \ref{eq:qsoc_log_redshift_variance}, effectively follows an approximately Gaussian distribution of median 0.007 and standard deviation (extrapolated from the inter-quartile range) of $1.053$ if observations with $|\Delta z| < 0.1$ are considered. If only observations for which {\tt flags\_qsoc = 0} are considered, $\Delta Z / \sigma_Z$ have a median of $0.002$ and standard deviation of $1.14$.

\subsection{Use of \qsoc results} \label{subsec:qsoc_filtering}

\begin{figure}
    \centering
    \includegraphics[width=0.49\textwidth]{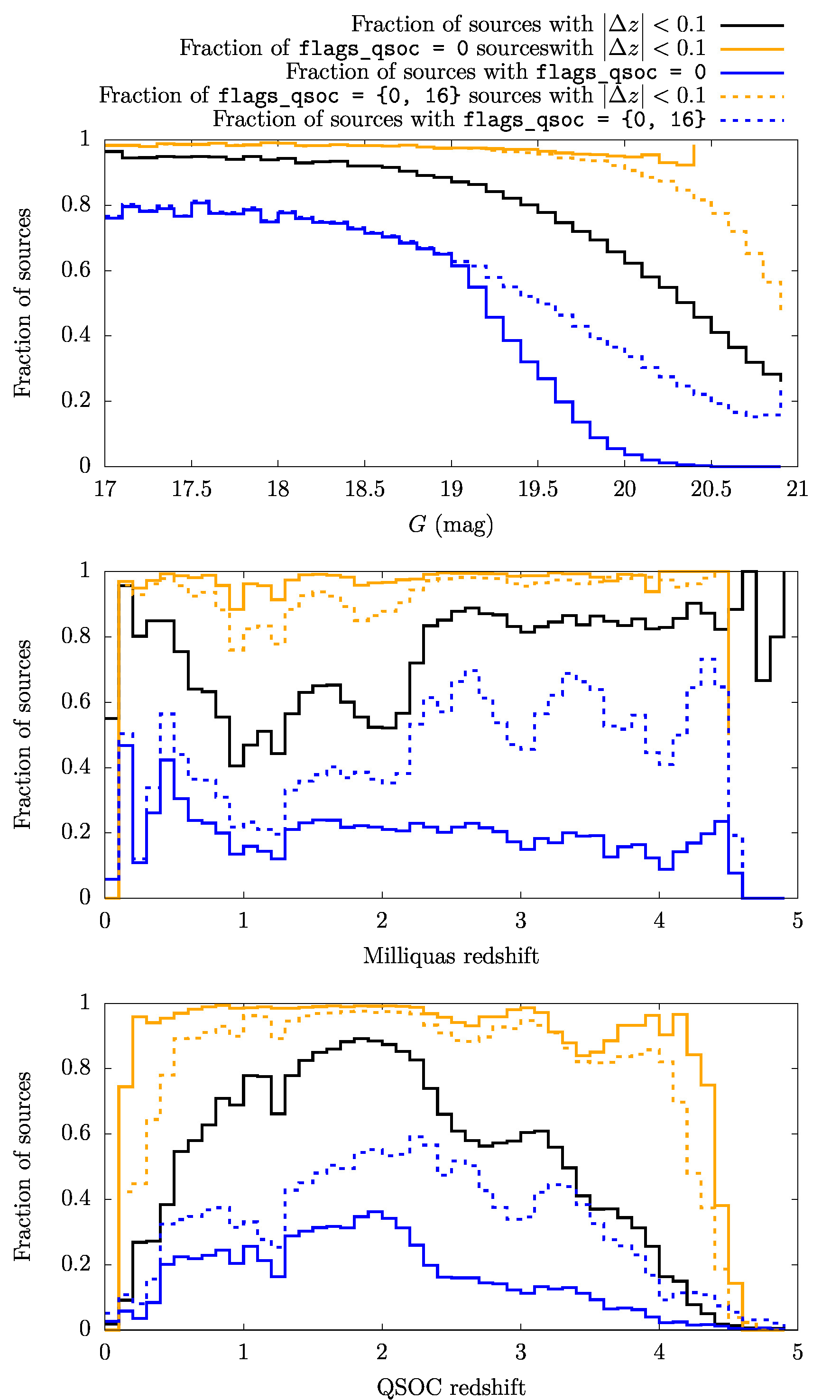}
    \caption{Fraction of successful and reliable \qsoc predictions computed over 439\,127 sources contained in the Milliquas 7.2 catalogue with respect to $G$ magnitude (top), Milliquas redshift (middle), and QSOC redshift (bottom). Black line: Fraction of observations with an absolute error of the predicted redshift, $|\Delta z|$, lower than 0.1. Orange line: Fraction of {\tt flags\_qsoc = 0} sources with $|\Delta z| < 0.1$. Blue line: Fraction of observations with {\tt flags\_qsoc = 0}. Orange and blue dotted lines correspond to their solid counterpart while considering {\tt (flags\_qsoc = 0 or flags\_qsoc = 16)} observations instead of {\tt flags\_qsoc = 0} observations.  Fractions are computed with respect to the number of sources in magnitude and redshift bins of $0.1$.}
    \label{fig:qsoc_z}
\end{figure}

In \gdr{3}, \qsoc systematically publish redshift predictions for which \linktoEGParam{qso_candidates}{classprob_dsc_combmod_quasar} $\geq 0.01$ and \linktoEGParam{qso_candidates}{flags_qsoc} $\leq 16$, leading to 1\,834\,118 sources that are published according to these criteria (see \linktoEGParam{qso_candidates}{source_selection_flags} for more information on the selection procedure). Nevertheless, for the sake of completeness, we also publish redshift estimates for all sources with \linktoEGParam{qso_candidates}{classprob_dsc_combmod_quasar} $\geq 0.01$ that are contained in the \linktoEGTable{qso_candidates} table, yielding $4\,540\,945$ additional sources for which \linktoEGParam{qso_candidates}{flags_qsoc} $> 16$. However, these last predictions are of lower quality as, for example, a comparison with the Milliquas spectroscopic redshift shows that $39.6\%$ of the \linktoEGParam{qso_candidates}{flags_qsoc} $> 16$ sources have $| \Delta z | < 0.1$, compared to $87\%$ for sources with \linktoEGParam{qso_candidates}{flags_qsoc} $\leq 16$.

Of the source parameters published in the \gdr{3}, the $G$-band magnitude, \linktoMainParam{gaia_source}{phot_g_mean_mag}, has a particularly strong impact on the quality of the \qsoc predictions; it shows a clear correlation with the S/N of the \bporrp spectra, as does the number of \bporrp spectral transits to a lesser extent. From the top panel of Figure \ref{fig:qsoc_z}, we see that more than $89\%$ of the sources with $G \leq 19$ mag have $|\Delta z| < 0.1$ (black line) while the same fraction is obtained for spectra with $19.9 < G < 20$ mag only for sources with \linktoEGParam{qso_candidates}{flags_qsoc}$ = 0$ (orange solid line). However, these correspond to a very small fraction ($5.5\%$) of the sources in this magnitude range (blue solid line). A less stringent cut, \linktoEGParam{qso_candidates}{flags_qsoc} = 0 or \linktoEGParam{qso_candidates}{flags_qsoc} = 16, where we encounter no processing issue (i.e.\ flag bits $1$--$4$ are not set) even when the \bporrp spectra are unreliable (i.e. flag bit $5$ can be set), still leads to $92\%$ of the sources with $|\Delta z| < 0.1$ (orange dotted line) while retaining $36.5\%$ of the sources in this magnitude range (blue dotted line). The same cut concurrently retains $22\%$ of the $20.4 < G < 20.5$ mag observations where $81.5\%$ of the predictions have $|\Delta z| < 0.1$ and is accordingly recommended for users dealing with sources at $G > 19$ mag.

Besides the aforementioned recommendations on the \linktoEGParam{qso_candidates}{flags_qsoc} and $G$ magnitude, we should point out an important limitation of the \gaia \bporrp spectro-photometers regarding the identification and characterisation of quasars, namely the fact that the \ion{Mg}{ii} emission line is often the sole detectable emission line in the \bporrp spectra of $0.9 < z < 1.3$ quasars in the moderate-S/N regime of $G \gtrsim 19$ mag spectra. Indeed, despite the broad $325$--$1050$ nm coverage of the \bporrp spectrophotometers, quasar emission lines are often significantly damped in the observed wavelength regions $\lambda < 430$ nm and $\lambda > 950$ nm, owing to the low instrumental response in these ranges \citep[see for example][Figure 10]{DR3-DPACP-101}. As a result, the H$\beta$ and \ion{C}{iii]} emission lines surrounding the \ion{Mg}{ii} line\footnote{The H$\gamma$ emission line being intrinsically weak, it is often not seen in the \bporrp spectra of quasars and is accordingly not considered here.} only enter the \bporrp spectra at $z = 0.95$ and $z = 1.25$, respectively. Nevertheless, we consider a range of $0.9 < z < 1.3$ in order to take into account low-S/N spectra where these lines, although present, are often lost in the noise. The sole presence of the \ion{Mg}{ii} emission line has the deleterious effect of increasing the rate of mismatches between this line and mainly the Ly$\alpha$ and H$\beta$ emission lines, as seen in Figure \ref{fig:qsoc_Zerr}. Another issue also arises for $z \approx 1.3$ quasars, where the \ion{C}{iii]} emission line enters the BP spectrum while the \ion{Mg}{ii} line now lies on the peak of the BP spectrum, which complicates its detection by the algorithm leading to mismatches between \ion{C}{iii]} and the Ly$\alpha$ or \ion{Mg}{ii} emission lines. These effects are clearly visible in the middle panel of Figure \ref{fig:qsoc_z} at $0.9 < z < 1.3$, along with the previously discussed misidentification of the \ion{C}{iv} line as Ly$\alpha$ at $z \approx 2$. Appropriate cuts on \linktoEGParam{qso_candidates}{flags_qsoc} allow both of these shortcomings to be alleviated, as seen in \figref{fig:qsoc_z}.

In the bottom panel of Figure \ref{fig:qsoc_z}, we see that the fraction of sources with $| \Delta z | < 0.1$ amongst very low- and high-redshift sources, as predicted by \qsoc, is low ($7.25\%$ for $z < 0.2$ sources and $2.66\%$ for $z > 4$ sources). The explanation is that these very low- and high-$z$ quasars are rare in our sample, such that any erroneous prediction towards these loosely populated regions is largely reflected in the final fraction of predictions (i.e. the `purity' in these regions becomes very low). Again, cuts on the \linktoEGParam{qso_candidates}{flags_qsoc} allow us to recover about  90\% of sources with $| \Delta z | < 0.1$ in the range $0.1 < z < 4.4$. Concentrating on the drop at $z < 0.1$, we note that only 69 sources have a Milliquas redshift in this range, while only 31 have $0.0826 < z < 0.1$ (i.e.\ in the predictable \qsoc redshift range). Amongst these 69 sources, 38 have $| \Delta z | < 0.1$ while 4 have \linktoEGParam{qso_candidates}{flags_qsoc}$ = 0$ but these are unfortunately erroneously predicted. These low numbers, along with the fact that \qsoc predicts 2\,154 sources in this redshift range (i.e. 0.5\% of the total predictions) explains the drop at $z < 0.1$ in the middle and bottom panels of Figures \ref{fig:qsoc_z}, even when \linktoEGParam{qso_candidates}{flags_qsoc}$ = 0$. Regarding the $z > 4.4$ quasars, only 76 of them have redshifts in both \gaia and Milliquas, while only 10 have {\tt flags\_qsoc = 0} and 9 of these also have $| \Delta z | < 0.1$. There are 18\,959 sources with \qsoc redshift predictions in this range, although only 101 (i.e. $0.5\%$) of them have \linktoEGParam{qso_candidates}{flags_qsoc}$ = 0$. This leads to a rather poor fraction of $9/101$ of the sources with $| \Delta z | < 0.1$ and {\tt flags\_qsoc = 0} in this redshift range.

In conclusion, we should insist first on the fact that \qsoc is designed to process Type-I/core-dominated quasars with broad emission lines in the optical and accordingly yields only poor predictions on galaxies, type-II AGN, and BL Lacertae/blazar objects. Secondly, \smsgen does not provide covariance matrices on the integrated flux \citep{DR3-DPACP-157}, meaning that the computed $\chi^2$ from Equation \ref{eq:qsoc_chi2} is systematically underestimated and is consequently not published in \gdr{3}. The computed redshift and associated confidence intervals, $z_{\rm low}$ and $z_{\rm up}$ from Equation \ref{eq:qsoc_redshift_confidence_interval}, though appropriately re-scaled, might also sporadically suffer from this limitation.

\section{Unresolved galaxy classifier (UGC)}
\label{sec:ugc}

\subsection{Objectives} \label{subsec:ugc_objective}

The Unresolved Galaxy Classifier (\ugc) module estimates the redshift, $z$, of the sources with $G < 21$ mag that are classified as galaxies by \dsc-Combmod with a probability of 0.25 or more (see \secref{sec:dsc} for details). \ugc infers redshifts in the range $0 \leq z \leq 0.6$ by using a combination of three support vector machines \citep[SVMs,][]{CortesVapnik95}, all taking as input the \bporrp spectra of the sources as sampled by \smsgen \citep[Section 2.3.2]{DR3-DPACP-157}. The SVMs are trained on a set of \bporrp spectra of galaxies that are spectroscopically confirmed in the SDSS DR16 archive \citep{2020ApJS..249....3A}. \ugc further applies filtering criteria for selecting redshifts to be published in \gdr{3}, as described in \secref{subsec:ugc_method}.

\subsection{Method} \label{subsec:ugc_method}

\ugc is based on the LIBSVM library of \cite{CC01a}, from which three SVM models are built: (i) \emph{t-SVM}, the \emph{total-redshift range} SVM model, which computes the published redshift, \linktoEGParam{galaxy_candidates}{redshift_ugc}, and associated SVM prediction intervals, \linktoEGParam{galaxy_candidates}{redshift_ugc_lower} and \linktoEGParam{galaxy_candidates}{redshift_ugc_upper}, (ii)  \emph{r-SVM,} and (iii) \emph{c-SVM}, which are respectively regression and classification SVM models applied to discretised versions of the redshift and used exclusively for the internal validation of the redshift produced by the t-SVM model. All SVM models use common training and test sets, which we describe below.

\subsubsection{Training and test sets}
\label{subsubsec:ugc_svm_training_test_sets}

The sources in the training and test sets were selected from the SDSS DR16 archive \citep{2020ApJS..249....3A}, which provide position, redshift, magnitudes in the $u$-, $g$-, $r$-, $i$-, $z$-bands, photometric sizes (we used here the Petrosian radius), and  interstellar extinction for each  spectroscopically
confirmed galaxy. There are 2\,787\,883 objects in SDSS DR16 that are spectroscopically classified as galaxies, but we rejected sources with poor or missing photometry, size, or redshift, thus reducing the number of galaxies to 2\,714\,637. Despite the known lack of uniformity of the SDSS DR16 redshift distribution due to the BOSS target selection\footnote{\href{https://www.sdss.org/dr16/algorithms/boss_target_selection/}{https://www.sdss.org/dr16/algorithms/boss\_target\_selection/}}, this survey still provides the largest existing  database of accurate spectroscopic redshifts of galaxies that can be used as target values in the SVM  training and test sets.

The selected galaxies were cross-matched to the \gdr{3} sources prior to their filtering by CU9 using a search radius of 0.54\arcsec, which resulted in 1\,189\,812 cross-matched sources. Amongst these, 711\,600 have \bporrp spectra, though not all of them are published in \gdr{3}. Because the inclusion of high-redshift galaxies would lead to a very unbalanced training set (i.e.\ very few high-redshift galaxies), we further imposed an upper limit on the SDSS DR16 redshift of $z \leq 0.6$, leaving 709\,449 sources that constitute our \emph{base set}.

For the preparation of the training set, a number of conditions were further imposed on the sources in the base set: 
(i) $\gmag \leq 21.0$ mag;
(ii) \bporrp spectra must be composed of a minimum of six epochs of observations;
(iii) the mean flux in the blue and red parts of the \bporrp spectra, as computed by \ugc, must lie in the ranges $0.3\leq bpSpecFlux\leq 100$ e$^-$s$^{-1}$ and $0.5\leq{}rpSpecFlux\leq 200$ e$^-$s$^{-1}$, respectively, in order to exclude potentially poor-quality spectra; 
(iv) the image size, as characterised by the Petrosian radius, must be in the range $0.5\arcsec\leq{}petroRad50\_r\leq5\arcsec$ in order to exclude suspiciously compact or significantly extended galaxies; 
(v) the interstellar extinction in the $r$-band must be below the  upper limit of $extinction\_r\leq0.5$ mag to avoid highly reddened sources; and 
(vi) the redshift must be larger than $0.01$ in order to exclude nearby extended galaxies. After applying all these cuts, 377\,875 sources remained, which we refer to as the \emph{clean set}. Of these, 6\,000 sources were randomly selected in order to construct the \emph{training set}, the redshift distribution of which is given in \tabref{tab:ugc_traintest}. The imbalance of this training set is clearly visible in this table, and is caused by the small number of high-redshift galaxies present in the clean set.
 
\begin{table*} 
\small
\centering
\caption{Distribution of the sources in the \ugc data sets according to their SDSS redshifts.}    
\label{tab:ugc_traintest}
\begin{tabular}{lrrrrrrr}
\hline
 & \multicolumn{6}{c}{Redshift ranges} & \\
Data set name & $0.0$--$0.1$ & $0.1$--$0.2$ & $0.2$--$0.3$ & $0.3$--$0.4$ & $0.4$--$0.5$ & $0.5$--$0.6$ & Total \\
\hline \hline
Base set                         & 224\,264 & 292\,968 & 118\,248 & 65\,912 & 7\,055 & 1\,002 & 709\,449 \\
\hspace{0.5cm} Clean set         & 152\,564 & 192\,675 &  29\,145 &  2\,490 &    724 &    327 & 377\,875 \\
\hspace{1cm} Clean test set$^a$      & 150\,964 & 191\,025 &  28\,045 &  1\,590 &    224 &     27 & 371\,875 \\
\hspace{1cm} Training set        &   1\,600 &   1\,600 &   1\,100 &     900 &    500 &    300 &   6\,000 \\
\hspace{0.5cm} Base test set$^a$ & 222\,664 & 291\,368 & 117\,148 & 65\,012 & 6\,555 &    702 & 703\,449 \\
\hline
\multicolumn{8}{p{12.5cm}}{$^a$ The base test set and clean test set are respectively composed of sources in the base set and clean set that are not contained in the training set.}
\end{tabular}
\end{table*}

\begin{table*} 
\small
\caption{Galactic coordinates and colour--colour regions from which \ugc results are filtered out. Those correspond to regions where extragalactic objects are not expected: Magellanic clouds (LMC, SMC) and an area (CNT) close to the Galactic centre.} 
\centering 
\label{tab:ugc_galactic_areas}
\begin{tabular}{lcccc}
\hline
\multirow{2}{*}{Area} & \multicolumn{2}{c}{Galactic coordinates range} & Colour-colour box A &  Colour-colour box B \\
 & longitude [\deg]& latitude [\deg] &[mag] & [mag] \\
\hline
\multirow{2}{*}{CNT} & \multirow{2}{*}{$0.0\pm15.0$} & \multirow{2}{*}{$-5.0\pm5.0$} & $-0.5<\gmag-\gbp<0.5$ & $-0.5<\gmag-\gbp<3.0$\\
& &     & $0.4<\gbp-\grp<1.3$ & $-0.2<\gbp-\grp<1.4$ \\
\hline
\multirow{2}{*}{LMC} & \multirow{2}{*}{$279.5\pm4.0$} & \multirow{2}{*}{$-33.25\pm3.25$} & $-3.0<\gmag-\gbp<-1.5$& $-0.7<\gmag-\gbp<2.0$\\
 & & & $-0.4<\gbp-\grp<1.0$ & $-0.8<\gbp-\grp<1.4$ \\
\hline
\multirow{2}{*}{SMC} & \multirow{2}{*}{$303.0\pm1.0$} & \multirow{2}{*}{$-44.0\pm1.0$} & $-3.0<\gmag-\gbp<-1.5$& $-0.7<\gmag-\gbp<2.0$\\
 & & & $-0.4<\gbp-\grp<1.0$ & $-0.8<\gbp-\grp<1.4$ \\
\hline
\end{tabular}
\end{table*}
 
The conditions described in the previous paragraph were not imposed for the test set. Instead, all 703\,449 spectra in the base set that were not used for training were included in the \emph{base test set}, whose redshift distribution is shown in \tabref{tab:ugc_traintest}. Additionally, a purest test sample, the \emph{clean test set}, was derived from the clean set by removing the training data it contains. 

\subsubsection{Support vector machine models}
\label{subsubsec:ugc_method_svm}

The input of all SVM models are \bporrp spectra. The spectra are first truncated by removing the first 34 and the last 6 samples in BP, and the first 4 and the last 10 samples in RP, in order to avoid regions of low S/N. These cuts result in the definition of the usable wavelength ranges for the BP and the RP parts of the spectrum, namely 366--627 nm and 620--996 nm, respectively. Each pair of truncated spectra is then concatenated to form the SVM input vector of 186 fluxes.

A common setup was implemented for the SVM model preparation (see  LIBSVM\footnote{\href{https://www.csie.ntu.edu.tw/~cjlin/libsvm/}{https://www.csie.ntu.edu.tw/~cjlin/libsvm}} for details): The  Standardization Unbiased method was selected to scale the target data and the vector elements to the range $[-1.0,1.0]$; the radial basis function (RBF)  $K(\mathbf{x_{i}},\mathbf{x_{j}})=\exp(-\gamma|\mathbf{x_{i}}-\mathbf{x_{j}}|^{2})$ was chosen as the kernel function, and the tolerance of the termination criterion is set to $e=0.001$; shrinking heuristics are used to speed up the training process; a four-folded tuning (cross-validation) is applied to determine the optimal $\gamma$ kernel parameter and the penalty parameter $C$ of the error term in the optimisation problem.

The \ugc redshifts are estimated by t-SVM, which implements a $\epsilon$-SVR regression model trained for redshifts in the range $0.0\leq{}z\leq{}0.6$. The two other SVM models, c-SVM and r-SVM, use the \bporrp spectra as input but are trained to predict a discretised version of the redshifts and are used solely for the purpose of redshift validation (\secref{subsubsec:ugc_method_source_filtering}). The c-SVM model is a C-SVC classification model trained on six different classes corresponding to the redshift ranges $0\leq{}z<0.1$, $0.1\leq{}z<0.2$, $0.2\leq{}z<0.3$, $0.3\leq{}z<0.4$, $0.4\leq{}z<0.5,$ and $0.5\leq{}z<0.6$. The output of the c-SVM model is a class-probability vector. The element of the vector with the highest value above 0.5 is taken as the selected class. If there is no element with probability larger than 0.5, then the source is marked as unclassified. The r-SVM model implements the $\epsilon$-SVR regression model of LIBSVM ---similarly to the t-SVM model--- but it is trained on six discrete target values $(0.05, 0.15, \dots, 0.55)$. As only the first decimal is retained for the predictions, the output of the r-SVM model is directly comparable to the classes used by the c-SVM model.

\subsubsection{Source filtering}
\label{subsubsec:ugc_method_source_filtering}

Two sets of criteria are used to select the \ugc outputs to be published in \gdr{3}. The first set applies to specific properties of the processed sources, while the second concerns the redshift validity. An output is included in \gdr{3} only if all the criteria of the two sets are satisfied.

Although \ugc processes all $G < 21$ mag sources for which the \dsc Combmod galaxy probability is higher than or equal to $0.25$, additional criteria were imposed for selecting the purest sample of results. First, we require that the number of spectral transits in both BP and RP is higher than or equal to ten. Second, we require that the mean flux in the blue and red parts of the  \bporrp spectra lies in the ranges  set in  \secref{subsubsec:ugc_svm_training_test_sets}. Third, we decided to only publish  redshifts  for sources with $G > 17$ mag, so as to exclude  bright and possibly extended sources, for which it is likely that only part of the galaxy has been recorded. Fourth,  we require $G-\gbp > 0.25$ mag in order to reduce the number of sources with true $z > 0.6$ (which lie outside the range of the training data) by  as much as possible. The fifth and final condition is related to the location of blended sources that are erroneously classified as galaxies in high-density regions in the sky (see also \secref{subsec:dsc_results}). Indeed, the positional distribution of the sources processed by \ugc shows a high concentration of galaxies in three small areas where extragalactic objects are not expected in large numbers: a region below the Galactic centre, and two areas centred on the Magellanic Clouds (see \tabref{tab:ugc_galactic_areas}). Almost 9\% of the total number of processed sources originate in these three areas. Sources in these areas also occupy a specific region of the $G-\gbp, \gbp-\grp$ colour--colour diagram that is distinct from the locus of the  remaining sources. This distinction has been used to define colour cuts (shown in \tabref{tab:ugc_galactic_areas}) which, in combination with the coordinates of the three areas, allowed us to clean the suspicious clumps of galaxies and to remove a large number of potentially misclassified sources in these three areas. Nonetheless, conditions listed in \tabref{tab:ugc_galactic_areas} are not applied if the \dsc Combmod probability for the source to be a galaxy is equal to one. 

The comparison of the redshifts produced by the t-SVM model to those of the r-SVM and c-SVM models allows us to internally validate the \ugc redshifts. The implementation of the filtering involves first the rejection of sources for which at least one of the SVM models has not produced an output (either because there is no prediction or because the source is marked as unclassified). Second, the three computed redshifts are required to span at most two adjacent bins of redshift, similar to those defined for the c-SVM and r-SVM models. The largest absolute difference between the t-SVM redshift and the central value of the c-SVM and r-SVM redshift bins is $0.08$. The redshifts of sources not satisfying  one of these criteria are not published in \gdr{3}.

\subsection{Performance} \label{subsec:ugc_performance}

\begin{figure*}
\centering
\includegraphics[width=0.48\textwidth]{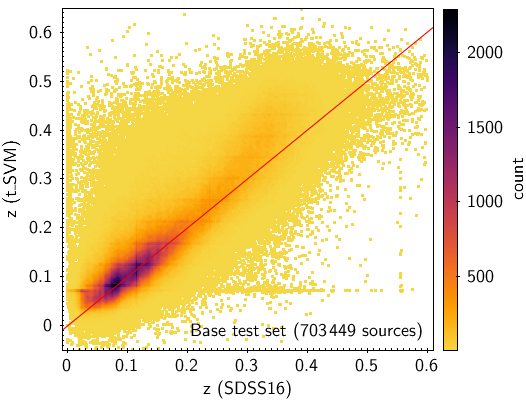}
\includegraphics[width=0.48\textwidth]{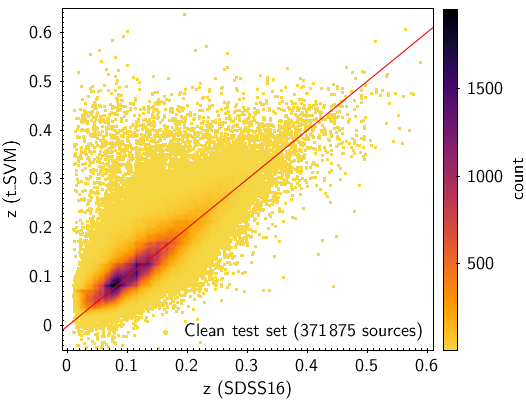}
\caption[\ugc t-SVM model performance]{Comparison of the \ugc redshifts, as estimated from the t-SVM model with SDSS DR16 redshifts for the base test  set (left) and for the clean test set (right), as identified in \secref{subsubsec:ugc_svm_training_test_sets}.}
\label{fig:ugc_t-svm3_test_perf}
\end{figure*}

\begin{figure*}
\centering
\includegraphics[width=0.33\textwidth]{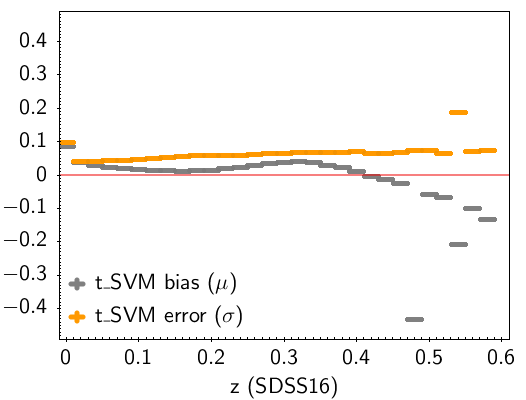}
\includegraphics[width=0.32\textwidth]{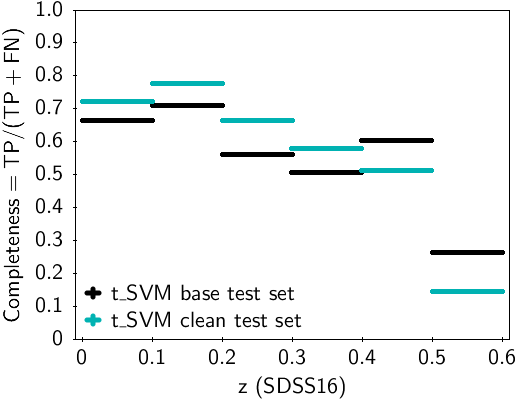}
\includegraphics[width=0.32\textwidth]{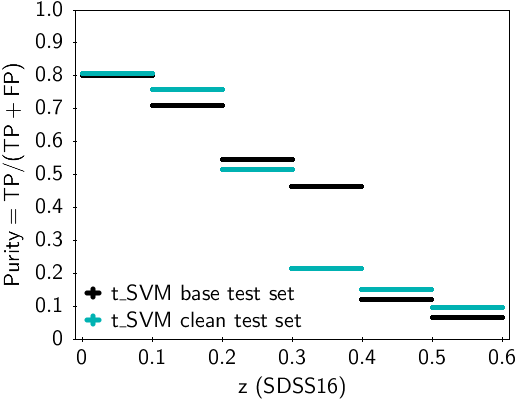}
\caption{
Left panel: Mean ($\mu_i$) and standard deviation ($\sigma_i$) of the difference between the \ugc redshifts, from the t-SVM model, and associated SDSS redshifts for sources contained in the \ugc base test set and averaged over redshift bins of size $0.02$. Completeness (middle panel) and purity (right panel) as a function of redshift, evaluated on the \ugc test set 
(black) and clean set (cyan). The bin size is equal to 0.1.}
\label{fig:ugc_t-svm3_performance}
\end{figure*}
 
The overall performance of the t-SVM model is given by the mean ($\mu$) and the standard deviation ($\sigma$) of the difference between the estimated and the real (target) redshifts. The internal test, applied to  the training set itself, yields $\sigma=0.047$ and $\mu=-0.003$. The external test, which is performed on all 703\,449 spectra in the base test set, yields $\sigma=0.053$ and $\mu=0.020$ (\figref{fig:ugc_t-svm3_test_perf}, left panel). These values indicate that the performance is worse for the base test set, as expected. If the clean test set of 371\,875 spectra is used the performance is improved significantly, with $\sigma=0.037$ and $\mu=0.008$ (\figref{fig:ugc_t-svm3_test_perf}, right panel). 

The performance varies with redshift. To quantify this,  the base test set was divided into SDSS redshift bins of size $0.02$. The mean, $\mu_i$, and the standard deviation, $\sigma_i$, of the differences between the redshift predicted by t-SVM and the real  (SDSS) redshifts were determined for each one of these bins, as shown in \figref{fig:ugc_t-svm3_performance} (left panel). Generally, there are three regions with different performance. For $z<0.02,$ the error and the bias are relatively large indicating that the t-SVM is ineffective for redshifts close to zero. The performance is good in the range of $0.02<z<0.26$; however, for larger redshifts, the bias changes significantly  from almost zero to positive and then to negative values, while the error progressively increases. For $z>0.5,$ both $\mu_i$ and $\sigma_i$ show large scatter, probably due to the fact that large redshifts are under-represented in the t-SVM training set.

In addition, the performance of the t-SVM model as a function of redshift was investigated by constructing a confusion matrix, as in classification problems. To this effect,  a different class has been assigned to each redshift bin, $z_{\rm bin}$,  both for the real (SDSS) and the predicted (t-SVM) redshifts. In this case, the bin size was $0.1$. The confusion matrix presents the total number of cases for each real and each predicted class  (see for details the \linktosec{cu8par}{apsis}{ugc}).

For a given redshift bin, $z_{\rm bin}$, the numbers of true-positive $TP$, false-negative $FN$, and false-positive $FP$ predictions are used to evaluate the sensitivity, or completeness, $TP/(TP+FN),$ and the precision, or purity, $TP/(TP+FP)$.  \figref{fig:ugc_t-svm3_performance} (middle and right panels) show the t-SVM completeness and purity for the redshift bins of the base and clean test sets in bins of redshift. Both  completeness and purity for the base and clean test sets are very good  up to a redshift of $z=0.2$. The purity is moderate ($\sim$ 0.5) for the two test sets for the redshift bin 0.2--0.3 and fails at larger redshifts. The completeness is moderate in the  0.3--0.5 bin and fails for the last bin. Generally, good performance can be expected for  redshifts $z\leq 0.2$.


\subsection{Results}
\label{subsubsec:ugc_results}

\begin{figure*}
\centering
\includegraphics[width=\textwidth]{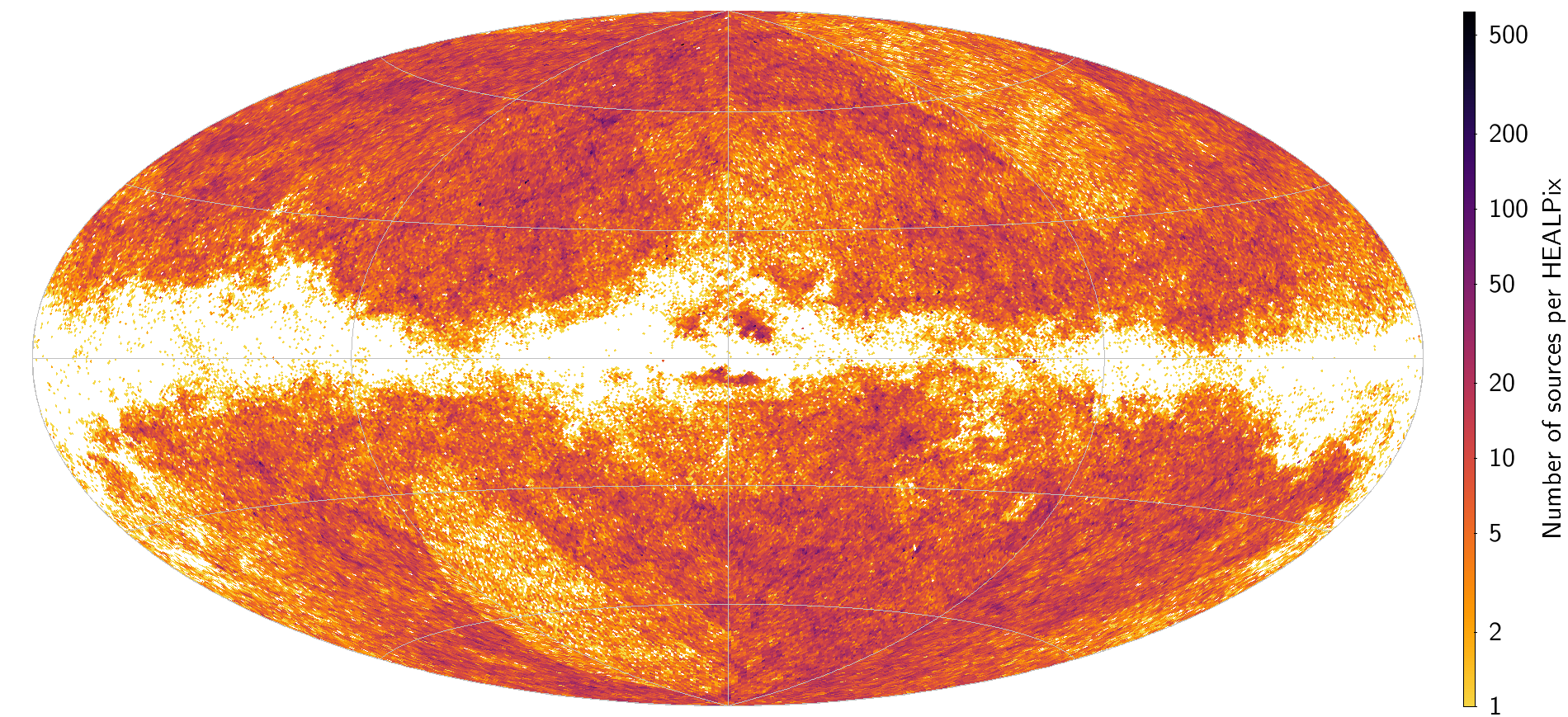} 
\caption[\ugc sources Galactic sky]{Galactic sky distribution of the number of sources with redshifts estimated by \ugc. The plot is shown at HEALPix level 7 (0.210 \sqdeg).}
\label{fig:ugc_galsky}
\end{figure*}

\begin{figure*}
\centering
\includegraphics[width=0.32\textwidth]{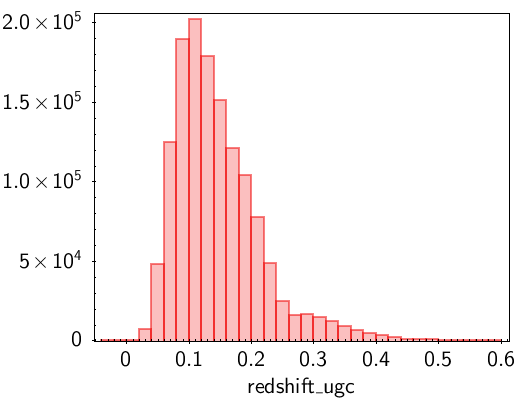} 
\includegraphics[width=0.32\textwidth]{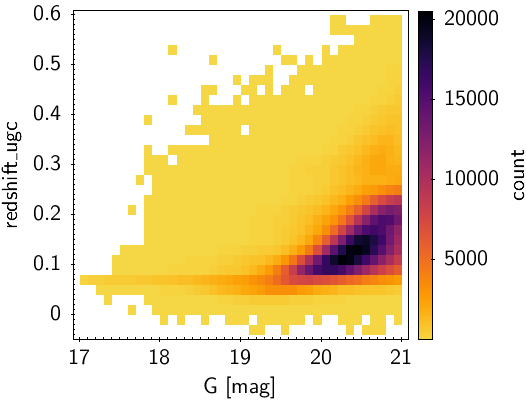} 
\includegraphics[width=0.32\textwidth]{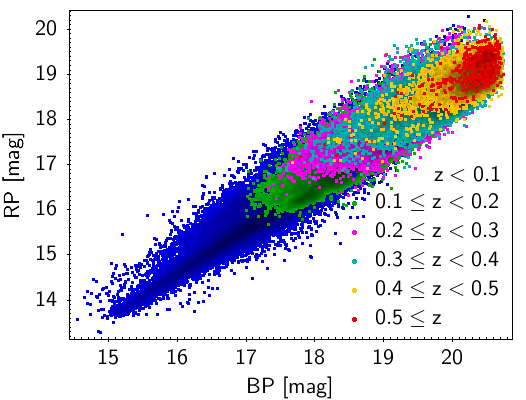} \caption[\ugc\ estimated redshifts distribution]{Distribution of the \ugc redshifts. (Left) Histogram of the estimated redshift in bins of size 0.02. (Middle) \ugc redshifts as a function of \gmag magnitude. (Right) Distribution of the sources with \ugc redshifts on a \bporrp magnitude diagram where different colours correspond to different redshift ranges.} 
\label{fig:ugc_redshifts_distr}
\end{figure*}

The \ugc output is included in the \linktoEGTable{galaxy_candidates} table. There are 1\,367\,153 sources for which \ugc provides a redshift value as estimated by t-SVM  (\secref{subsubsec:ugc_method_svm}), \linktoEGParam{galaxy_candidates}{redshift_ugc}, along with the corresponding lower and upper limits of the SVM prediction interval, \linktoEGParam{galaxy_candidates}{redshift_ugc_lower} and \linktoEGParam{galaxy_candidates}{redshift_ugc_upper}, respectively. The parameter \texttt{redshift\_ugc\_lower} is defined as \texttt{redshift\_ugc}$-\mu_{i}-\sigma_{i}$, where $i$ corresponds to the $i$th redshift range identified in the previous section, and $\mu_i$ and $\sigma_i$ are the associated bias and standard deviation computed on the base test set. Similarly, the parameter \texttt{redshift\_ugc\_upper} is defined as \texttt{redshift\_ugc}$-\mu_{i}+\sigma_{i}$. The value of $($\texttt{redshift\_ugc\_upper}$-$\texttt{redshift\_ugc\_lower}$)/2$ can therefore be used as an estimate of the 1-$\sigma$ uncertainty on \texttt{redshift\_ugc}.  

Apart from the Galactic plane, the sources with \ugc redshifts are almost uniformly distributed on the sky, as seen in \figref{fig:ugc_galsky}, although there are two strips (lower-left and upper-right) of relatively lower density  displaying residual patterns. These are regions that have been observed fewer times by Gaia and thus many of the sources in them do not appear in  the \ugc output because of the filters applied on the number of transits (see Figure~\ref{fig:dsc_number_skyplots}).

The distribution of the estimated \texttt{redshift\_ugc} values shown in the left panel of \figref{fig:ugc_redshifts_distr} has a maximum at $z\simeq$ 0.1, while almost 91\% of the redshifts are within $0.05\leq{}z<0.25$. About 7\% of the sources have redshifts larger than 0.25. The lowest and the highest redshifts reported  are $z_{min}=-0.036$ and $z_{max}=0.598$, respectively. There are 33 sources with negative redshifts, although most of these values are very close to zero (with median value of -0.0054). 

The dependence of the  \texttt{redshift\_ugc} values on \gmag magnitude is shown in the middle panel of \figref{fig:ugc_redshifts_distr}. As expected, sources with higher  redshift are fainter (e.g. $z>0.4$ sources are mostly found at $\gmag>19$ mag, while $z>0.5$ sources are found at $\gmag>20$ mag). The dependence of the estimated redshift on the source magnitude is also evident in the \bporrp magnitude--magnitude diagram shown in the right panel of \figref{fig:ugc_redshifts_distr}, where different redshift ranges are represented with different colours.  

\begin{figure*}
\centering
\includegraphics[width=0.32\textwidth]{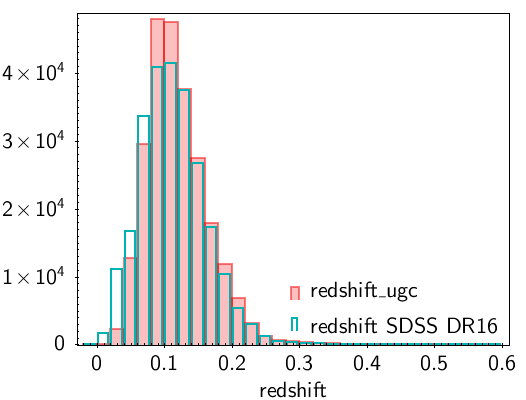}
\includegraphics[width=0.32\textwidth]{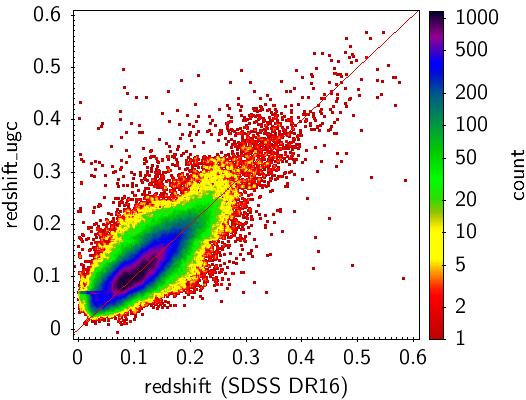} 
\includegraphics[width=0.32\textwidth]{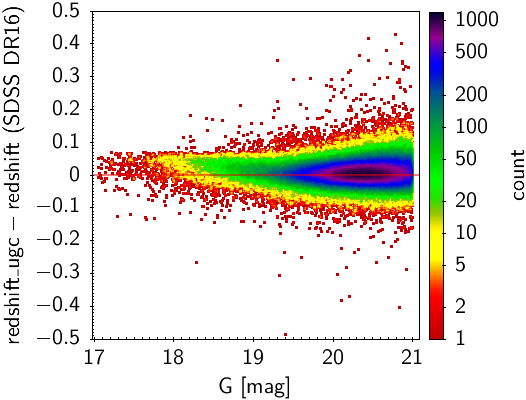}
\caption[Redshift SDSS and \ugc\ redshifts comparison]{Comparison of the \ugc estimated and the actual (SDSS DR16) redshifts for the 248\,356  sources in common (not shown are 67 sources with actual redshift greater than 0.6). Left panel: Distributions of the \ugc redshifts and SDSS DR16 redshifts indicates that \ugc tends to overestimate the small redshifts. Middle panel: Comparison of the \ugc redshifts and SDSS DR16 redshifts. The unit line is shown in red. A small horizontal branch at \texttt{redshift\_ugc}=0.07 is discussed in the text. Right panel: Differences between the \ugc and SDSS DR16 redshifts as a function of $\gmag$ magnitude. The red horizontal line designates perfect agreement.}
\label{fig:ugc_redshift_sdss_compare}
\end{figure*}

There are 248\,356 sources with published  \texttt{redshift\_ugc} in common  with those spectroscopically classified as \texttt{'GALAXY'} or \texttt{'QSO'} in the SDSS DR16 (using a radius of $0.54$\arcsec, as before). The differences between the \texttt{redshift\_ugc} and the SDSS redshifts have a mean and standard deviation of $\mu=0.006$ and $\sigma=0.054$, respectively. If the 67 sources with SDSS redshifts greater than 0.6 are excluded, the standard deviation is reduced to $0.029$. \figref{fig:ugc_redshift_sdss_compare} (left panel) compares the distributions of the two redshift estimates. There is a clear excess in the number of sources with  \ugc redshifts around 0.1 compared to the SDSS redshifts. At the same time, there is a deficit in the lower redshift bins for \ugc. The observed differences are probably due to an overestimation by \ugc of lower  SDSS redshifts. These effects are better demonstrated in  \figref{fig:ugc_redshift_sdss_compare} (middle panel). Most of the sources follow the unit line, albeit with significant scatter. However, there is a small bias which tends to be positive for $z \approx 0.1$.  

We also see in \figref{fig:ugc_redshift_sdss_compare} (middle panel) a short dense horizontal feature of sources with \texttt{redshift\_ugc} around 0.07, while the corresponding SDSS redshifts span a range of values from $\simeq$ 0 to 0.07. We see that the majority of these problematic values occur at $0.07 < $\texttt{redshift\_ugc}$ < 0.071$, with 5178 sources with  redshift values in the range 0.070822--0.070823. Detailed analysis (see the \linktosec{cu8par}{apsis}{ugc}) indicates that this peak contains a relatively large fraction of very bright sources (with $\gmag < 17.5$, $\gbp <16$ and $\grp<15$ mag), suggesting  that the SVM models, which are not trained at all for bright, nearby galaxies, tend to make constant redshift predictions for such objects.

\begin{figure}
\centering
\includegraphics[width=0.48\textwidth]{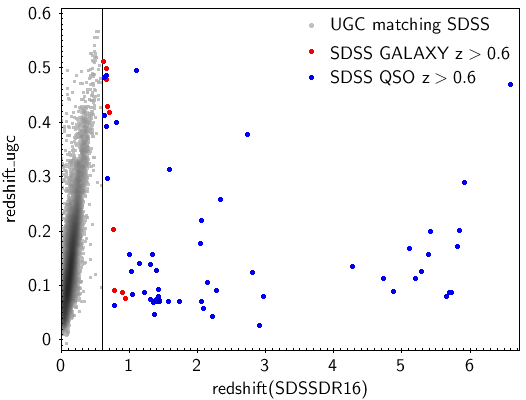}
\caption[\ugc output includes high-redshift sources]{
\ugc sources with high redshift from the SDSS DR16. Blue and red points are sources that are spectroscopically classified as \texttt{`QSO'} and \texttt{`GALAXY'} in the SDSS DR16, respectively.}
\label{fig:ugc_matched_contamin_highz}
\end{figure}

\begin{figure}
\centering
\includegraphics[width=0.48\textwidth]{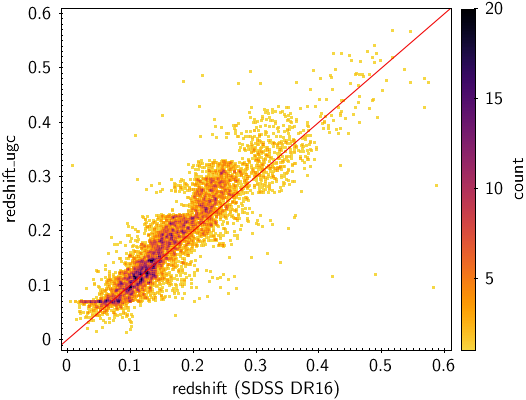} 
\caption[\ugc output includes QSOs]{
Comparison of the \ugc redshifts for sources classified as \texttt{`QSO'} in the SDSS DR16, with actual redshift lower than 0.6.}
\label{fig:ugc_matched_contamin_qso}
\end{figure}

\figref{fig:ugc_redshift_sdss_compare} (right panel) shows the difference between \texttt{redshift\_ugc} and the actual SDSS redshift, as a function of $\gmag$ magnitude. As expected, the performance of the \ugc redshift estimator is poorer for fainter sources as indicated by the larger dispersion seen at faint $\gmag$ magnitudes. The positive bias of the very bright and nearby galaxies is also clearly seen.  


\subsection{Use of \ugc results} \label{subsec:ugc_filtering}

\ugc selects sources that have a \dsc probability of being a galaxy of   \linktoAPParam{astrophysical_parameters}{classprob_dsc_combmod_galaxy} $\geq 0.25$. This is a relatively low threshold, and so the final \ugc  galaxy catalogue is expected to include some misclassified quasars. Indeed, 5170 sources, or $\simeq 2\%$ of the sources in common with the SDSS DR16, have a SDSS spectroscopic class \texttt{`QSO'} while 58 of them also have SDSS redshifts $z>0.6$, i.e. higher than the \ugc limit. There are also 9 high-redshift sources spectroscopically classified as \texttt{`GALAXY'} by the SDSS. \figref{fig:ugc_matched_contamin_highz} shows a comparison between \texttt{redshift\_ugc} and SDSS redshifts for high-redshift sources. As expected, the \ugc predictions are unreliable for these sources. However, as seen in \figref{fig:ugc_matched_contamin_qso}, the agreement between \texttt{redshift\_ugc} and SDSS redshifts of QSOs with redshifts below 0.6 is good, despite the fact that the SVM was not trained for quasars. 

The \ugc performance varies with redshift. As a consequence, redshifts larger than $0.4$ and lower than  $0.02$ are less reliable. A suspiciously large peak of sources also appears in the redshift bin $0.070 < ${\tt redshift\_ugc}$ < 0.071$, where about $17\,000$ sources are found. It is estimated that most of the sources in this peak are some of the brightest in the \ugc output and have SDSS redshifts below 0.04. About 40\% of these can be discarded by applying the previously mentioned cuts to sources with $0.070 < ${\tt redshift\_ugc}$ < 0.071$: $\gmag > 17.5$, $\gbp > 16.2,$ and $\grp > 15.0$ mag (see the \linktosec{cu8par}{apsis}{ugc} for details).

\section{Total Galactic extinction (TGE) map}
\label{sec:tge}
\subsection{Objectives} \label{subsec:tge_objective}

To support extragalactic studies, it was decided to use the extinction determinations obtained for single stars based on their astrometry and spectrophotometry \citep{DR3-DPACP-156} to estimate the total extinction from the Milky Way as a function of sky position, that is, the full cumulative foreground extinction by the Milky Way on distant extragalactic sources. 
Taking advantage of the HEALPix encoded in the \linktoMainParam{gaia_source}{source_id}, a series of HEALPix maps of the total Galactic extinction are provided 
using a selected subset of sources in each HEALPix, which are referred to as extinction tracers. 

All-sky HEALPix maps of the total Galactic extinction are delivered in two tables at various resolutions (i.e.\ HEALPix levels).  These are the tables  \linktoAPTable{total_galactic_extinction_map} and  \linktoAPTable{total_galactic_extinction_map_opt}, described below. The first of these tables contains HEALPix maps at levels 6 through 9 (corresponding to pixel sizes of 0.839 to 0.013 \sqdeg), with extinction estimates for all HEALPixes that have at least three extinction tracers, while the second map is a reduced version of this first map where a subset of the pixels is used to construct a map at variable resolution, using the smallest HEALPix available with at least ten tracers for HEALPix levels 7 through 9. 

This extinction map is the first of its kind, as reported values are based on sources beyond the interstellar medium (ISM) in the disc of the Milky Way.  This differs from previous 2D extinction maps where it is not clear to what distance the extinction is integrated to, while for extant 3D maps, not every line of sight contains tracers beyond the ISM layer of the Galactic disc. As such, it is well suited for extra-galactic studies and comparisons with line-of-sight-integrated observations such as dust emission  or diffuse gamma-ray emission.

\begin{figure*}[t]
    \centering
    \includegraphics[width=\textwidth]{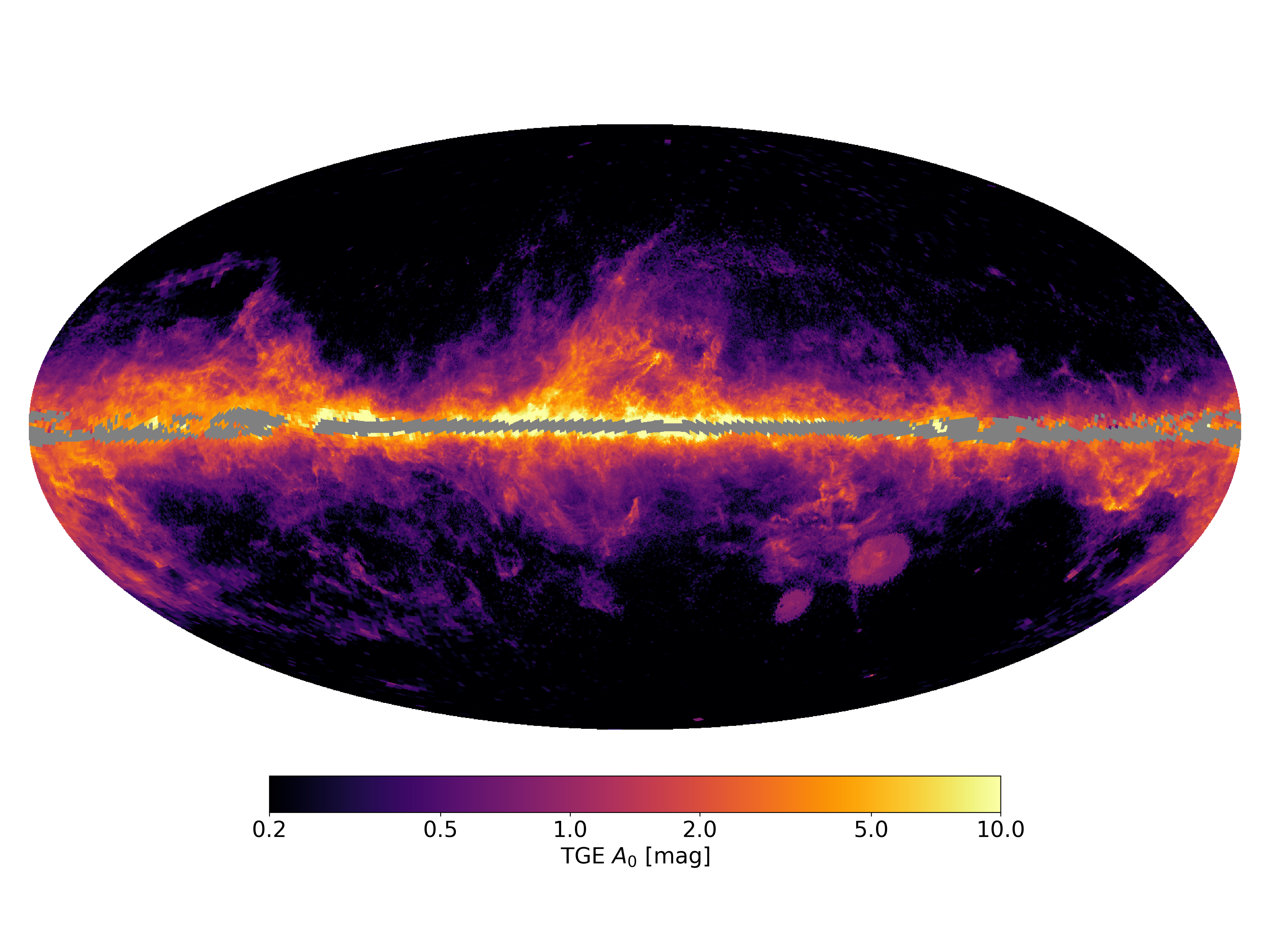}
    \caption{HEALPix map of the total Galactic extinction, built from HEALPixes between levels 6 and 9 (0.839 to 0.013 \sqdeg), which are identified as being at the optimum resolution over their field of view.}
    \label{fig:tge_opt}
\end{figure*}

\subsection{Method} \label{subsec:tge_method}

 To estimate the extinction in each HEALPix, sources that are classified as stars by DSC (i.e. sources with  \linktoMainParam{gaia_source}{classprob_dsc_combmod_star} $>0.5$; see Section \ref{sec:dsc}) and with stellar parameters consistent with being giants (as provided by the set of \gspphot APs from the `best' library from \cite{DR3-DPACP-156} and provided in the main \linktoMainTable{gaia_source} table) are used as extinction tracers.  Giant stars are used as they are intrinsically bright and numerous outside the ISM layer of the Galactic disc.  The selection of these tracers is done based on \gspphot effective temperatures (\linktoMainParam{gaia_source}{teff_gspphot}) $3000 < \teff < 5700$K, and absolute magnitudes (\linktoAPParam{astrophysical_parameters}{mg_gspphot}) $4 > M_G > -10$. Given these criteria, the extinction parameters from the \gspphot best library come from those based on either the MARCS or Phoenix spectral libraries. From an analysis of extinction estimates from two different libraries, no significant systematic trends are found when comparing the extinctions from the two libraries on a per HEALPix basis \citep{DR3-DPACP-160}.
 
 In addition, extinction tracers are required to be at least 300 pc above or below the Galactic plane ($b = 0$), or with a Galactocentric radius of $R > 16$ kpc. To establish these criteria, the distance to the source provided by \gspphot (\linktoMainParam{gaia_source}{distance_gspphot}) is used. 

Once the extinction tracers for a given HEALPix are selected, if three or more tracers are available, the median \a0 of the tracers\footnote{\a0 is the extinction parameter from the adopted Fitzpatrick extinction law \citep{1999PASP..111...63F}, defined as the monochromatic extinction at 541.4nm. See the \linktosec{cu8par}{data}{xp} for details.} ---as given by the \gspphot parameter \linktoMainParam{gaia_source}{azero_gspphot}--- is taken as the estimate of the total Galactic extinction
(\linktoAPParam{total_galactic_extinction_map}{a0})
for the HEALPix, while the uncertainty of the total Galactic extinction (\linktoAPParam{total_galactic_extinction_map}{a0_uncertainty}) is taken as the standard error of the sample mean of \a0 of the tracers. This latter is a choice of convenience, as the small number of tracers in most of the HEALPixes prevents a meaningful estimate of quantiles. Both the median and uncertainty are estimated after a 3-$\sigma$ cut about the median of the unclipped sample in order to remove outliers; this was done principally  to remove outliers that were otherwise strongly impacting our estimate of the uncertainty. HEALPixes with fewer than three tracers have no extinction value assigned to them. A diagnostic flag  \linktoAPParam{total_galactic_extinction_map}{status} is provided which is set to zero if the number of tracers is three or greater, while a non-zero value gives an indication as to why an insufficient number of tracers were found. 

The uncertainty of the TGE extinction is generally much smaller than the dispersion of the individual extinction measures of the tracers in the HEALPix, which can be dominated by intrinsic variation of extinction in the field defined by the HEALPix, especially at lower Galactic latitudes with significant extinction. To recover the standard deviation of the distribution of \a0 measures of the tracers in a HEALPix, one should multiply the given uncertainty by the square root of the number of tracers used (\linktoAPParam{total_galactic_extinction_map}{num_tracers_used}). The full range of \a0 extinction measures of the tracers  (\linktoAPParam{total_galactic_extinction_map}{a0_min}, \linktoAPParam{total_galactic_extinction_map}{a0_max}) is also provided. 


The first table, \linktoAPTable{total_galactic_extinction_map}, contains HEALPix maps at four different HEALPix levels, from level 6 (49\,152 HEALPixes with an area of 0.84 \sqdeg) to level 9 (3\,145\,728 HEALPixes with an area of 0.013 \sqdeg)
, with the HEALPix level indicated with the parameter \linktoAPParam{total_galactic_extinction_map}{healpix_level}.
This range of HEALPix levels ensures that a minimum number of tracers per HEALPix will be found at high Galactic latitudes, where the sky density of tracers is low, while allowing a higher resolution in areas of the sky where the density of tracers is high. (At level 9 only 1\% of the sky has more than 40 tracers per HEALPix.)  

For any given direction we determine the  optimum HEALPix level, that is, the set of the smallest HEALPixes with at least ten tracers to ensure a reliable estimate of the extinction and its uncertainties. However, as the base resolution is HEALPix level 6, all HEALPixes with fewer than ten tracers at this level are tagged as `optimum'. As in the level 6 map, the optimum map has full sky coverage at $|b| > 5\deg$ (i.e. all HEALPixes at $|b| > 5\deg$ have at least three tracers, so an \a0 value is reported for each of them).  In the HEALPix scheme, each HEALPix at level $n$ contains four sub-HEALPixes at level $n+1$, meaning that each of the four sub-HEALPixes must have at least ten tracers to allow all four to be tagged as optimum. This algorithm is repeated iteratively over each level, starting at the base level 6, until the lack of tracers in a sub-HEALPix prevents further subdivision, or until level 9 is reached. In the table \linktoAPTable{total_galactic_extinction_map}, the optimum HEALPixes are flagged as such with the boolean flag \linktoAPParam{total_galactic_extinction_map}{optimum_hpx_flag}. This algorithm ensures that the subset of optimum HEALPixes do not overlap with one another, yet cover the entire sky. 

The second table, \linktoAPTable{total_galactic_extinction_map_opt}, is a single optimum HEALPix map at level 9 provided for convenience, where each HEALPix adopts the extinction value of the optimum HEALPix \linktoAPTable{total_galactic_extinction_map}  coincident with or containing the HEALPix.  
That is, if a HEALPix at level 6 is tagged as optimum in \linktoAPTable{total_galactic_extinction_map}, then all 64 of its level-9 sub-HEALPixes in the  \linktoAPTable{total_galactic_extinction_map_opt} map will be assigned the \linktoAPParam{total_galactic_extinction_map}{a0} value of the level 6 HEALPix. The parameter \linktoAPParam{total_galactic_extinction_map_opt}{optimum_hpx_level} in this table indicates, for each HEALPix, the HEALPix level of the optimum HEALPix from which its \linktoAPParam{total_galactic_extinction_map_opt}{a0} value is based.

\subsection{Performance} \label{subsec:tge_performances}

\begin{figure}
    \centering
    \includegraphics[width=.48\textwidth]{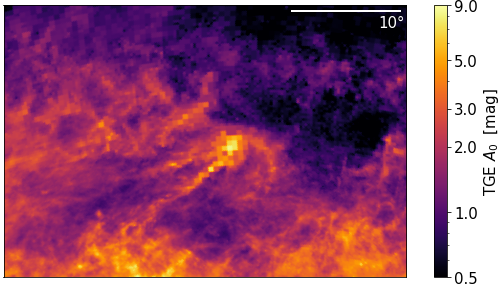}
    \caption{
    \a0 towards Rho Ophiuchi from the TGE optimum HEALPix map (Fig.\,\ref{fig:tge_opt}) centred at $(l,b)=(-5\deg,18\deg)$.  The solid white line in the upper right corner provides the angular scale of the image. The variable resolution of the optimum HEALPix map is particularly obvious towards the middle of the figure.}
    \label{fig:tge_rho-oph}
\end{figure}

At the base level 6, only 2.8\% of the sky (1379 out of 49152 HEALPixes) close to the Galactic plane (with $|b| < 5\deg$) has no \linktoAPParam{total_galactic_extinction_map}{a0} values because of an insufficient number of tracers. The fraction of HEALPixes with an insufficient number of tracers increases at the higher HEALPix levels as the HEALPixes become smaller: 5.2\% at level 7, 30.4\% at level 8, and 66.3\% at level 9. The average number of tracers for the HEALPixes with \a0 estimates is 268.3 at level 6, but only 10.7 at level 9, while the average number of tracers for the optimum HEALPix map is 30.3. The optimum HEALPix map, \linktoAPTable{total_galactic_extinction_map_opt}, shown in Figure \ref{fig:tge_opt}, has the same sky coverage as the level 6 map, but is of higher resolution when a sufficient number of tracers are available. To better demonstrate this, we show a zoom into the Rho Ophiuchi region in Figure \ref{fig:tge_rho-oph}.
Over the whole sky, only about 1\% of the HEALPixes at level 9 have more than 40 tracers, and thus the potential to be mapped at higher resolution. 
Figures showing the individual all-sky maps at levels 6 through 9 can be found in the \linktosec{cu8par}{apsis}{tge}, along with maps of the \linktoAPParam{total_galactic_extinction_map}{a0_uncertainty}. 
We note that the \linktoAPParam{total_galactic_extinction_map}{a0_uncertainty} is smallest in HEALPix level 6 with a mean value of 0.03 mag; this is due to the larger number of tracers contained in the HEALPixes at this level, whereas the mean \linktoAPParam{total_galactic_extinction_map}{a0_uncertainty} of the HEALPixes in \linktoAPTable{total_galactic_extinction_map} tagged as optimum (\linktoAPParam{total_galactic_extinction_map}{optimum_hpx_flag} $=1$) is of 0.06 mag, as they cover various HEALPix levels.
 
\begin{figure}
    \centering
    \includegraphics[width=.49\textwidth]{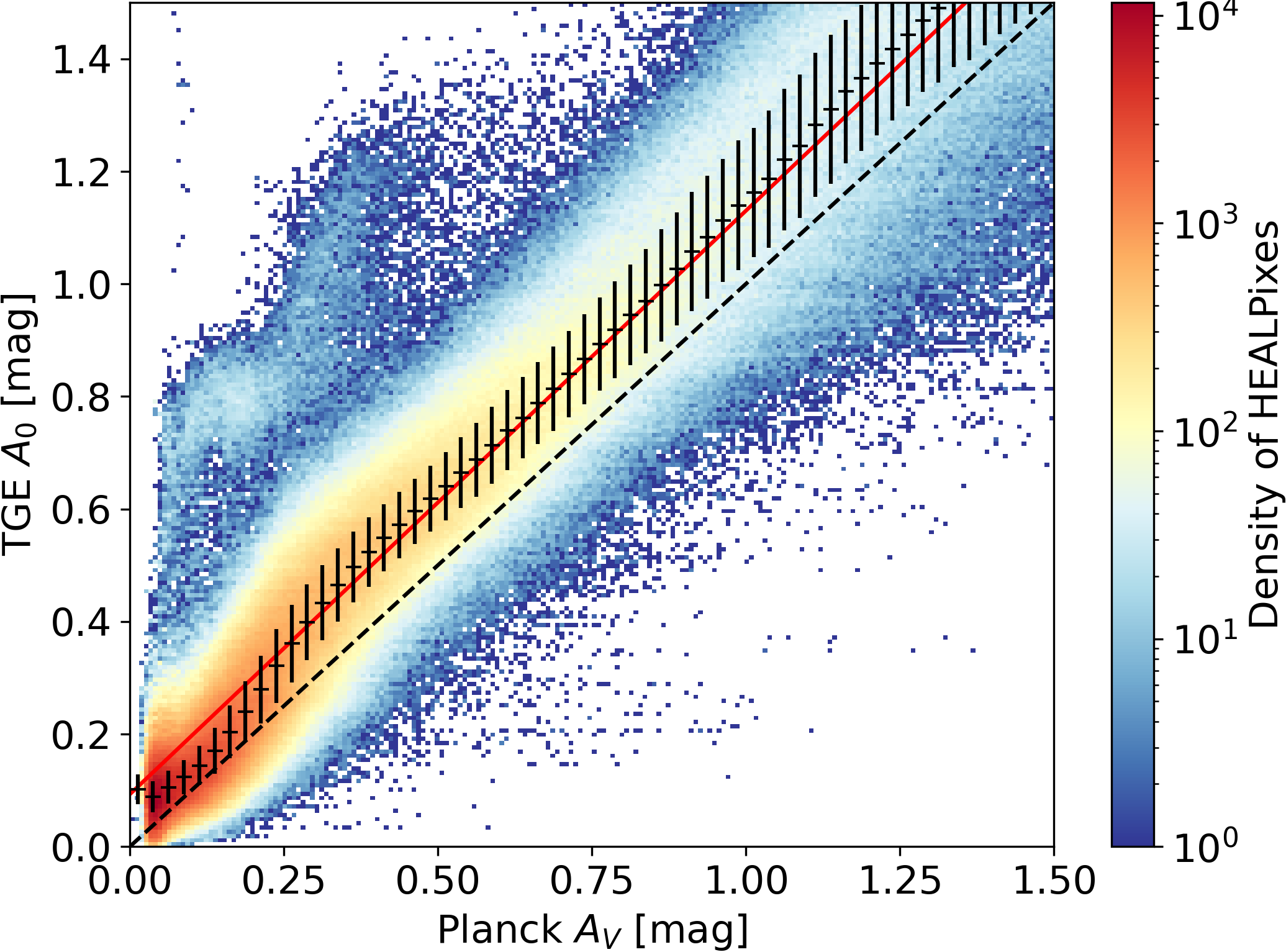}
    \caption{Extinction comparison between the TGE \a0 optimum HEALPix map and the Planck $A_V$ HEALPix level 9 map at small extinction values. The colour scale shows the density of HEALPixes, the red dashed line represents unity, and the points with error bars are the median \a0 and average absolute deviation computed in $A_V$ bins of width 0.025 mag. The red line is the result of a linear fit to the points. }
    \label{fig:tge_vs_planck}
\end{figure}

In Fig.\,\ref{fig:tge_vs_planck}, the TGE $\a0$ estimate at the optimum HEALPix level 9 is plotted against the dust optical depth expressed as $A_V$ from  \citet{2016A&A...596A.109P}\footnote{The Planck collaboration reports $E(B-V)$ that we convert to $A_V$ via $A_V=R_V E(B-V)$ and $R_V=3.1$. See the Planck Legacy Archive (\href{http://pla.esac.esa.int}{http://pla.esac.esa.int}) for details.}, once re-binned at the same HEALPix level. 
We see good agreement, as a linear fit using the median points with $0.2\le A_V \le 3$ results in a slope of $1.04 \pm 0.05$,  albeit with an offset of 0.09 $\pm 0.05$. It should be noted that the ratio of $A_V / A_0$ for giants (stars with effective temperature $3000 < \teff < 5700$K) is $\sim 0.98$ (see the \linktosec{cu8par}{data}{xp}), meaning that the slope of TGE (converted to $A_V$) over Planck($A_V$) is $1.04 \times 0.98 = 1.02$. Also worth bearing in mind is that there are a number of Planck maps of the dust distribution available on the Planck Legacy Archive; for example, using the map described in \citet{2016A&A...586A.132P} we find a slope of 0.90$\pm0.04$  and an offset of 0.05$\pm0.04$.

Performing a linear fit in the same extinction range between TGE \a0 and \citet{1998ApJ...500..525S} $A_V$ results in a slope of $0.98 \pm 0.04$  (offset: 0.10$\pm0.04$, in agreement with the $1.04 \pm 0.05$ obtained using Planck. However, the same linear fit performed between TGE and the Bayestar's map \citep{2019ApJ...887...93G} results in a slope of $1.20 \pm 0.04$ (offset: 0.01$\pm0.04$), suggesting that the Bayestar map is systematically underestimating the extinction with respect to other extinction maps; see discussion in \citet{DR3-DPACP-156}.

Towards the limit where the extinction measured by Planck tends to zero, the TGE \a0 tends to a non-zero value. This offset is found empirically by fitting a third-order polynomial to the median points for \a0$ < 0.4$ and obtaining the TGE \a0 value at Planck $A_V=0$. The resulting offset is  $0.10 \pm 0.03$ mag and starts to become evident at $A_V < 0.1$\,mag.
The existence of this offset is likely due to the fact that the \gspphot extinction prior forces its extinction estimate to be non-negative, which creates a statistical bias at very low extinction values. Indeed, this \a0 offset is of the order expected if the true uncertainty of the \a0 estimates per source were 0.1 magnitude. See \citet{DR3-DPACP-156} for further discussion.

\begin{figure}
    \centering
    \includegraphics[width=.55\textwidth]{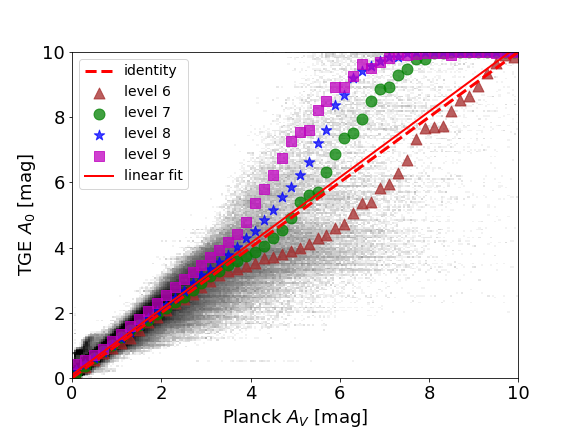}
    \caption{Comparison of the extinction between the TGE \a0 optimum HEALPix map and the Planck $A_V$ HEALPix level 9 map for extinctions up to 10 mag. 
    The background grey scale is a density plot of the entire optimal HEALPix TGE map (comprising the optimal HEALPixes at several HEALPix levels).
    The dashed red line represents unity and the solid red line is a linear fit of the medians of all HEALPixes in the optimum HEALPix map with $0.5\le A_V \le 3$. Coloured symbols refer to the median \a0 computed in $A_V$ bins of width 0.2 mag for various HEALPix levels that are used to assign the \a0 value.}
    \label{fig:tge_vs_planck_per_hpx_level}
\end{figure}
Comparing TGE \a0 to Planck $A_V$ over a larger interval 
highlights a possible bias at extinctions $A_V \ge 4$ mag.
In Fig.\ref{fig:tge_vs_planck_per_hpx_level}, TGE is plotted versus Planck over an interval of ten magnitudes. 
A large dispersion in \a0 is observed for the optimal map for $A_V > 4$ mag, and it can be seen that the different HEALPix levels do not behave in the same way. The coarser resolutions (levels 6 and 7) initially predict less extinction than Planck (for $4\le A_0 \le 5$ mag) whereas the finer resolutions either agree or predict higher extinction. Above an $A_V$ of 5 mag, only level 6 predicts less extinction than Planck, while the others predict more. Even for $A_V<4$ mag, where TGE and Planck are in very good agreement, a difference can be seen where the lower resolutions predict lower extinction. This is likely due to a selection effect where in a given HEALPix with variable extinction, more stars will be observed where the extinction is smaller. This will bias the extinction estimate for the HEALPix to lower values, and will be more obvious for larger HEALPixes. 

\begin{figure}
    \centering
    \includegraphics[width=.48\textwidth]{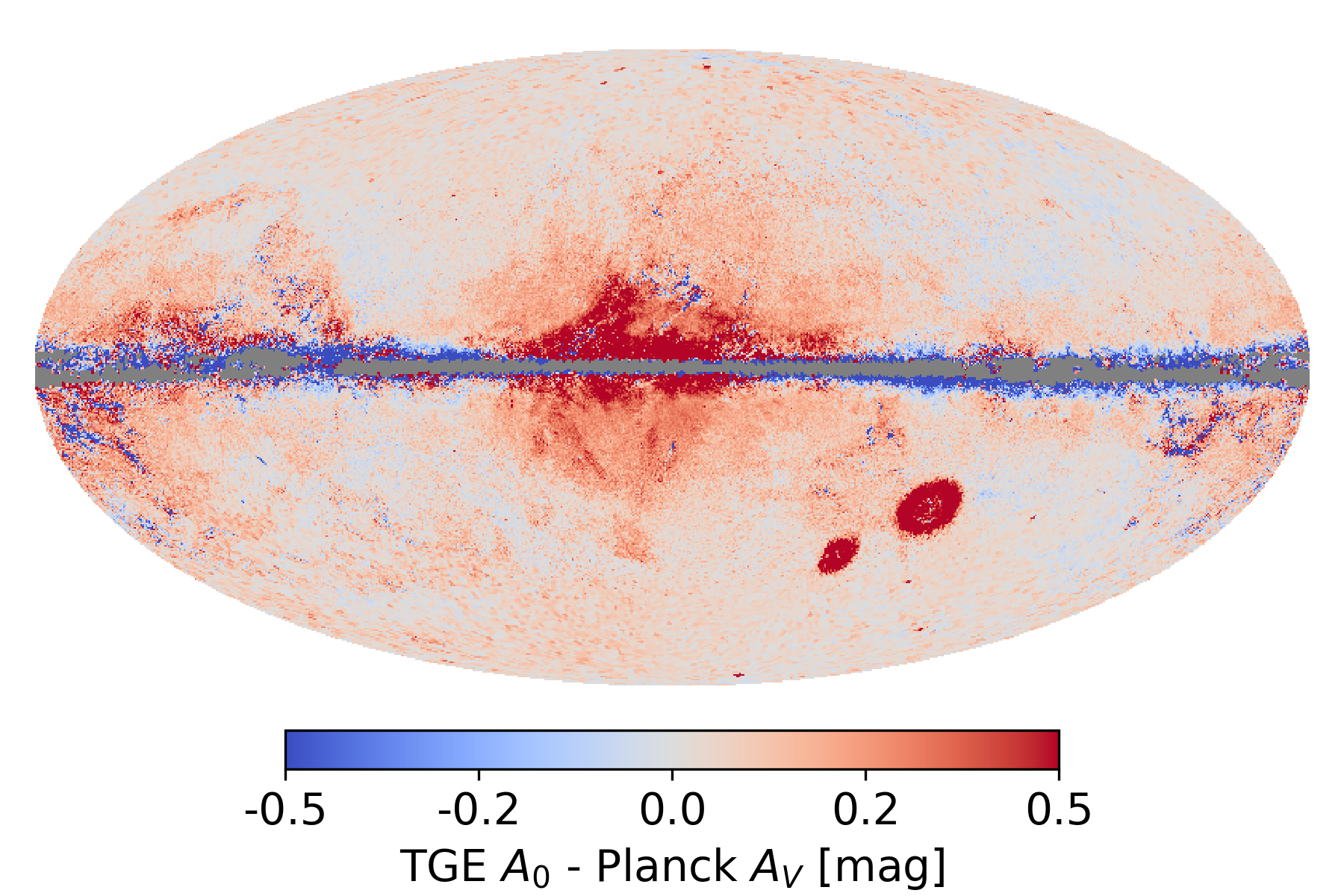}
    \caption{Residual sky map of  TGE $A_0$ minus Planck $A_V$, using the  optimum HEALPix level 9 map. Red values show regions where TGE predicts more extinction than Planck, whereas blue values show the opposite. }
    \label{fig:tge_vs_planck_map}
\end{figure}

Finally in Fig. \ref{fig:tge_vs_planck_map} the residual map of TGE $A_0$ minus Planck $A_V$ is shown. TGE underestimates extinction with respect to Planck toward molecular clouds, where dust emission remains optically thin but where TGE estimates may be biased toward smaller values as unresolved areas with below average extinction are oversampled, as mentioned above; see further discussion regarding high-extinction regions in the following section.
Meanwhile, within about 30\deg towards the Galactic centre, TGE shows more extinction than Planck, apart from the foreground molecular complexes we just mentioned. 

\subsection{Use of \tge results} \label{subsec:tge_filtering}
The \tge extinction maps estimate the total Galactic extinction \a0 from the Milky Way ISM toward extragalactic sources, where \a0 is the monochromatic extinction at 541.4nm. As mentioned above, $A_V / A_0$ is approximately equal to 0.98 for cool stars at $A_0 < 3$mag. However, in general, the effective extinction in a passband depends on the SED of the source; see the \linktosec{cu8par}{data}{xp} for a discussion on how to derive the extinction from \a0 for any passband.

\begin{figure}
    \centering
    \includegraphics[width=.48\textwidth]{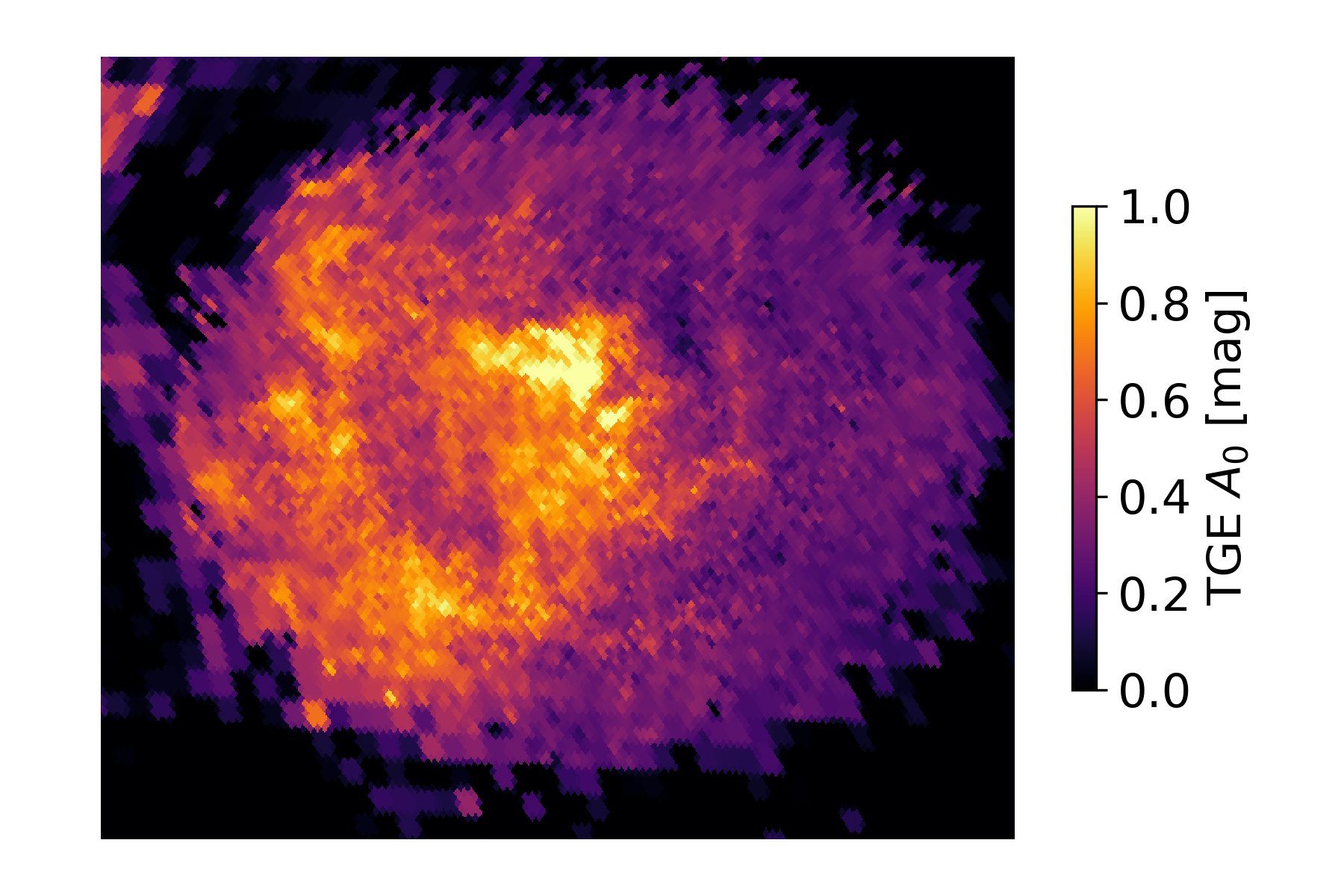}
    \caption{\a0 towards the LMC from the TGE Optimum HEALPix map (Fig.\,\ref{fig:tge_opt}), centred at $(l,b)=(280.0\deg,-33.0\deg)$. The estimated offset of \a0=0.6  mag has been subtracted.  The solid white line in the bottom left corner provides the angular scale of the image.}
    \label{fig:tge_LMC}
\end{figure}

As the selected extinction tracers were required to be beyond a certain minimum distance to ensure that they were outside the ISM layer of the Milky Way's disc, sources in nearby galaxies may also be selected as tracers. This means that the extinction towards the LMC and SMC will be a combination of Galactic extinction, inter-galactic extinction, and extinction in the Magellanic clouds (although the latter will be the dominant contribution). Another factor that will influence the amount of reported extinction in these directions stems from the distance prior used in \gspphot, which assumes that the sources are Galactic. As such, the extinction will be overestimated. An evaluation of this overestimation can be obtained via a comparison with an external data set. 
Indeed, in Fig. \ref{fig:tge_vs_planck}, there is a cloud of points with a locus stretching from around $A_V$=0.2, $\a0$=0.8 to  $A_V$=0.4, $\a0$=1.2 that consists entirely of lines of sight towards the Magellanic clouds. Comparing the median TGE $A_0$ (1.0 mag) to the median Planck $A_V$ (0.4 mag) towards the LMC reveals a difference of 0.6 mag. These values are both higher than the extinction found using near-infrared observations \citep[$A_V$ = 0.3 mag; ][]{2007ApJ...662..969I} and in the visible 
\citep[$A_V$ = 0.24 mag; ][]{2013MNRAS.431.1565W}.
This difference is likely not only due to the \gspphot distance prior, but also to variations in dust properties in the LMC/SMC. Although the absolute level of extinction in these Galactic satellites needs to be interpreted with caution, the relative variations evidencing structured patterns are most certainly real (see Fig. \ref{fig:tge_LMC}).

Because extinction tracers are required to be outside the dust layer of the Milky Way, they must be at greater distances at lower Galactic latitudes. This, together with the effect of increasing extinction and \gaia's magnitude limit, means that at very low latitudes it is not possible to find a sufficient number of tracers outside the ISM layer of the Milky Way with which to make a reliable estimate of the total Galactic extinction. This explains the band of HEALPixes at $b \approx 0$ with no extinction values. Indeed we recommend that the map should not be used for latitudes $|b| < 5\deg$. 
Also, \gspphot sets an upper limit of ten magnitudes on its estimate of \a0 per source, and so any HEALPixes with an extinction near this value should be interpreted as a lower bound. However, as suggested by figure \ref{fig:tge_vs_planck_per_hpx_level}, our maps may instead be over-estimating extinction toward these lines of sight with respect to Planck, though we point out that HEALPixes with $\a0 > 4$mag are at low Galactic latitude and make up only 2\% of the sky. Furthermore, Planck estimates towards the Galactic plane may be underestimated as a consequence of assuming a single mean dust temperature for the whole line of sight. Further details of the TGE data products are documented in the \linktosec{cu8par}{apsis}{tge}.

\section{Beyond Gaia DR3}
\label{sec:beyond_gdr3}

We present the non-stellar and classification modules from CU8 in their present status, as for \gdr{3}. However, they are in constant evolution and changes are already planned for \gdr{4} and later, which we summarise for each module in this section.

Although the intrinsic performance of \dsc is very good, once we take into account class prior ---as we do for all results shown in this paper--- the purities of the classified samples are modest. 
In preparation for \gdr{4,} we will aim to improve this, for example by optimising the feature set in Allosmod and how this is used. We will also reconsider the class definitions and the training data, in particular for white dwarfs and physical binaries.
As Specmod uses the entire \bporrp spectrum, we expected better performance (compared to Allosmod), and so we will investigate improving the classifier. We may also introduce filters to remove the classifications of the lowest quality data (which are the main determinant of the low purities).

OA will be upgraded by implementing its own outlier detector, which will be mostly based on unsupervised clustering algorithms. Additionally, we will improve the statistical description and the templates that were used for \gdr{3}. The functionality offered by the GUASOM visualisation tool will be extended in order to allow the user to perform and explore their own clustering analysis.

QSOC will use epoch \bporrp spectra re-sampled into logarithmic wavelength bins in order to overcome the issues we encountered while using the Hermite spline polynomials associated with the internal representation of the \bporrp spectra. This internal representation effectively tends to produce wiggles whose strength can be comparable to those of quasar emission lines in faint $G \geq 19$ mag spectra \citep{DR3-DPACP-157}. This solution will concurrently allow us to use sampled \bporrp spectra with uncorrelated noise on their flux, as the algorithm described in \cite{2016MNRAS.460.2811D} is not optimised to deal with full covariance matrices.

The performance of the \ugc redshift estimator strongly depends on the training set used. As more epochs are incorporated in the \bporrp spectra, we expect to have more (and generally fainter) sources with redshifts above 0.4 available for inclusion in the training set, thus improving the performance especially for higher redshifts.  We will also investigate optimisation of the SVM model parameters in order to reduce the large variability in the performance with redshift and to minimise the positive bias for bright, low-redshift objects.  

In future data releases, we can expect the \tge maps to improve with future improvements of \gspphot \citep{DR3-DPACP-156}.  In particular, we expect that the number of sources with stellar parameters will increase, which will improve the reliability of the \tge maps, and possibly allow for maps at a resolution higher than HEALPix level 9. 


\section*{Acknowledgements\label{sec:acknowl}}
\addcontentsline{toc}{chapter}{Acknowledgements}

This work presents results from the European Space Agency (ESA) space mission \gaia. \gaia\ data are being processed by the \gaia\ Data Processing and Analysis Consortium (DPAC). Funding for the DPAC is provided by national institutions, in particular the institutions participating in the \gaia\ MultiLateral Agreement (MLA). The \gaia\ mission website is \url{https://www.cosmos.esa.int/gaia}. The \gaia\ archive website is \url{https://archives.esac.esa.int/gaia}.
Acknowledgements are given in Appendix~\ref{ssec:appendixA}.

\bibliographystyle{aa}
\bibliography{dpac,bibliography}

\appendix

\section{}\label{ssec:appendixA}
This work presents results from the European Space Agency (ESA) space mission \gaia. \gaia\ data are being processed by the \gaia\ Data Processing and Analysis Consortium (DPAC). Funding for the DPAC is provided by national institutions, in particular the institutions participating in the \gaia\ MultiLateral Agreement (MLA). The \gaia\ mission website is \url{https://www.cosmos.esa.int/gaia}. The \gaia\ archive website is \url{https://archives.esac.esa.int/gaia}.

The \gaia\ mission and data processing have financially been supported by, in alphabetical order by country:
\begin{itemize}
\item the Algerian Centre de Recherche en Astronomie, Astrophysique et G\'{e}ophysique of Bouzareah Observatory;
\item the Austrian Fonds zur F\"{o}rderung der wissenschaftlichen Forschung (FWF) Hertha Firnberg Programme through grants T359, P20046, and P23737;
\item the BELgian federal Science Policy Office (BELSPO) through various PROgramme de D\'{e}veloppement d'Exp\'{e}riences scientifiques (PRODEX) grants, the Research Foundation Flanders (Fonds Wetenschappelijk Onderzoek) through grant VS.091.16N, the Fonds de la Recherche Scientifique (FNRS), and the Research Council of Katholieke Universiteit (KU) Leuven through grant C16/18/005 (Pushing AsteRoseismology to the next level with TESS, GaiA, and the Sloan DIgital Sky SurvEy -- PARADISE);
\item the Brazil-France exchange programmes Funda\c{c}\~{a}o de Amparo \`{a} Pesquisa do Estado de S\~{a}o Paulo (FAPESP) and Coordena\c{c}\~{a}o de Aperfeicoamento de Pessoal de N\'{\i}vel Superior (CAPES) - Comit\'{e} Fran\c{c}ais d'Evaluation de la Coop\'{e}ration Universitaire et Scientifique avec le Br\'{e}sil (COFECUB);
\item the Chilean Agencia Nacional de Investigaci\'{o}n y Desarrollo (ANID) through Fondo Nacional de Desarrollo Cient\'{\i}fico y Tecnol\'{o}gico (FONDECYT) Regular Project 1210992 (L.~Chemin);
\item the National Natural Science Foundation of China (NSFC) through grants 11573054, 11703065, and 12173069, the China Scholarship Council through grant 201806040200, and the Natural Science Foundation of Shanghai through grant 21ZR1474100;  
\item the Tenure Track Pilot Programme of the Croatian Science Foundation and the \'{E}cole Polytechnique F\'{e}d\'{e}rale de Lausanne and the project TTP-2018-07-1171 `Mining the Variable Sky', with the funds of the Croatian-Swiss Research Programme;
\item the Czech-Republic Ministry of Education, Youth, and Sports through grant LG 15010 and INTER-EXCELLENCE grant LTAUSA18093, and the Czech Space Office through ESA PECS contract 98058;
\item the Danish Ministry of Science;
\item the Estonian Ministry of Education and Research through grant IUT40-1;
\item the European Commission’s Sixth Framework Programme through the European Leadership in Space Astrometry (\href{https://www.cosmos.esa.int/web/gaia/elsa-rtn-programme}{ELSA}) Marie Curie Research Training Network (MRTN-CT-2006-033481), through Marie Curie project PIOF-GA-2009-255267 (Space AsteroSeismology \& RR Lyrae stars, SAS-RRL), and through a Marie Curie Transfer-of-Knowledge (ToK) fellowship (MTKD-CT-2004-014188); the European Commission's Seventh Framework Programme through grant FP7-606740 (FP7-SPACE-2013-1) for the \gaia\ European Network for Improved data User Services (\href{https://gaia.ub.edu/twiki/do/view/GENIUS/}{GENIUS}) and through grant 264895 for the \gaia\ Research for European Astronomy Training (\href{https://www.cosmos.esa.int/web/gaia/great-programme}{GREAT-ITN}) network;
\item the European Cooperation in Science and Technology (COST) through COST Action CA18104 `Revealing the Milky Way with \gaia\ (MW-\gaia)';
\item the European Research Council (ERC) through grants 320360, 647208, and 834148 and through the European Union’s Horizon 2020 research and innovation and excellent science programmes through Marie Sk{\l}odowska-Curie grant 745617 (Our Galaxy at full HD -- Gal-HD) and 895174 (The build-up and fate of self-gravitating systems in the Universe) as well as grants 687378 (Small Bodies: Near and Far), 682115 (Using the Magellanic Clouds to Understand the Interaction of Galaxies), 695099 (A sub-percent distance scale from binaries and Cepheids -- CepBin), 716155 (Structured ACCREtion Disks -- SACCRED), 951549 (Sub-percent calibration of the extragalactic distance scale in the era of big surveys -- UniverScale), and 101004214 (Innovative Scientific Data Exploration and Exploitation Applications for Space Sciences -- EXPLORE);
\item the European Science Foundation (ESF), in the framework of the \gaia\ Research for European Astronomy Training Research Network Programme (\href{https://www.cosmos.esa.int/web/gaia/great-programme}{GREAT-ESF});
\item the European Space Agency (ESA) in the framework of the \gaia\ project, through the Plan for European Cooperating States (PECS) programme through contracts C98090 and 4000106398/12/NL/KML for Hungary, through contract 4000115263/15/NL/IB for Germany, and through PROgramme de D\'{e}veloppement d'Exp\'{e}riences scientifiques (PRODEX) grant 4000127986 for Slovenia;  
\item the Academy of Finland through grants 299543, 307157, 325805, 328654, 336546, and 345115 and the Magnus Ehrnrooth Foundation;
\item the French Centre National d’\'{E}tudes Spatiales (CNES), the Agence Nationale de la Recherche (ANR) through grant ANR-10-IDEX-0001-02 for the `Investissements d'avenir' programme, through grant ANR-15-CE31-0007 for project `Modelling the Milky Way in the \gaia\ era’ (MOD4\gaia), through grant ANR-14-CE33-0014-01 for project `The Milky Way disc formation in the \gaia\ era’ (ARCHEOGAL), through grant ANR-15-CE31-0012-01 for project `Unlocking the potential of Cepheids as primary distance calibrators’ (UnlockCepheids), through grant ANR-19-CE31-0017 for project `Secular evolution of galaxies' (SEGAL), and through grant ANR-18-CE31-0006 for project `Galactic Dark Matter' (GaDaMa), the Centre National de la Recherche Scientifique (CNRS) and its SNO \gaia\ of the Institut des Sciences de l’Univers (INSU), its Programmes Nationaux: Cosmologie et Galaxies (PNCG), Gravitation R\'{e}f\'{e}rences Astronomie M\'{e}trologie (PNGRAM), Plan\'{e}tologie (PNP), Physique et Chimie du Milieu Interstellaire (PCMI), and Physique Stellaire (PNPS), the `Action F\'{e}d\'{e}ratrice \gaia' of the Observatoire de Paris, the R\'{e}gion de Franche-Comt\'{e}, the Institut National Polytechnique (INP) and the Institut National de Physique nucl\'{e}aire et de Physique des Particules (IN2P3) co-funded by CNES;
\item the German Aerospace Agency (Deutsches Zentrum f\"{u}r Luft- und Raumfahrt e.V., DLR) through grants 50QG0501, 50QG0601, 50QG0602, 50QG0701, 50QG0901, 50QG1001, 50QG1101, 50\-QG1401, 50QG1402, 50QG1403, 50QG1404, 50QG1904, 50QG2101, 50QG2102, and 50QG2202, and the Centre for Information Services and High Performance Computing (ZIH) at the Technische Universit\"{a}t Dresden for generous allocations of computer time;
\item the Hungarian Academy of Sciences through the Lend\"{u}let Programme grants LP2014-17 and LP2018-7 and the Hungarian National Research, Development, and Innovation Office (NKFIH) through grant KKP-137523 (`SeismoLab');
\item the Science Foundation Ireland (SFI) through a Royal Society - SFI University Research Fellowship (M.~Fraser);
\item the Israel Ministry of Science and Technology through grant 3-18143 and the Tel Aviv University Center for Artificial Intelligence and Data Science (TAD) through a grant;
\item the Agenzia Spaziale Italiana (ASI) through contracts I/037/08/0, I/058/10/0, 2014-025-R.0, 2014-025-R.1.2015, and 2018-24-HH.0 to the Italian Istituto Nazionale di Astrofisica (INAF), contract 2014-049-R.0/1/2 to INAF for the Space Science Data Centre (SSDC, formerly known as the ASI Science Data Center, ASDC), contracts I/008/10/0, 2013/030/I.0, 2013-030-I.0.1-2015, and 2016-17-I.0 to the Aerospace Logistics Technology Engineering Company (ALTEC S.p.A.), INAF, and the Italian Ministry of Education, University, and Research (Ministero dell'Istruzione, dell'Universit\`{a} e della Ricerca) through the Premiale project `MIning The Cosmos Big Data and Innovative Italian Technology for Frontier Astrophysics and Cosmology' (MITiC);
\item the Netherlands Organisation for Scientific Research (NWO) through grant NWO-M-614.061.414, through a VICI grant (A.~Helmi), and through a Spinoza prize (A.~Helmi), and the Netherlands Research School for Astronomy (NOVA);
\item the Polish National Science Centre through HARMONIA grant 2018/30/M/ST9/00311 and DAINA grant 2017/27/L/ST9/03221 and the Ministry of Science and Higher Education (MNiSW) through grant DIR/WK/2018/12;
\item the Portuguese Funda\c{c}\~{a}o para a Ci\^{e}ncia e a Tecnologia (FCT) through national funds, grants SFRH/\-BD/128840/2017 and PTDC/FIS-AST/30389/2017, and work contract DL 57/2016/CP1364/CT0006, the Fundo Europeu de Desenvolvimento Regional (FEDER) through grant POCI-01-0145-FEDER-030389 and its Programa Operacional Competitividade e Internacionaliza\c{c}\~{a}o (COMPETE2020) through grants UIDB/04434/2020 and UIDP/04434/2020, and the Strategic Programme UIDB/\-00099/2020 for the Centro de Astrof\'{\i}sica e Gravita\c{c}\~{a}o (CENTRA);  
\item the Slovenian Research Agency through grant P1-0188;
\item the Spanish Ministry of Economy (MINECO/FEDER, UE), the Spanish Ministry of Science and Innovation (MICIN), the Spanish Ministry of Education, Culture, and Sports, and the Spanish Government through grants BES-2016-078499, BES-2017-083126, BES-C-2017-0085, ESP2016-80079-C2-1-R, ESP2016-80079-C2-2-R, FPU16/03827, PDC2021-121059-C22, RTI2018-095076-B-C22, and TIN2015-65316-P (`Computaci\'{o}n de Altas Prestaciones VII'), the Juan de la Cierva Incorporaci\'{o}n Programme (FJCI-2015-2671 and IJC2019-04862-I for F.~Anders), the Severo Ochoa Centre of Excellence Programme (SEV2015-0493), and MICIN/AEI/10.13039/501100011033 (and the European Union through European Regional Development Fund `A way of making Europe') through grant RTI2018-095076-B-C21, the Institute of Cosmos Sciences University of Barcelona (ICCUB, Unidad de Excelencia `Mar\'{\i}a de Maeztu’) through grant CEX2019-000918-M, the University of Barcelona's official doctoral programme for the development of an R+D+i project through an Ajuts de Personal Investigador en Formaci\'{o} (APIF) grant, the Spanish Virtual Observatory through project AyA2017-84089, the Galician Regional Government, Xunta de Galicia, through grants ED431B-2021/36, ED481A-2019/155, and ED481A-2021/296, the Centro de Investigaci\'{o}n en Tecnolog\'{\i}as de la Informaci\'{o}n y las Comunicaciones (CITIC), funded by the Xunta de Galicia and the European Union (European Regional Development Fund -- Galicia 2014-2020 Programme), through grant ED431G-2019/01, the Red Espa\~{n}ola de Supercomputaci\'{o}n (RES) computer resources at MareNostrum, the Barcelona Supercomputing Centre - Centro Nacional de Supercomputaci\'{o}n (BSC-CNS) through activities AECT-2017-2-0002, AECT-2017-3-0006, AECT-2018-1-0017, AECT-2018-2-0013, AECT-2018-3-0011, AECT-2019-1-0010, AECT-2019-2-0014, AECT-2019-3-0003, AECT-2020-1-0004, and DATA-2020-1-0010, the Departament d'Innovaci\'{o}, Universitats i Empresa de la Generalitat de Catalunya through grant 2014-SGR-1051 for project `Models de Programaci\'{o} i Entorns d'Execuci\'{o} Parallels' (MPEXPAR), and Ramon y Cajal Fellowship RYC2018-025968-I funded by MICIN/AEI/10.13039/501100011033 and the European Science Foundation (`Investing in your future');
\item the Swedish National Space Agency (SNSA/Rymdstyrelsen);
\item the Swiss State Secretariat for Education, Research, and Innovation through the Swiss Activit\'{e}s Nationales Compl\'{e}mentaires and the Swiss National Science Foundation through an Eccellenza Professorial Fellowship (award PCEFP2\_194638 for R.~Anderson);
\item the United Kingdom Particle Physics and Astronomy Research Council (PPARC), the United Kingdom Science and Technology Facilities Council (STFC), and the United Kingdom Space Agency (UKSA) through the following grants to the University of Bristol, the University of Cambridge, the University of Edinburgh, the University of Leicester, the Mullard Space Sciences Laboratory of University College London, and the United Kingdom Rutherford Appleton Laboratory (RAL): PP/D006511/1, PP/D006546/1, PP/D006570/1, ST/I000852/1, ST/J005045/1, ST/K00056X/1, ST/\-K000209/1, ST/K000756/1, ST/L006561/1, ST/N000595/1, ST/N000641/1, ST/N000978/1, ST/\-N001117/1, ST/S000089/1, ST/S000976/1, ST/S000984/1, ST/S001123/1, ST/S001948/1, ST/\-S001980/1, ST/S002103/1, ST/V000969/1, ST/W002469/1, ST/W002493/1, ST/W002671/1, ST/W002809/1, and EP/V520342/1.
\end{itemize}

The \gaia\ project and data processing have made use of:
\begin{itemize}
\item the Set of Identifications, Measurements, and Bibliography for Astronomical Data \citep[SIMBAD,][]{2000AAS..143....9W}, the `Aladin sky atlas' \citep{2000A&AS..143...33B,2014ASPC..485..277B}, and the VizieR catalogue access tool \citep{2000A&AS..143...23O}, all operated at the Centre de Donn\'{e}es astronomiques de Strasbourg (\href{http://cds.u-strasbg.fr/}{CDS});
\item the National Aeronautics and Space Administration (NASA) Astrophysics Data System (\href{http://adsabs.harvard.edu/abstract_service.html}{ADS});
\item the SPace ENVironment Information System (SPENVIS), initiated by the Space Environment and Effects Section (TEC-EES) of ESA and developed by the Belgian Institute for Space Aeronomy (BIRA-IASB) under ESA contract through ESA’s General Support Technologies Programme (GSTP), administered by the BELgian federal Science Policy Office (BELSPO);
\item the software products \href{http://www.starlink.ac.uk/topcat/}{TOPCAT}, \href{http://www.starlink.ac.uk/stil}{STIL}, and \href{http://www.starlink.ac.uk/stilts}{STILTS} \citep{2005ASPC..347...29T,2006ASPC..351..666T};
\item Matplotlib \citep{Hunter:2007};
\item IPython \citep{PER-GRA:2007};  
\item Astropy, a community-developed core Python package for Astronomy \citep{2018AJ....156..123A};
\item R \citep{RManual};
\item the HEALPix package \citep[][\url{http://healpix.sourceforge.net/}]{2005ApJ...622..759G};
\item Vaex \citep{2018A&A...618A..13B};
\item the \hip-2 catalogue \citep{2007A&A...474..653V}. The \hip and \tyc catalogues were constructed under the responsibility of large scientific teams collaborating with ESA. The Consortia Leaders were Lennart Lindegren (Lund, Sweden: NDAC) and Jean Kovalevsky (Grasse, France: FAST), together responsible for the \hip Catalogue; Erik H{\o}g (Copenhagen, Denmark: TDAC) responsible for the \tyc Catalogue; and Catherine Turon (Meudon, France: INCA) responsible for the \hip Input Catalogue (HIC);  
\item the \tyctwo catalogue \citep{2000A&A...355L..27H}, the construction of which was supported by the Velux Foundation of 1981 and the Danish Space Board;
\item The Tycho double star catalogue \citep[TDSC,][]{2002A&A...384..180F}, based on observations made with the ESA \hip astrometry satellite, as supported by the Danish Space Board and the United States Naval Observatory through their double-star programme;
\item data products from the Two Micron All Sky Survey \citep[2MASS,][]{2006AJ....131.1163S}, which is a joint project of the University of Massachusetts and the Infrared Processing and Analysis Center (IPAC) / California Institute of Technology, funded by the National Aeronautics and Space Administration (NASA) and the National Science Foundation (NSF) of the USA;
\item the ninth data release of the AAVSO Photometric All-Sky Survey (\href{https://www.aavso.org/apass}{APASS}, \citealt{apass9}), funded by the Robert Martin Ayers Sciences Fund;
\item the first data release of the Pan-STARRS survey \citep{panstarrs1,panstarrs1b,panstarrs1c,panstarrs1d,panstarrs1e,panstarrs1f}. The Pan-STARRS1 Surveys (PS1) and the PS1 public science archive have been made possible through contributions by the Institute for Astronomy, the University of Hawaii, the Pan-STARRS Project Office, the Max-Planck Society and its participating institutes, the Max Planck Institute for Astronomy, Heidelberg and the Max Planck Institute for Extraterrestrial Physics, Garching, The Johns Hopkins University, Durham University, the University of Edinburgh, the Queen's University Belfast, the Harvard-Smithsonian Center for Astrophysics, the Las Cumbres Observatory Global Telescope Network Incorporated, the National Central University of Taiwan, the Space Telescope Science Institute, the National Aeronautics and Space Administration (NASA) through grant NNX08AR22G issued through the Planetary Science Division of the NASA Science Mission Directorate, the National Science Foundation through grant AST-1238877, the University of Maryland, Eotvos Lorand University (ELTE), the Los Alamos National Laboratory, and the Gordon and Betty Moore Foundation;
\item the second release of the Guide Star Catalogue \citep[GSC2.3,][]{2008AJ....136..735L}. The Guide Star Catalogue II is a joint project of the Space Telescope Science Institute (STScI) and the Osservatorio Astrofisico di Torino (OATo). STScI is operated by the Association of Universities for Research in Astronomy (AURA), for the National Aeronautics and Space Administration (NASA) under contract NAS5-26555. OATo is operated by the Italian National Institute for Astrophysics (INAF). Additional support was provided by the European Southern Observatory (ESO), the Space Telescope European Coordinating Facility (STECF), the International GEMINI project, and the European Space Agency (ESA) Astrophysics Division (nowadays SCI-S);
\item the eXtended, Large (XL) version of the catalogue of Positions and Proper Motions \citep[PPM-XL,][]{2010AJ....139.2440R};
\item data products from the Wide-field Infrared Survey Explorer (WISE), which is a joint project of the University of California, Los Angeles, and the Jet Propulsion Laboratory/California Institute of Technology, and NEOWISE, which is a project of the Jet Propulsion Laboratory/California Institute of Technology. WISE and NEOWISE are funded by the National Aeronautics and Space Administration (NASA);
\item the first data release of the United States Naval Observatory (USNO) Robotic Astrometric Telescope \citep[URAT-1,][]{urat1};
\item the fourth data release of the United States Naval Observatory (USNO) CCD Astrograph Catalogue \citep[UCAC-4,][]{2013AJ....145...44Z};
\item the sixth and final data release of the Radial Velocity Experiment \citep[RAVE DR6,][]{2020AJ....160...83S,rave6a}. Funding for RAVE has been provided by the Leibniz Institute for Astrophysics Potsdam (AIP), the Australian Astronomical Observatory, the Australian National University, the Australian Research Council, the French National Research Agency, the German Research Foundation (SPP 1177 and SFB 881), the European Research Council (ERC-StG 240271 Galactica), the Istituto Nazionale di Astrofisica at Padova, the Johns Hopkins University, the National Science Foundation of the USA (AST-0908326), the W.M.\ Keck foundation, the Macquarie University, the Netherlands Research School for Astronomy, the Natural Sciences and Engineering Research Council of Canada, the Slovenian Research Agency, the Swiss National Science Foundation, the Science \& Technology Facilities Council of the UK, Opticon, Strasbourg Observatory, and the Universities of Basel, Groningen, Heidelberg, and Sydney. The RAVE website is at \url{https://www.rave-survey.org/};
\item the first data release of the Large sky Area Multi-Object Fibre Spectroscopic Telescope \citep[LAMOST DR1,][]{LamostDR1};
\item the K2 Ecliptic Plane Input Catalogue \citep[EPIC,][]{epic-2016ApJS..224....2H};
\item the ninth data release of the Sloan Digitial Sky Survey \citep[SDSS DR9,][]{SDSS9}. Funding for SDSS-III has been provided by the Alfred P. Sloan Foundation, the Participating Institutions, the National Science Foundation, and the United States Department of Energy Office of Science. The SDSS-III website is \url{http://www.sdss3.org/}. SDSS-III is managed by the Astrophysical Research Consortium for the Participating Institutions of the SDSS-III Collaboration including the University of Arizona, the Brazilian Participation Group, Brookhaven National Laboratory, Carnegie Mellon University, University of Florida, the French Participation Group, the German Participation Group, Harvard University, the Instituto de Astrof\'{\i}sica de Canarias, the Michigan State/Notre Dame/JINA Participation Group, Johns Hopkins University, Lawrence Berkeley National Laboratory, Max Planck Institute for Astrophysics, Max Planck Institute for Extraterrestrial Physics, New Mexico State University, New York University, Ohio State University, Pennsylvania State University, University of Portsmouth, Princeton University, the Spanish Participation Group, University of Tokyo, University of Utah, Vanderbilt University, University of Virginia, University of Washington, and Yale University;
\item the thirteenth release of the Sloan Digital Sky Survey \citep[SDSS DR13,][]{2017ApJS..233...25A}. Funding for SDSS-IV has been provided by the Alfred P. Sloan Foundation, the United States Department of Energy Office of Science, and the Participating Institutions. SDSS-IV acknowledges support and resources from the Center for High-Performance Computing at the University of Utah. The SDSS web site is \url{https://www.sdss.org/}. SDSS-IV is managed by the Astrophysical Research Consortium for the Participating Institutions of the SDSS Collaboration including the Brazilian Participation Group, the Carnegie Institution for Science, Carnegie Mellon University, the Chilean Participation Group, the French Participation Group, Harvard-Smithsonian Center for Astrophysics, Instituto de Astrof\'isica de Canarias, The Johns Hopkins University, Kavli Institute for the Physics and Mathematics of the Universe (IPMU) / University of Tokyo, the Korean Participation Group, Lawrence Berkeley National Laboratory, Leibniz Institut f\"ur Astrophysik Potsdam (AIP),  Max-Planck-Institut f\"ur Astronomie (MPIA Heidelberg), Max-Planck-Institut f\"ur Astrophysik (MPA Garching), Max-Planck-Institut f\"ur Extraterrestrische Physik (MPE), National Astronomical Observatories of China, New Mexico State University, New York University, University of Notre Dame, Observat\'ario Nacional / MCTI, The Ohio State University, Pennsylvania State University, Shanghai Astronomical Observatory, United Kingdom Participation Group, Universidad Nacional Aut\'onoma de M\'{e}xico, University of Arizona, University of Colorado Boulder, University of Oxford, University of Portsmouth, University of Utah, University of Virginia, University of Washington, University of Wisconsin, Vanderbilt University, and Yale University;
\item the second release of the SkyMapper catalogue \citep[SkyMapper DR2,][Digital Object Identifier 10.25914/5ce60d31ce759]{2019PASA...36...33O}. The national facility capability for SkyMapper has been funded through grant LE130100104 from the Australian Research Council (ARC) Linkage Infrastructure, Equipment, and Facilities (LIEF) programme, awarded to the University of Sydney, the Australian National University, Swinburne University of Technology, the University of Queensland, the University of Western Australia, the University of Melbourne, Curtin University of Technology, Monash University, and the Australian Astronomical Observatory. SkyMapper is owned and operated by The Australian National University's Research School of Astronomy and Astrophysics. The survey data were processed and provided by the SkyMapper Team at the Australian National University. The SkyMapper node of the All-Sky Virtual Observatory (ASVO) is hosted at the National Computational Infrastructure (NCI). Development and support the SkyMapper node of the ASVO has been funded in part by Astronomy Australia Limited (AAL) and the Australian Government through the Commonwealth's Education Investment Fund (EIF) and National Collaborative Research Infrastructure Strategy (NCRIS), particularly the National eResearch Collaboration Tools and Resources (NeCTAR) and the Australian National Data Service Projects (ANDS);
\item the \gaia-ESO Public Spectroscopic Survey \citep[GES,][]{GES_final_release_paper_1,GES_final_release_paper_2}. The \gaia-ESO Survey is based on data products from observations made with ESO Telescopes at the La Silla Paranal Observatory under programme ID 188.B-3002. Public data releases are available through the \href{https://www.gaia-eso.eu/data-products/public-data-releases}{ESO Science Portal}. The project has received funding from the Leverhulme Trust (project RPG-2012-541), the European Research Council (project ERC-2012-AdG 320360-\gaia-ESO-MW), and the Istituto Nazionale di Astrofisica, INAF (2012: CRA 1.05.01.09.16; 2013: CRA 1.05.06.02.07).
\end{itemize}

The GBOT programme uses observations collected at (i) the European Organisation for Astronomical Research in the Southern Hemisphere (ESO) with the VLT Survey Telescope (VST), under ESO programmes
092.B-0165,
093.B-0236,
094.B-0181,
095.B-0046,
096.B-0162,
097.B-0304,
098.B-0030,
099.B-0034,
0100.B-0131,
0101.B-0156,
0102.B-0174, and
0103.B-0165;
%
%
and (ii) the Liverpool Telescope, which is operated on the island of La Palma by Liverpool John Moores University in the Spanish Observatorio del Roque de los Muchachos of the Instituto de Astrof\'{\i}sica de Canarias with financial support from the United Kingdom Science and Technology Facilities Council, and (iii) telescopes of the Las Cumbres Observatory Global Telescope Network.

In case of errors or omissions, please contact the \href{https://www.cosmos.esa.int/web/gaia/gaia-helpdesk}{\gaia\ Helpdesk}.

\section{Combining probabilities for DSC-Combmod}\label{app:combmod_definition}

Combmod in DSC combines the posterior probabilities from Specmod and Allosmod into a new posterior probability, taking care to ensure that the global prior is only counted once. If Specmod and Allosmod used the same classes, and operated on independent data, then combining their probabilities would be simple.  
However, Specmod has three classes (star, white dwarf, physical binary star) that correspond to the single star class in Allosmod. It is also possible that Specmod or Allosmod provides no result.
The combination method is therefore a bit more complicated. The basic idea is that a fraction of the Allosmod probability for the single `superclass' is taken to correspond to each subclass in Specmod, with that fraction equal to the prior. We assume that Specmod and Allosmod are independent, which is not quite true as the colours in Allosmod are derived from the \bporrp spectra used by Specmod.

\begin{itemize}[leftmargin=0.5cm]
\item Let \prob{m}{k} be the posterior probability from classifier $m$ for class $k$. 
\item Let \prior{m}{k} be the prior probability used in classifier $m$ for class $k$. 
\item For Specmod, $m=s$ and $k=1\ldots 5$ corresponding to quasar, galaxy, star, white dwarf, physical binary star respectively. 
\item For Allosmod, $m=a$ and $k=1\ldots 3$ corresponding to quasar, galaxy, star, respectively.
\item For each classifier, classes are disjoint and exhaustive, so the probabilities sum to one.
\item The priors for the two classifiers are consistent, so $\prior{a}{1}=\prior{s}{1}$,  $\prior{a}{2}=\prior{s}{2}$, and $\prior{a}{3}=\sum_{k=3}^5 \prior{s}{k}$.
\end{itemize}

For the classes that correspond one-to-one, the combined posterior probability is obtained by multiplying the likelihoods (the posterior divided by the prior, to within a normalisation factor) and then multiplying by the prior. This is
\begin{equation}
\prob{c}{k} \,=\, a\,
\frac{\prob{s}{k}}{\prior{s}{k}}
\frac{\prob{a}{k}}{\prior{a}{k}}
\prior{a}{k} \,=\, a\,
\prob{s}{k}\prob{a}{k}\,\frac{1}{\prior{s}{k}}
\hspace{1em} k \in \{1,2\}
,\end{equation}
where $a$ is a data-dependent but class-independent normalisation factor.
For each of the three stellar classes in Specmod, we assume that a fraction $\prior{s}{k}/\prior{a}{3}$ for $k \in \{3,4,5\}$ of the posterior probability $\prob{a}{3}$ is the Allosmod posterior probability for that class. Thus the combined probability for each of these three classes is
\begin{equation}
\prob{c}{k} = a\, 
\frac{\prob{s}{k}}{\prior{s}{k}}
\frac{\prob{a}{3}}{\prior{a}{3}}
\frac{\prior{s}{k}}{\prior{a}{3}}
\prior{s}{k} \,=\, a\,
\prob{s}{k}\prob{a}{3}\,
\frac{\prior{s}{k}}{(\prior{a}{3})^2}
\hspace{1em} k \in \{3,4,5\} \ .
\end{equation}
If Specmod probabilities are not available (missing), the combined posterior probability 
for the classes that correspond one-to-one is equal to the Allosmod probabilities:
\begin{equation}
\prob{c}{k} \,=\, \prob{a}{k}
\hspace{1em} k \in \{1,2\} \hspace{1em}\text{(no Specmod results)} \ .
\end{equation}
For the three stellar classes, we distribute the corresponding Allosmod probability to these classes in proportion to the priors, i.e.\
\begin{equation}
\prob{c}{k} = 
\prob{a}{3}\,\frac{\prior{s}{k}}{\sum_{k=3}^{k=5}\prior{s}{k}} 
\hspace{1em} k \in \{3,4,5\} \hspace{1em}\text{(no Specmod probabilities)} \ .
\end{equation}
If Allosmod probabilities are not available, we simply copy the Specmod probabilities:
\begin{equation}
\prob{c}{k} \,=\, \prob{s}{k}
\hspace{1em} k \in \{1,2,3,4,5\} \hspace{1em}\text{(no Allosmod probabilities)} \ .
\end{equation}
If neither the Specmod nor the Allosmod probabilities are available, the Combmod probabilities will be empty.

The above equations run the risk of divide by zero if probabilities are exactly zero. To avoid this we `soften' the Specmod and Allosmod probabilities prior to combination by adding $10^{-8}$. This is only done in the combination: the Specmod and Allosmod probabilities written to the catalogue are not modified.

The above probability combination is not complicated conceptually, but it can lead to counter-intuitive results. \cite{LL:CBJ-094} works through various examples to demonstrate and explain this.

\section{Adjusting the DSC probabilities to accommodate a new prior}\label{sec:cu8par_apsis_dsc_adjusting_probabilities}

All \dsc probabilities are posterior probabilities that have taken into account the class priors listed in \tabref{tab:cu8par_apsis_dsc_classprior}. Posteriors are equal to the product of a likelihood and a prior that has then been normalized.
It is therefore simple to adjust the \dsc probabilities to reflect a different prior probability: we simply divide each output by the prior used (to strip this off), multiply by the new prior, and then normalise the resulting probability vector. That is, if
\prob{d}{k} is the \dsc probability in the catalogue (for any of its classifiers) for class $k$, and if \prior{d}{k} is the corresponding catalogue prior (\tabref{tab:cu8par_apsis_dsc_classprior}), then the new posterior probabilities corresponding to a new prior \prior{{\rm new}}{k} are
\begin{equation}
\frac{\prob{d}{k}}{\prior{d}{k}}\prior{{\rm new}}{k} \bigg/ 
\sum_{{k'}}\left( \frac{\prob{d}{{k'}}}{\prior{d}{{k'}}}\prior{{\rm new}}{{k'}} \right) .
\end{equation}

\end{document}